\documentclass[english,prl,twocolumn,superscriptaddress]{revtex4-1}
\usepackage[T1]{fontenc}
\usepackage{amsmath}
\usepackage{color}
\usepackage{rotating}
\usepackage{textcomp}
\usepackage{multirow}
\usepackage{amsbsy}
\usepackage{amssymb}
\usepackage{graphicx}
\usepackage{esint}
\usepackage{color}
\usepackage[percent]{overpic}
\newcommand{\uw}[0]{\uparrow}
\newcommand{\dw}[0]{\downarrow}  

\newcommand{\uth}[0]{$^{\mbox{\underline{th}}}$}
\newcommand{\kfsn}[4]{
        \left|
                \begin{array}{c}
                        \begin{array}{cc}
                                #1 & #2 
                        \end{array}
                \\
                        \begin{array}{cc}
                                #4 & #3 
                        \end{array}
                \end{array}
        \right\rangle
}

\newcommand{\kfsb}[2]{
        \left|
                \begin{array}{c}
                        \begin{array}{cc}
                                #1 & #2 
                        \end{array}
                \\
                        \boxed{
                                \begin{array}{cc}
                                        \, & \,
                                \end{array}
                        }
                \end{array}
        \right\rangle
}

\newcommand{\kfst}[2]{
        \left|
                \begin{array}{c}
                        \boxed{
                                \begin{array}{cc}
                                        \, & \,
                                \end{array}
                        }
                \\
                        \begin{array}{cc}
                                #2 & #1 
                        \end{array}
                \end{array}
        \right\rangle
}

\newcommand{\kfsl}[2]{
        \left|     \begin{array}{cc}
                        \boxed{
                                \begin{array}{c}
                                        \, \\ \,
                                \end{array}
                        }
                &
                        \begin{array}{c}
                                #1 \\ #2 
                        \end{array}
                \end{array}
                \right\rangle
}

\newcommand{\kfsr}[2]{
        \left|     \begin{array}{cc}
                        \begin{array}{c}
                                #2 \\ #1 
                        \end{array}
                &
                        \boxed{
                                \begin{array}{c}
                                        \, \\ \,
                                \end{array}
                        }
                \end{array}
                \right\rangle
}

\newcommand{\ket}[1]{\left| #1 \right\rangle}

\makeatletter

\@ifundefined{textcolor}{}
{%
 \definecolor{BLACK}{gray}{0}
 \definecolor{WHITE}{gray}{1}
 \definecolor{RED}{rgb}{1,0,0}
 \definecolor{LIGHTRED}{rgb}{1,0.7,0.7}
 \definecolor{GREEN}{rgb}{0,1,0}
 \definecolor{BLUE}{rgb}{0,0,1}
 \definecolor{CYAN}{cmyk}{1,0,0,0}
 \definecolor{MAGENTA}{cmyk}{0,1,0,0}
 \definecolor{YELLOW}{cmyk}{0,0,1,0}
}

\usepackage{draftwatermark}
\SetWatermarkText{}
\SetWatermarkScale{1}
\makeatother

\usepackage{babel}

\begin{document}

\title{Control of entanglement transitions in quantum spin clusters}

\begin{abstract}
Quantum spin clusters provide a new platform for the experimental study of many-body entanglement. Here we address a simple model of a single-molecule nano-magnet featuring $N$ interacting spins in a transverse field. The field can control an entanglement transition (ET).
We calculate the magnetisation, low-energy gap and neutron-scattering cross-section and find that the ET has distinct signatures, detectable at temperatures as high as 5\% of the interaction strength. %Unlike a quantum critical point, the signatures of the ET that we discuss here 
The signatures are stronger for \emph{smaller} clusters. 
\end{abstract}

\author{Hannah R. Irons}

\affiliation{SEPnet and Hubbard Theory Consortium, University of Kent, Canterbury, CT2 7NH, U.K.}
\affiliation{ISIS Facility, STFC Rutherford Appleton Laboratory, Harwell Oxford, Didcot, OX11 0QX, U.K.}

\author{Jorge Quintanilla}

\email[Email address: ]{j.quintanilla@kent.ac.uk}

\affiliation{SEPnet and Hubbard Theory Consortium, University of Kent, Canterbury, CT2 7NH, U.K.}
\affiliation{ISIS Facility, STFC Rutherford Appleton Laboratory, Harwell Oxford, Didcot, OX11 0QX, U.K.}

\author{Toby G. Perring}

\affiliation{ISIS Facility, STFC Rutherford Appleton Laboratory, Harwell Oxford, Didcot, OX11 0QX, U.K.}

\author{Luigi Amico}

\affiliation{Centre for Quantum Technologies, National University of Singapore, 3 Science Drive 2, Singapore 117543}
\affiliation{Dipartimento di Fisica e Astronomia, Universit\'a Catania, Via S. Sofia 64, 95123 Catania, Italy}
\affiliation{CNR-IMM  UOS  Universit\`a  (MATIS), Consiglio  Nazionale  delle  Ricerche $\&$  INFN,  Sezione  di  Catania, Via  Santa  Sofia  64,  95123  Catania,  Italy}
\affiliation{LANEF {\it 'Chaire d'excellence'}, Universit\`e Grenoble-Alpes \& CNRS, F-38000 Grenoble, France}

\author{Gabriel Aeppli}

\affiliation{Swiss Light Source, Paul Scherrer Institute, CH-5232 Villigen PSI, Switzerland}
\affiliation{Institut de Physique, \'Ecole Polytechnique F\'ed\'erale
de Lausanne, CH-1015 Lausanne, Switzerland}
\affiliation{Department of Physics, ETH Z\"urich, CH-8093 Z\"urich, Switzerland}

%\date{[Date]}

\maketitle 

\section{\label{sec:intro}Introduction}

Classical phase transitions, such as the melting of ice and
the boiling of liquid water, are usually driven by thermal fluctuations.
To understand quantum materials
the theory was extended to quantum phase transitions, which are ubiquitous  in systems with strong electron-electron correlations \cite{Mathur1998,Saxena2000,ColdeaE8,Anderson1987,Ruegg2003,Ruegg2014}. 
The paradigmatic models feature localised spins under applied magnetic
fields \cite{Sachdev2011}. As the field is increased, quantum fluctuations grow, eventually
\textquotedbl{}melting\textquotedbl{} a magnetically-ordered ground
state at a quantum critical point. This has clear experimental manifestations
in materials that realise such models \cite{ColdeaE8}. More recently,
quantum information theory has been applied to these and other models of many-body systems \cite{Amico2008}.
It has been found that, before the quantum critical point is reached,
another qualitative change takes place: a change in the type of spin-spin quantum entanglement. Until now, however, there
have been limited predictions of experimental phenomena resulting from
such so-called \textquotedbl{}entanglement transitions\textquotedbl{} \cite{Muller1982,Roscilde2004,Roscilde2005,Amico2006,Fubini2006,Giampaolo2009,Giampaolo2010}.
Here we predict qualitative changes in the magnetisation and the neutron-scattering cross-section
of clustered quantum magnets that take place exactly at the entanglement
transition. Our main results are the predicted neutron scattering cross-sections shown in Fig.~\ref{fig:Neutron-scattering-function} displaying an experimentally-detectable qualitative re-organisation of the spin-spin correlations that coincides with the ET. Our results suggest that the phenomenology of clustered quantum
materials is dominated by the entanglement transition.%, rather than the associated quantum critical point. 

Entanglement is a salient and pervasive feature of quantum
many-body systems. Spin-spin entanglement 
is apparent in 
many simple properties of magnets
such as the temperature-dependence of the susceptibility and specific heat \cite{Ghosh2003,Brukner2004,Bose2005} 
and correlation functions as measured with neutron scattering
\cite{Christensen2007}
%Schmidt2014}
. Indeed
very long-range entanglement has been established experimentally in some 
magnetic materials \cite{Sahling2015}. Thus it is perhaps not surprising that 
when quantum information theory, which focuses on entanglement as
the main property of interest, is applied to simple models of quantum
magnets it opens up much richer vistas \cite{Amico2008} than those offered
by more traditional quantum field theory approaches focusing on order
parameters and correlations \cite{Sachdev2011}. Of particular interest
is the entanglement transition \cite{Muller1982,Roscilde2004,Roscilde2005,Amico2006,Giampaolo2009,Giampaolo2010}: a qualitative change in the type of
entanglement present in a quantum magnet taking place at the point
in the phase diagram where the ground state factorises \cite{Muller1982,Giorgi2009,Giampaolo2009,Giampaolo2010}
and characterised by vanishing entanglement measures \cite{Roscilde2004,Roscilde2005}
and the divergence of the range of spin-spin entanglement \cite{Amico2006}. This
divergence occurs within the ordered phase, i.e. \emph{not }at the critical
point where the correlation length diverges. Indeed the entanglement
transition is not a change of thermodynamic phase. Studies of entanglement
transitions thus promise to take our understanding of correlated quantum
matter beyond the quantum-critical and renormalisation-group paradigms. %Moreover, naturally-occurring entanglement in materials could one day be used as a quantum-computing resource, towards which its experimental detection and characterisation is an essential first step. 

In principle, it is possible to extract entanglement measures from
measurements of correlation functions such as those performed using
neutron scattering, and therefore to establish the existence of the
entanglement transition in this way \cite{Marty2014}. Here we do not attempt to extract measures of entanglement. In contrast, we predict experimental phenomena that are concomitant with the entanglement transition. They occur in clustered quantum materials, 
i.e. those composed of separate, independent units with a few elementary
constituents each. Since each cluster is effectively an isolated, finite-size system, the entanglement transition can be studied here without the complications associated with quantum criticality. The results we present have been obtained for a simplified model of clustered 
magnets, where the constituents are localised spins. Such systems can be 
regarded as finite-size generalisations of the paradigmatic models of quantum 
criticality mentioned above \cite{Sachdev2011}. Experimental realisations of clustered quantum 
magnets abound and include single-molecule nano-magnets created by chemical synthesis \cite{Belik2007,Engelhardt2009,Baker2012,Baker2012b,Furrer2013,Timco2013}. We 
expect our main conclusion that the entanglement transition dominates the 
phase diagram of clustered systems %, instead of the quantum critical point, 
to be applicable to other clustered systems as well. These include nano-engineered atom clusters  on surfaces \cite{Wiesendanger2012,Heinrich2013,Feldman2017} and tunable networks of interacting trapped ions \cite{Kim2010} and atoms \cite{Simon2011}.

The paper is organised as follows: in Sec.~\ref{sec:model} we describe a simple model of clustered magnetic materials. In Secs.~\ref{sec:spectrum}, \ref{sec:mag} and \ref{sec:neutrons} we discuss the energy spectrum, magnetisation and neutron scattering cross-section. Sec.~\ref{sec:bulk} describes the approach to the bulk regime as the number of spins in our cluster becomes large. In Sec.~\ref{sec:con} we offer our conclusions.

\section{\label{sec:model}Model}

\begin{figure} %[h!]
    \centering
\includegraphics[width=0.90\columnwidth]{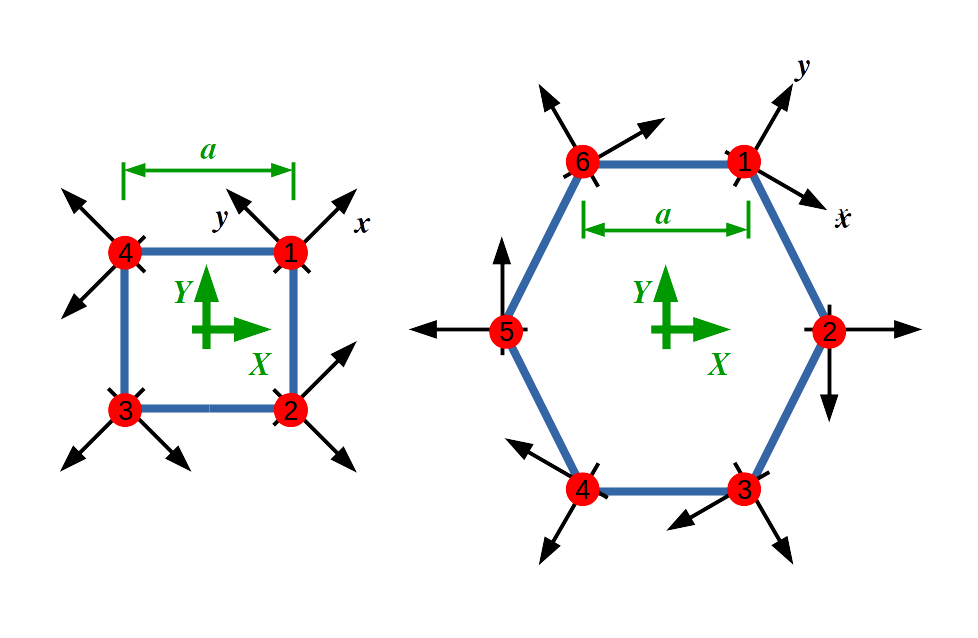}
    \caption{\label{fig:model-1}A simple model of a magnetic cluster
with $N=4$ (left) and $N=6$ (right) magnetic sites. The thick green arrows
represent the global $X,Y$ axes. The thin black arrows  represent the local $x, y$ axes with respect to which the three components of the spin $\hat{S}_j^x,\hat{S}_j^y,\hat{S}_j^z$ in Eq.~(\ref{eq:H}) are defined on each site $j=1,2,\ldots,N$. These local axes point tangentially and radially, respectively, on each site; by
convention the positive orientations correspond to the clockwise and
outward directions, respectively, as indicated. The rotation of the local easy axes from site to site respects
the $C_{N}$ symmetry of the molecule. The $Z$ and $z$ axes
point out of the page (not shown). The blue lines represent bonds along which
magnetic interactions occur. We assume the interactions between the spins are diagonal in the local axes and given by the Hamiltonian in Eq.~(\ref{eq:H}). 
%If one of the bonds is missing we obtain open boundary conditions, indicated by the blue dashed lines.
The distance ``$a$'' indicated on each plot is used as the unit
of length everywhere in this paper.}
\end{figure}

Numerical evaluations of measures of entanglement in models of finite-size
spin chains have shown that the entanglement transition does take
place in such systems at zero and finite temperature \cite{Campbell2013}.
Indeed the original argument by Kurmann et al. \cite{Muller1982}, applied to a small chain (where there is no broken symmetry), is quite
independent of the number of spins, $N$. Consider the spin-1/2
anisotropic Heisenberg model in a field,
\begin{eqnarray}
\hat{H} & = & -J\sum_{j=1}^{N}[\left(1+\gamma\right)\hat{S}_{j}^{x}\hat{S}_{j+1}^{x}+\left(1-\gamma\right)\hat{S}_{j}^{y}\hat{S}_{j+1}^{y}\nonumber \\
 &  & +\Delta\hat{S}_{j}^{z}\hat{S}_{j+1}^{z}]-h_{z}\sum_{j=1}^{N}\hat{S}_{j}^{z}.\label{eq:H}
\end{eqnarray}
Here $J$ sets the energy scale of nearest-neighbour spin-spin interactions,
$\gamma$ and $\Delta$ parametrise the anisotropy of those interactions
and $h_{z}$ the strength of the applied field (chosen, without loss
of generality, to point in the $z$ direction). For finite $N$ and
periodic boundary conditions ($\hat{S}_{N+1}^{\alpha}\equiv\hat{S}_{1}^{\alpha}$)
the above Hamiltonian, which is normally regarded as the archetype for a quantum spin chain, can be used instead to describe a single molecule with magnetic
moments located at the vertices of a regular polygon - see Fig.~\ref{fig:model-1}. In that case, since the orientation of the bonds joining nearest-neighbour sites is different at different sites, the axes $x,y,z$ with respect to which the three components of each spin are defined in Eq.~(\ref{eq:H}) must have a different orientation on each site with respect to some global axes $X,Y,Z$ defined by the overall orientation of the crystal. The choice shown in the figure is the only one compatible with the generic Hamiltonian in Eq.~(\ref{eq:H}) without breaking the $C_N$ symmetry of the molecule. With open boundary conditions ($\hat{S}_{N+1}^{\alpha}\equiv0$) a similar model
could describe such molecules where one of the bonds has been disrupted. %Note that the interactions in Eq.~(\ref{eq:H}) are diagonal in the local spin axes and all the anisotropy is in the interaction terms. 
%\footnote{More general models lead to very similar physics, albeit without exact factorisation in some cases.} 
The result is a simple, but quite generic model of a spin-1/2 clustered
magnet, which we will use in what follows, taking $\Delta=0$ for simplicity (relaxing this constraint does not alter any of our main conclusions; results for the more general anisotropic Heisenberg model are briefly outlined in Appendix \ref{sec:xyz}).

We now turn to the central question of this paper, namely the experimental
implications of the entanglement transition. We will see that, within our model, there are 
%clear signatures of the entanglement transition in the energy spectrum, magnetisation, and neutron scattering cross-section. Specifically, we describe below the occurrence of 
level crossings, magnetisation jumps and changes in the neutron-scattering cross-section 
%of our model 
as the value of the applied field is varied. 
Specifically, the latter reflect the 
%The question
%we wish to address is: are there any signatures of the 
change in quantum
correlations 
%observed specifically 
at the factorisation field.
% that result from the special nature
%of the state realised at that value of the field? As we will see the answer is affirmative.

\section{\label{sec:spectrum}Energy spectrum}

The general Kurmann et al. formula giving the value of the factorisation
field for our model is \cite{Barouch1971,Muller1982}
\begin{equation}
h_{f}=\sqrt{1-\gamma^{2}}.\label{eq:hf}
\end{equation}
This result is independent of $N$ and gives the value of the field at which a classical state is realised for any finite system (where there is no broken symmetry \cite{Tomasello2011}). For $N=2$ (quantum dimer) it
is easy to solve the problem analytically. The factorising field corresponds
to a level crossing where the ground state changes between two differently-entangled
states: for $h_{z} < h_{f}$ the ground state has zero magnetisation
and the spins have anti-parallel entanglement: $|\uparrow\downarrow\rangle-|\downarrow\uparrow\rangle$;
for $h > h_{f}$ the system magnetises but the spins remain entangled,
but in a parallel configuration: $|\uparrow\uparrow\rangle-\delta|\downarrow\downarrow\rangle$
(the constant $\delta$ is $0<\delta<1$ and tends to zero as $h_{z}\to\infty$).
Both of these states are evidently entangled in that no change of
basis can eliminate the inherent quantum superpositions. At $h=h_{f}$,
these two states are degenerate so any linear combination of them
is a valid ground state. It turns out that the mixing coefficient can be
chosen so as to create a completely non-entangled state. Details are given in Appendix \ref{sec:gs_wf}.

Quantum dimer
models and materials that realise them have been subject to intense
theoretical and experimental scrutiny \cite{Asoudeh2007,Sahling2015,Paulinelli2013,Hou2005,Ruegg2014,AndreZheludev}. The factorisation
field in our model evidently coincides with the dimerisation quantum phase transition
\cite{Ruegg2003}. When weak coupling between dimers is introduced, the
parallel-spins state becomes dispersive, forming magnons, and the dimerisation transition
is in the same universality class as Bose-Einstein condensation \cite{Giamarchi2008}.

\begin{figure} %[h]
   \centering
   \includegraphics[width=0.95\columnwidth]{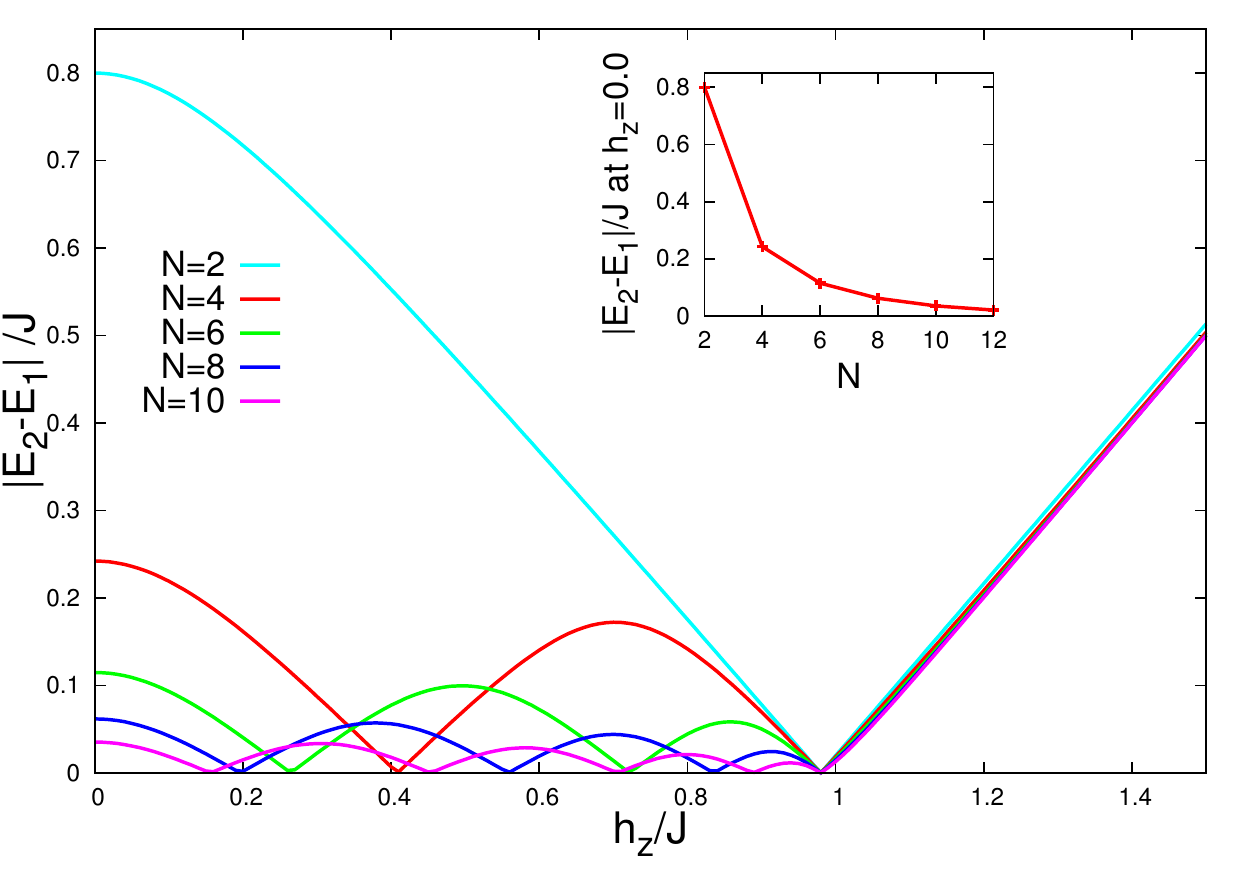}
   \includegraphics[width=0.95\columnwidth]{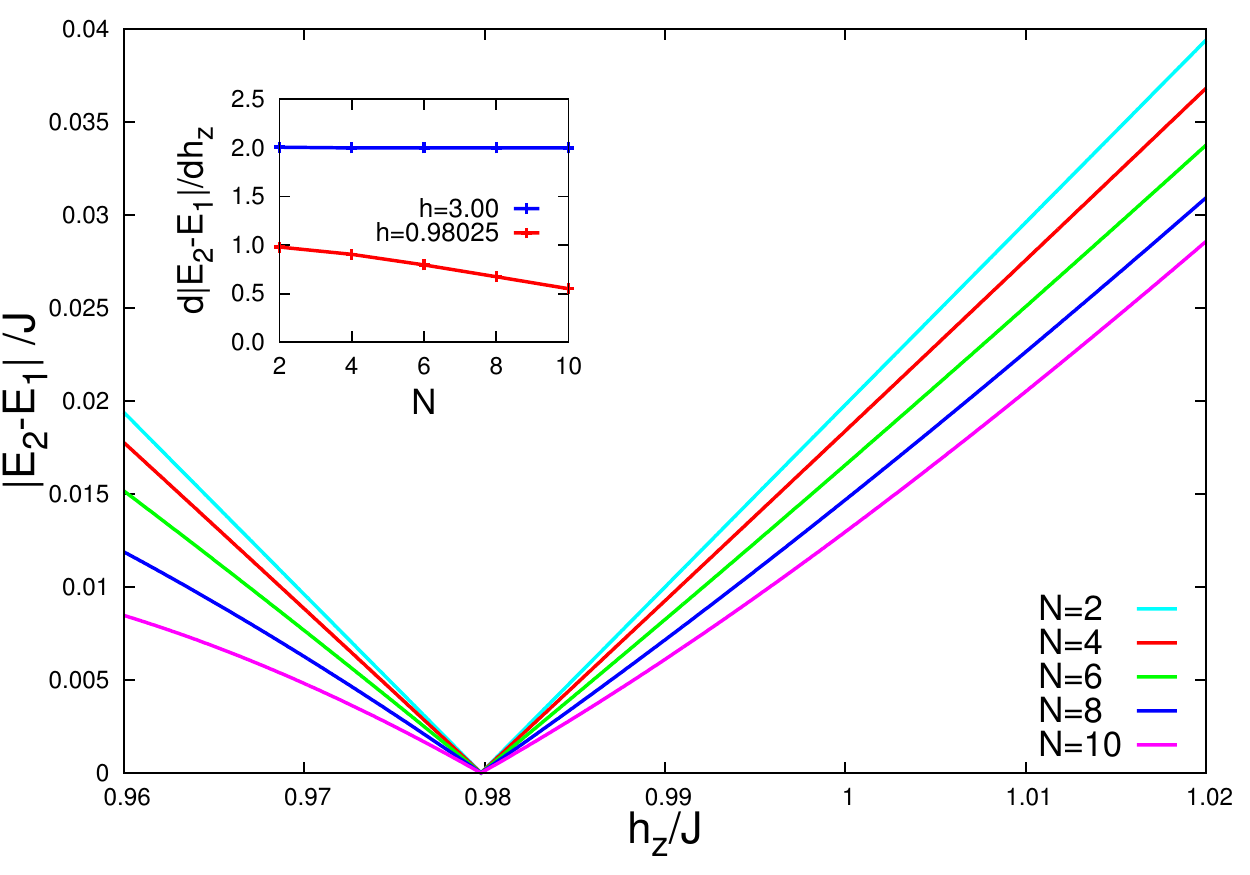}
   \caption{\label{fig:gap}Ground-state energy gap $\left|E_2-E_1\right|$ for a closed magnetic ring of $N=2,4,6,8,10$ sites, as indicated, as a function of the applied magnetic field. The results have been obtained by exact numerical diagonalisation of the model in Eq.~(\ref{eq:H}) with $\Delta=0$ and $\gamma=0.2$. The bottom panel is an expanded view of the field dependence near the factorisation field $h_f \approx 0.9798J$.    }
\end{figure}

More generally, for larger $N=4,6,\ldots$ (we restrict to even
$N$ to avoid additional complications due to frustration) we find
by numerical diagonalisation that there are two states that cross and constitute the ground and excited
state for any $h_{z}$. However, the number of crossings now is $N/2$, corresponding to successive changes of parity of the ground state \cite{Barwinkel2000,Barwinkel2003,Giorgi2009,DePasquale2009}. This is shown in Fig.~\ref{fig:gap} which shows the field-dependence of the gap for magnetic clusters of different sizes. 
As shown in the figure, the last crossing always occurs at the factorisation field $h_{f}$
of our model which is given by the Kurmann et al. formula (\ref{eq:hf}). Thus, in these finite-size systems factorisation coincides with an accidental ground-state degeneracy \cite{Fubini2006}. Inspection of
the numerically-obtained wave functions reveals that this ground state degeneracy corresponds
to a classical state in the same sense as in the dimer. The details of this analysis are given in Appendix \ref{sec:gs_wf}. The other
ground state degeneracies occur at lower fields $h$, $0<h<h_{f}$, which are different for different values of $N$. The
same numerical analysis shows that the state of the system does \emph{not} factorise at these additional crossing points (albeit it is closer
to factorisation than at other, intermediate fields) ---see Appendix \ref{sec:gs_wf}. 

It is important to note that the energy gap $\left|E_2-E_1\right|$ discussed here separates the non-degenerate ground state from the first excited state and exists only in finite-sized systems. In the thermodynamic limit, this gap closes and the ground state \textcolor{black}{becomes doubly-degenerate% and symmetry-breaking
.} A different, bulk gap emerges in this limit between this doubly-degenerate ground state and the lowest-energy excited states. That gap only closes at the critical point and separates the ground state from states higher in energy than those disucssed here. That bulk gap is not relevant to our discussion as the focus of the present work is on clustered (effectively finite) systems. We stress that all our discussions apply to spin-1/2 systems only; in particular, we do not consider the integer-spin case where a gap can appear due to quite different reasons \cite{Affleck1989}.
 
Interestingly, for open rings (i.e. open boundary conditions in our model) the level crossings occur at different values of the magnetic field. In particular, the last level crossing does \emph{not} occur at $h_f$ and moreover we do not find any factorised states. Factorisation thus seems to be\textcolor{black}{, for the very small clusters studied here,} a property that is dependent on the periodic boundary conditions.

Magnetic materials composed of spin-1/2 tetramers include,
for example, the spin-gap system Libethenite Cu$_{2}$PO$_{4}$OH
\cite{Belik2007}. Higher values of $N$ are realised in single-molecule
magnets \cite{Timco2013}. 
Indeed, level crossings of the type described here have
been known to occur for some time in single-molecule magnets and have
been extensively investigated theoretically \cite{Barwinkel2000,Waldmann2001,Barwinkel2003,Engelhardt2009,Giorgi2009,Cheng2010,Siloi2012} and experimentally, where the spectrum can be accessed directly using neutron scattering
\cite{Baker2012,Baker2012b,Furrer2013,Timco2013}. Our results indicate that tracking the field-dependence of the cluster energy gap $\left|E_2-E_1\right|$ to high-enough fields would enable the detection of the entanglement transition. This occurs at the highest among the sequence of fields $h_1$, $h_2$, $\ldots,h_{N/2}$ at which there is a closing of the gap. Also, in a comparison between samples with rings of different sizes (different values of $N$), $h_f$ is the only ground-state level-crossing field that occurs at the same value of $h_z/J$ for all $N$. Finally, because all the other ground state level-crossing fields are different for different $N$, in a sample with rings of different sizes (assuming they are all large enough that $J$ is approximately $N$-independent) there would only be one ground-state level-crossing field, and that would be $h_f$.

\section{\label{sec:mag}Magnetisation}

\begin{figure} %[h]
   \centering
   \includegraphics[width=0.95\columnwidth]{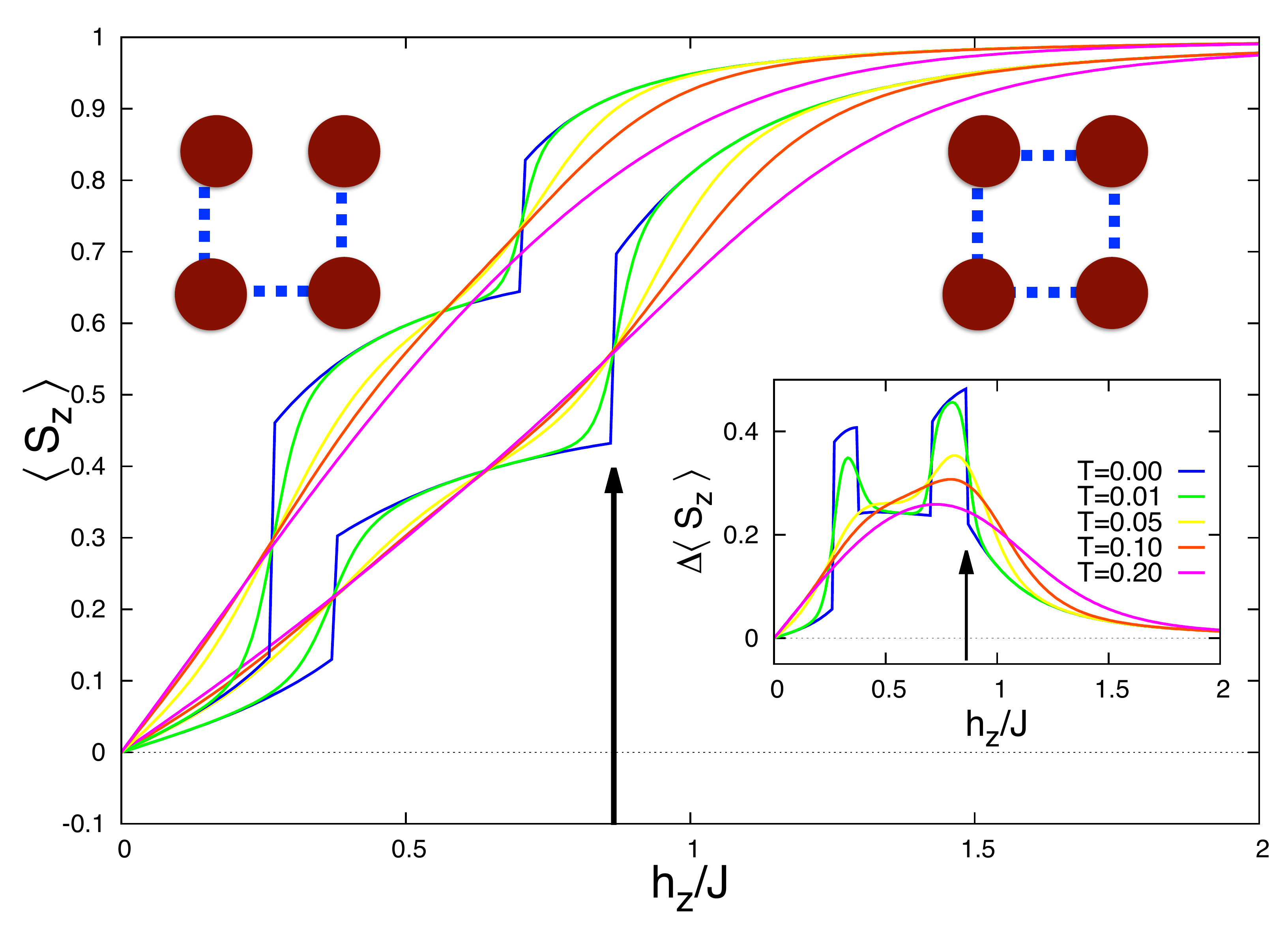}
   \caption{\label{fig:Difference-in-axis}
   Total $z$-axis
magnetisation per site as a function of applied field for a cluster with $N=4$ magnetic sites described by the model of Eq.~(\ref{eq:H}). The temperatures are as indicated, with the bottom set of curves corresponding to a closed ring, or plaquette, and the top set of curves to a small chain segment or, equivalently, a broken ring, as depicted. The anisotropy parameter is $\gamma = 0.5$ in all cases. The inset shows the difference between the chain and the ring. The arrow indicates the field at which
the ground state of the ring factorises exactly, where the largest jump in magnetisation takes place and also where the difference between the chain and ring magnetisations is largest.   
        }
\end{figure}

It is well-known theoretically and experimentally that in cluster magnets level crossings
like those described above coincide with jumps in total magnetisation \cite{Christou2000}. Fig.~\ref{fig:Difference-in-axis} shows the magnetisation of our model as a function of the applied field for $N=4$ open and closed clusters (the parameter values are given in the caption). $N/2$ jumps are seen, corresponding to each of the gap closings. For the closed rings, the last jump coincides with the entanglement transition.

The key feature of the state at $h_{f}$ in our model is that it is devoid of 
quantum entanglement \cite{Muller1982,Giampaolo2010,Roscilde2004,Roscilde2005,Amico2006}. 
One consequence of this is that, as in any classical
state, but unlike the states at higher and lower values
of $h_z$, all phase coherence between the wave functions of
individual spins is lost at $h_f$. At this particular value
of the applied field, therefore, the phase of the wave function of each individual spin can fluctuate independently of the others. We can thus consider the individual spin phases as a new %, diffusive 
degree of freedom that emerges as $h_z \to h_f$. This can, for example, contribute
to enhanced heat transport. In analogy with delocalisation transitions, such as the Anderson transition \cite{Edwards1972}, we might expect enhanced sensitivity to boundary conditions (open vs periodic). 
Experimentally, this could 
be accessed through measurements of magnetisation of
samples with different concentrations of open and closed
rings. The inset to Fig.~\ref{fig:Difference-in-axis} shows our prediction for such a 
measurement in the simplest, limiting case when one sample is
made up exclusively of open rings while the others are
all closed. Clearly, in the ground state the maximum
difference in magnetisation $\Delta \left\langle  S_z \right\rangle$ occurs quite precisely at the
factorisation field. The effect is smoothed by temperature, but it is clearly 
visible for $T \sim 5\%$ of $J$. Two sample values of $J/k_B$ for real cluster magnets are $17~{\rm K}$ for Cr$_8$ \cite{Baker2012b} and $138~{\rm K}$ for Cu$_2$PO$_4$OH \cite{Belik2007}. A smaller
peak is seen also at the field at which there is another level
crossing. This is what one would expect in view of the approximate factorisation at that field which we noted above. The enhanced value of $\Delta \left \langle S_z \right \rangle$ is due to the fact that the jump in magnetisation occurs at a different value of the field for an open ring, where the exactly factorised state is never realised.

\section{\label{sec:neutrons}Neutron scattering cross-section}

\begin{figure*}
\begin{centering}
%%%%%%%%%%%%%%%%%%%%%%%%%%%%%%%%%%%%%%%%%%%%%%%%%%%%%%%%%%%%%%%%%%%%%%
%
% NEW FIGURE COMMAND:
%
%	REQUIRES 1 X 1.0Mb FIGURE FILE.
%
%	RESULTS IN LOW-RESOLUTION FIGURE (ACCEPTABLE FOR arXiv).
%
    \includegraphics[width=2.0\columnwidth,angle=-90]{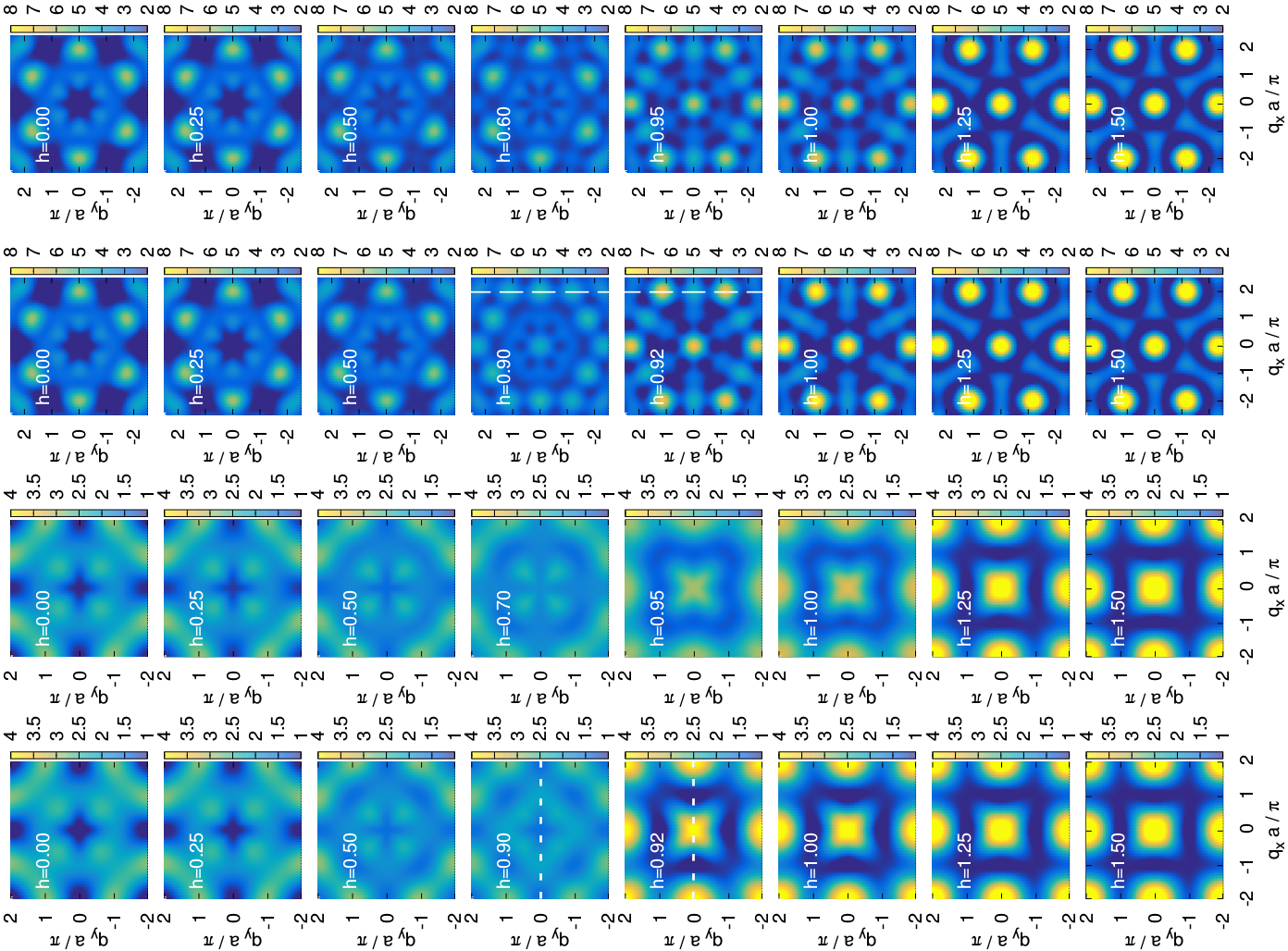} 
%
%%%%%%%%%%%%%%%%%%%%%%%%%%%%%%%%%%%%%%%%%%%%%%%%%%%%%%%%%%%%%%%%%%%%%%

%%%%%%%%%%%%%%%%%%%%%%%%%%%%%%%%%%%%%%%%%%%%%%%%%%%%%%%%%%%%%%%%%%%%%%
%
% ORIGINAL FIGURE COMMANDS: 
%
% 	REQUIRE 4 X 1.8Mb OF FIGURE FILES (NOT
% 	ACCEPTED BY THE arXiv).
%
% 	RESULT IN HIGH-RESOLUTION FIGURE.
%
%\begin{overpic}
%	[width=2\columnwidth,angle=-90,tics=10]
%	{HexGS001_v03-JQ_MULTIPLE-square-eps-converted-to.pdf}
% 	\put (3.8,45.05) {\textcolor{white}{\bf - - - - - - - -}}
% 	\put (3.8,56.84) {\textcolor{white}{\bf - - - - - - - -}}
%\end{overpic}
%\begin{overpic}
%	[width=2\columnwidth,angle=-90,tics=10]
%	{HexFT001_v03-JQ_MULTIPLE-square-eps-converted-to.pdf}
%\end{overpic}
%\begin{overpic}
%	[width=2\columnwidth,angle=-90,tics=10]
%	{HexGS001_v03-JQ_MULTIPLE-hexagon-eps-converted-to.pdf}
% 	\put (12.32,40.00) {
%		\textcolor{white}{\bf |}
%		}
% 	\put (12.32,42.5) {
%		\textcolor{white}{\bf |}
%		}
% 	\put (12.32,45) {
%		\textcolor{white}{\bf |}
%		}
% 	\put (12.32,47.5) {
%		\textcolor{white}{\bf |}
%		}
% 	\put (12.32,50.0) {
%		\textcolor{white}{\bf |}
%		}
% 	\put (12.32,51.5) {
%		\textcolor{white}{\bf |}
%		}
% 	\put (12.32,54.0) {
%		\textcolor{white}{\bf |}
%		}
% 	\put (12.32,56.5) {
%		\textcolor{white}{\bf |}
%		}
% 	\put (12.32,59.0) {
%		\textcolor{white}{\bf |}
%		}
% 	\put (12.32,61.5) {
%		\textcolor{white}{\bf |}
%		}
%\end{overpic}
%\begin{overpic}
%	[width=2\columnwidth,angle=-90,tics=10]
%	{HexFT001_v03-JQ_MULTIPLE-hexagon-eps-converted-to.pdf}
%\end{overpic}
%
%%%%%%%%%%%%%%%%%%%%%%%%%%%%%%%%%%%%%%%%%%%%%%%%%%%%%%%%%%%%%%%%%
\end{centering}
\caption{\label{fig:Neutron-scattering-function}
Frequency-integrated 
neutron scattering function $S\left(\mathbf{q}\right)$ as a function
of $q_{x}$ and $q_{y}$ for the model specified by Eq.~(\ref{eq:H})
and Fig.~\ref{fig:model-1}. The first and second columns show results for a system with $N=4$ spins and the third and fourth columns correspond to $N=6$. Columns 1 and 3 correspond to the ground state while columns 2 and 4 are for temperature $T=0.1 J$. Each panel corresponds to a different magnetic field $h_{z}$, as indicated.
The remaining Hamiltonian parameters are $\gamma=0.4$ and $\Delta=0$.
Note the values of $h_z$ are regularly-spaced except for two additional panels on each column, chosen to emphasise the sudden changes near the entanglement transition at $h_f \approx 0.917 J$. The white dashed lines indicate the directions of the scans shown in the insets to Figs.~\ref{fig:Ground-state-value-at} and \ref{fig:finite-T}. The calculation method is detailed in Appendix \ref{sec:neutron-formalism}. }
\end{figure*}
From a theoretical point of view, the most salient feature of our model at the factorisation
field $h_{f}$ is the transition between different types of quantum entanglement \cite{Amico2006} (see Appendix \ref{sec:gs_wf}). This suggests a strong effect on the correlation functions as measured by neutron scattering. 
Specifically, neutron scattering can be used to discriminate between antiferromagnetic
and ferromagnetic correlations and therefore we expect a significant
change in the magnetic neutron scattering cross-section at $h_{f}$.
The zero-field magnetic neutron scattering
spectrum of a system with $N=8$ has been investigated experimentally in detail \cite{Baker2012}. Fig.~\ref{fig:Neutron-scattering-function} shows the frequency-integrated in-plane magnetic structure function, $S({\bf q})$, for the model defined by our Hamiltonian (\ref{eq:H}) and the geometry shown in Fig.~\ref{fig:model-1} (with the transferred momentum ${\bf q}$ within the $xy$ plane). Results are shown for $N=4$ and $N=6$. The other model parameters are given in the figure caption and the details of the calculation are given in Appendix \ref{sec:neutron-formalism}. The top panels correspond to zero field and are clearly similar to the experimentally-determined low-energy scattering patterns in Ref.~\cite{Baker2012}: there is a deep minimum in scattering at the ferromagnetic wave vector ${\bf q}=0$ and $N$ sharp antiferromagnetic peaks with $|{\bf q}|=\frac{2\pi}{a}$ at angles $\phi=\frac{2\pi}{N}\left(\frac{1}{2}+n\right),~n=0,1,2,\ldots,N-1$ to the $q_y$ axis. A similar calculation for $N=8$ (not shown) confirms this close resemblance. The other panels show the changes we expect in such neutron scattering patterns as the field $h_z$ is increased.
%This is evidenced by Fig.~\ref{fig:Neutron-scattering-function}.
As the figure shows, each time a ground state degeneracy is encountered
there is a re-organisation of spectral weight. \textcolor{black}{At the} last degeneracy, i.e. at the factorisation field $h_f$, there is a large transfer of weight to ferromagnetic peaks that are not present in the zero-field state: one at ${\bf q}=0$ and $N$ more at $|{\bf q}|=\frac{2\pi}{a}\cos(\frac{\pi}{N})^{-1}$, $\phi=\frac{2\pi}{N}n$ with $n=0,1,2,N-1$. The peaks corresponding to anti-ferromagnetic
correlations between the spins get much weaker, as their spectral weight is transferred
to the new, purely ferromagnetic peaks. Thus the ground-state level-crossing fields (and especially the last one, corresponding in our model to exact factorisation) have clear signatures in the neutron scattering cross-section, indicating the re-organisations of correlations as such field values are crossed. Specifically, the neutron scattering functions shown in Fig.~\ref{fig:Neutron-scattering-function} allow us to discriminate the factorisation field, where entanglement vanishes, from the other ground-state level crossings where, as discussed in detail in Appendix \ref{sec:gs_wf}, it does not. This is in contrast to the energy gap and magnetisation measurements predicted earlier where the behaviour at all level crossings was similar.

\begin{figure*}
\begin{centering}
(a) \\ 
\includegraphics[width=0.42\columnwidth,angle=-90]{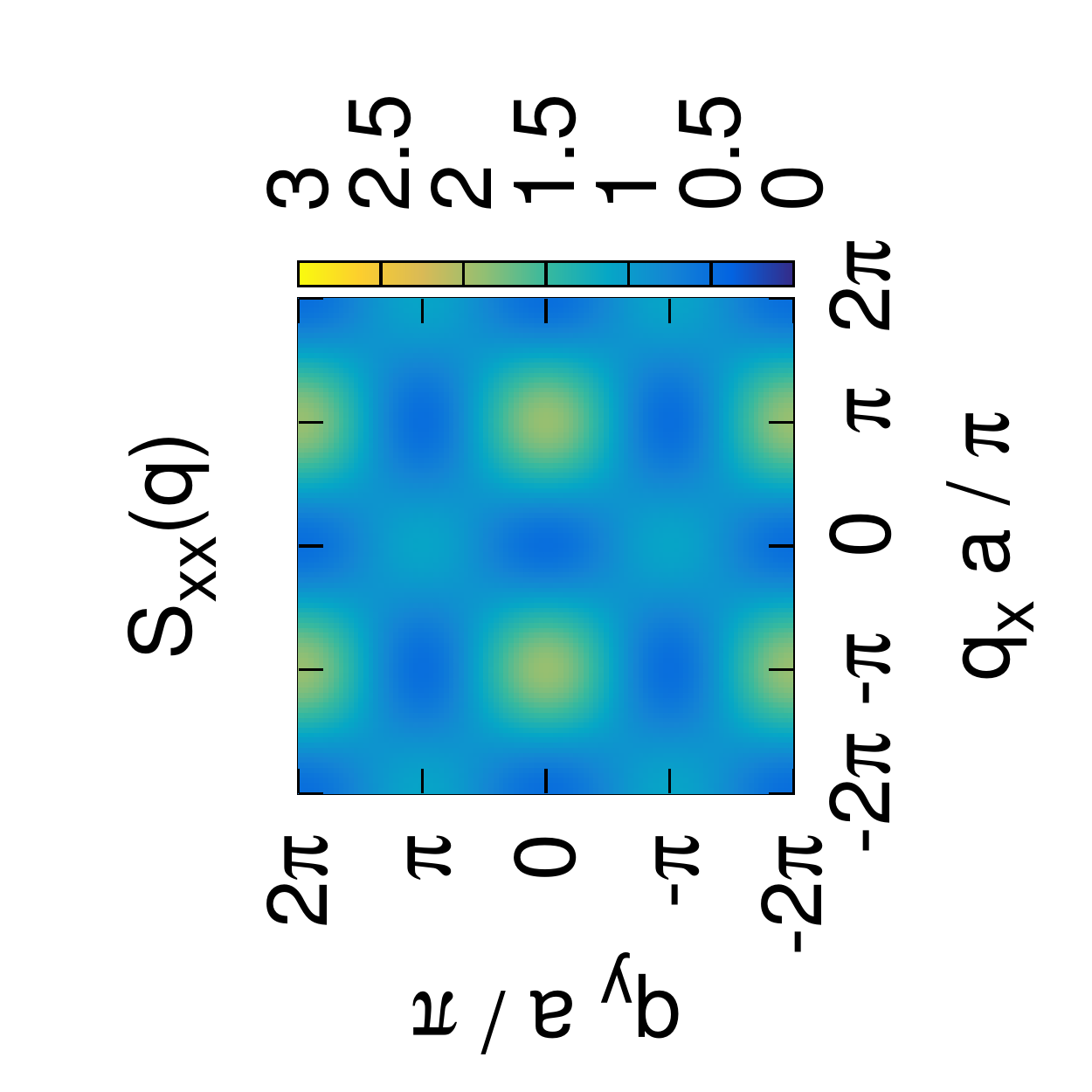}~
\includegraphics[width=0.42\columnwidth,angle=-90]{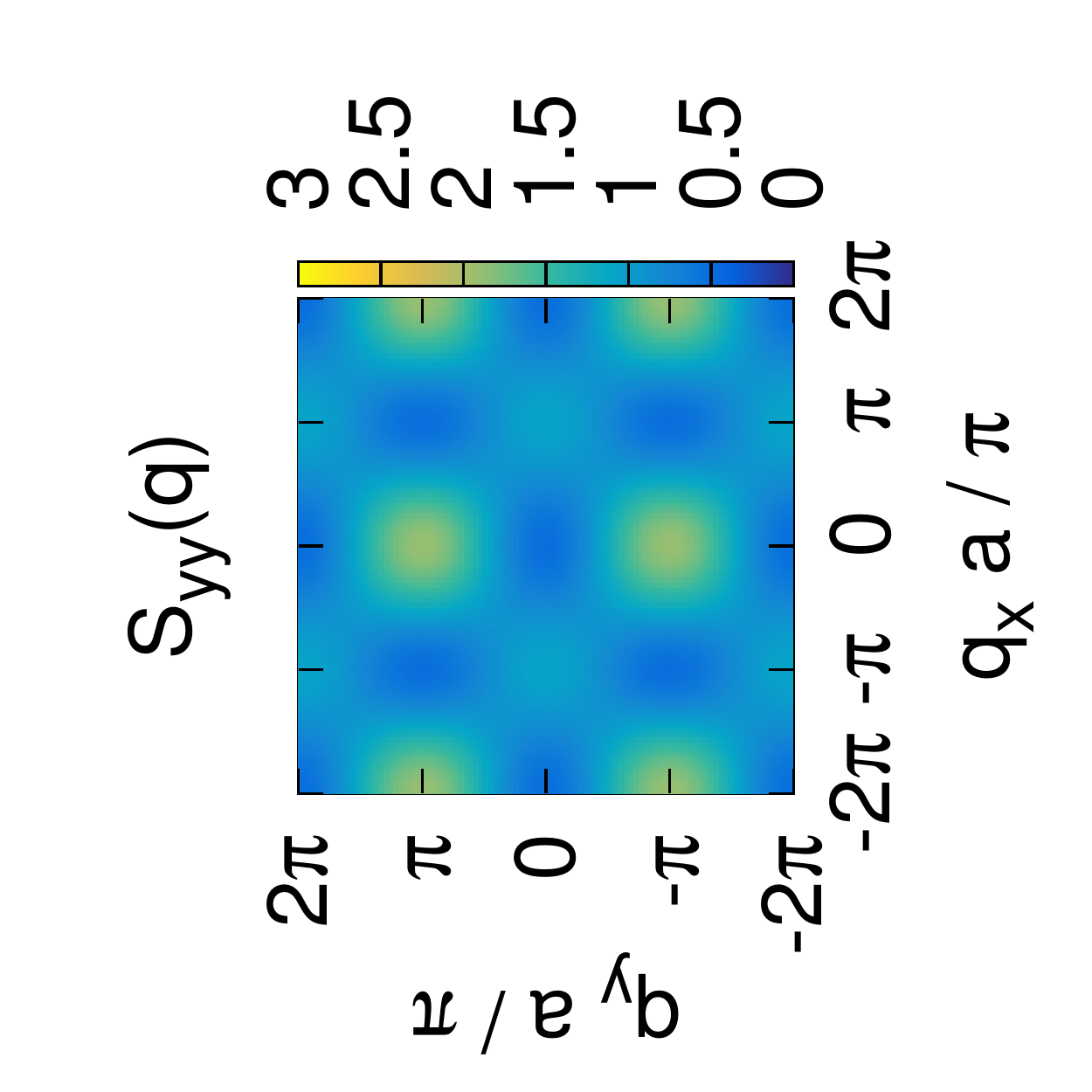}~
\includegraphics[width=0.42\columnwidth,angle=-90]{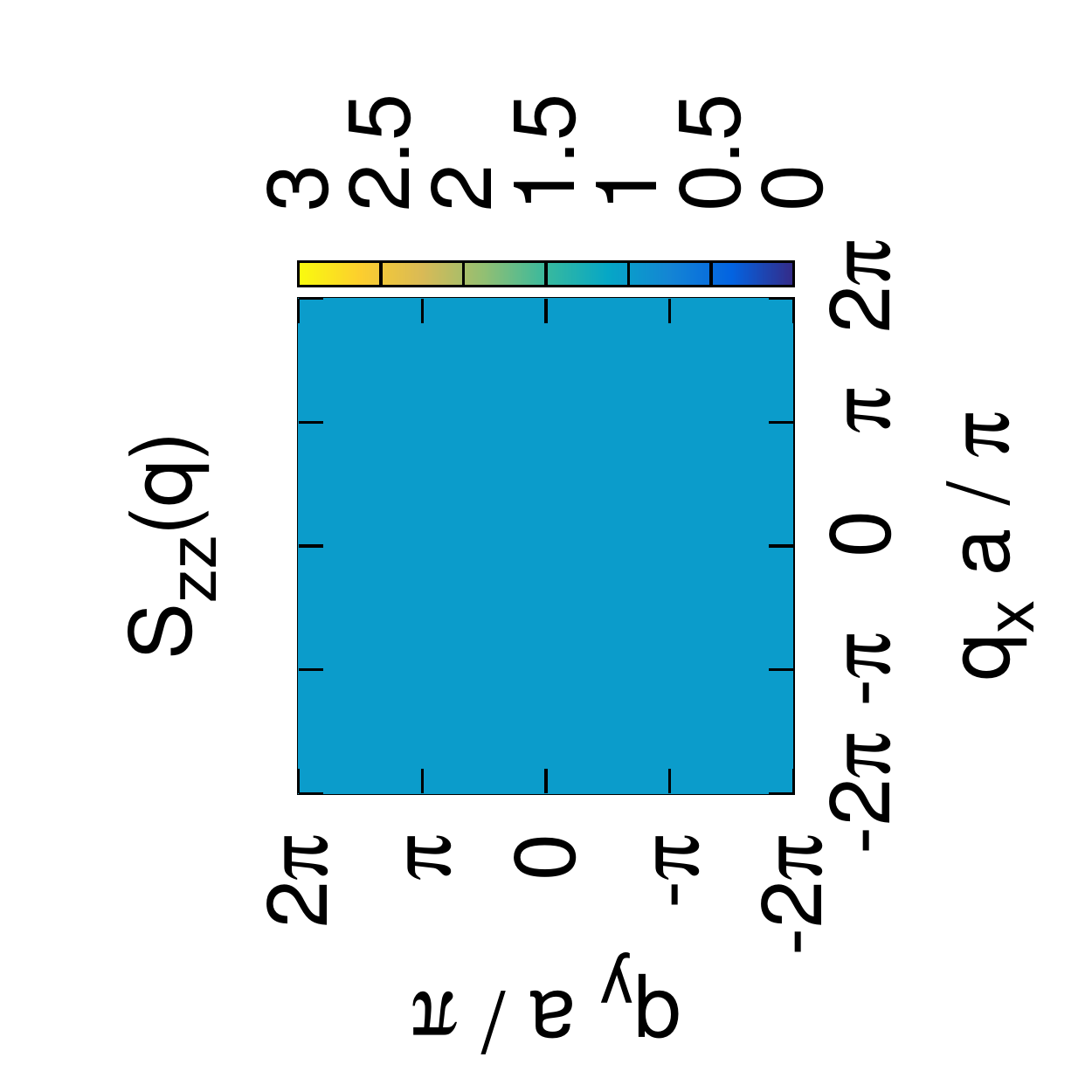}~
\includegraphics[width=0.42\columnwidth,angle=-90]{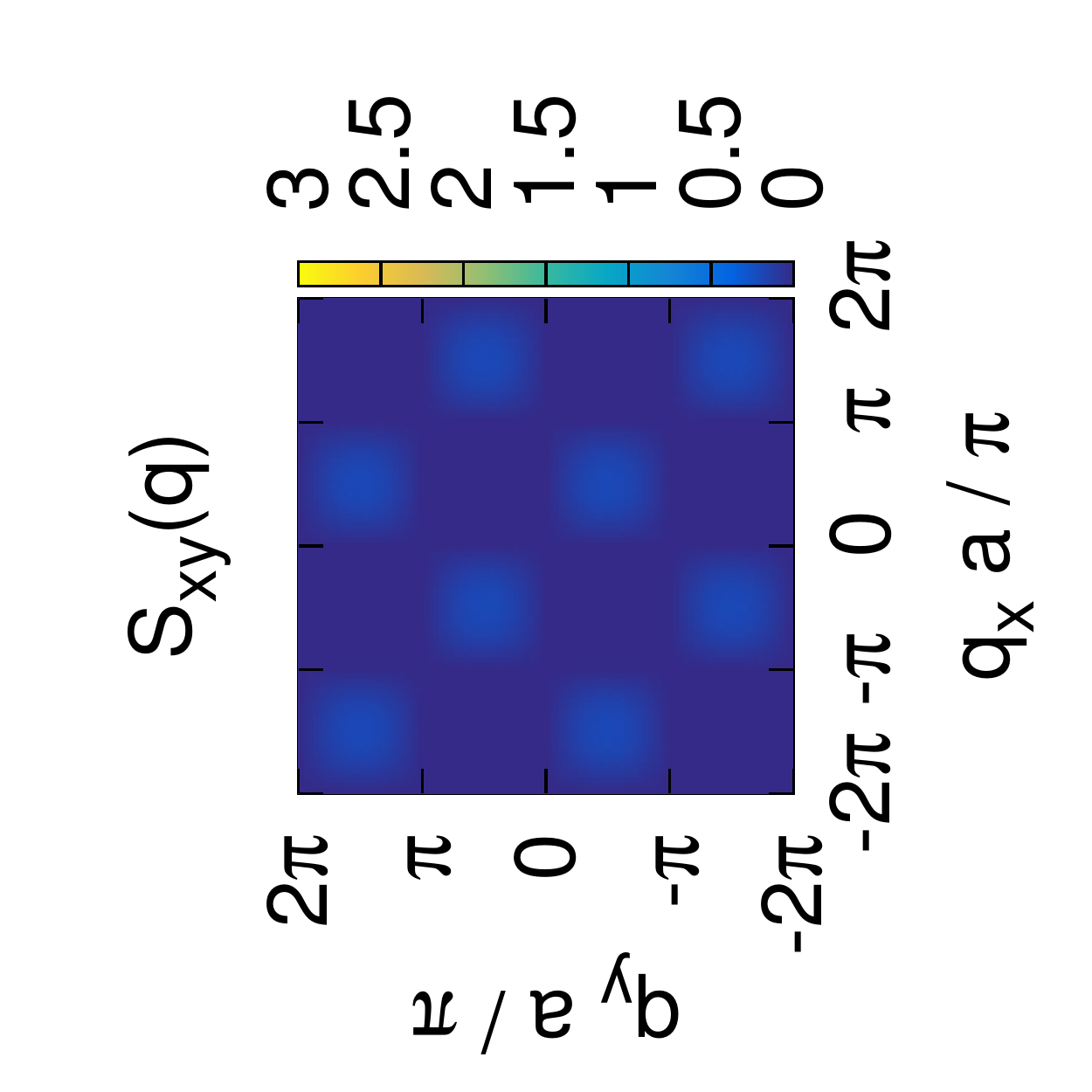}\\
\includegraphics[width=0.42\columnwidth,angle=-90]{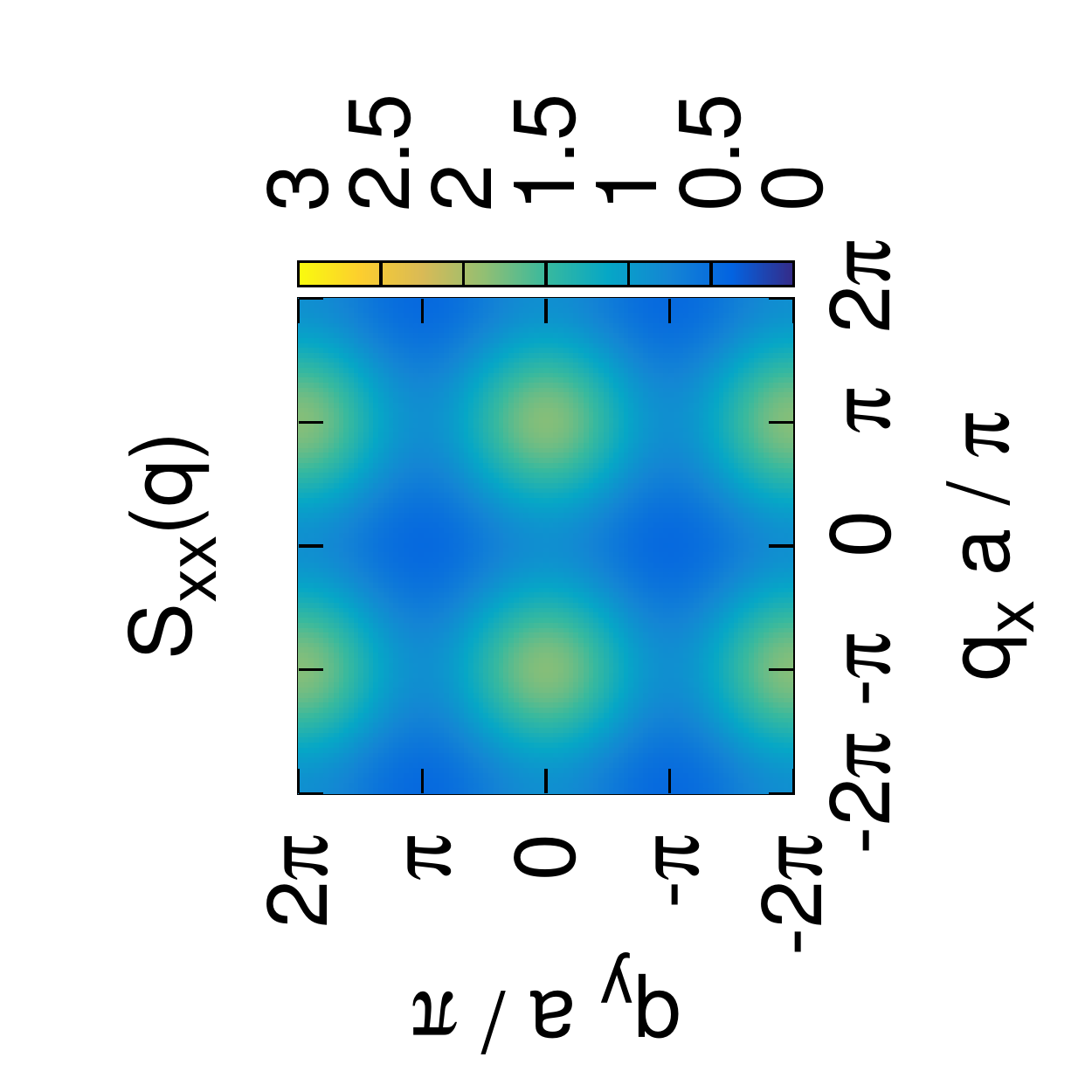}~
\includegraphics[width=0.42\columnwidth,angle=-90]{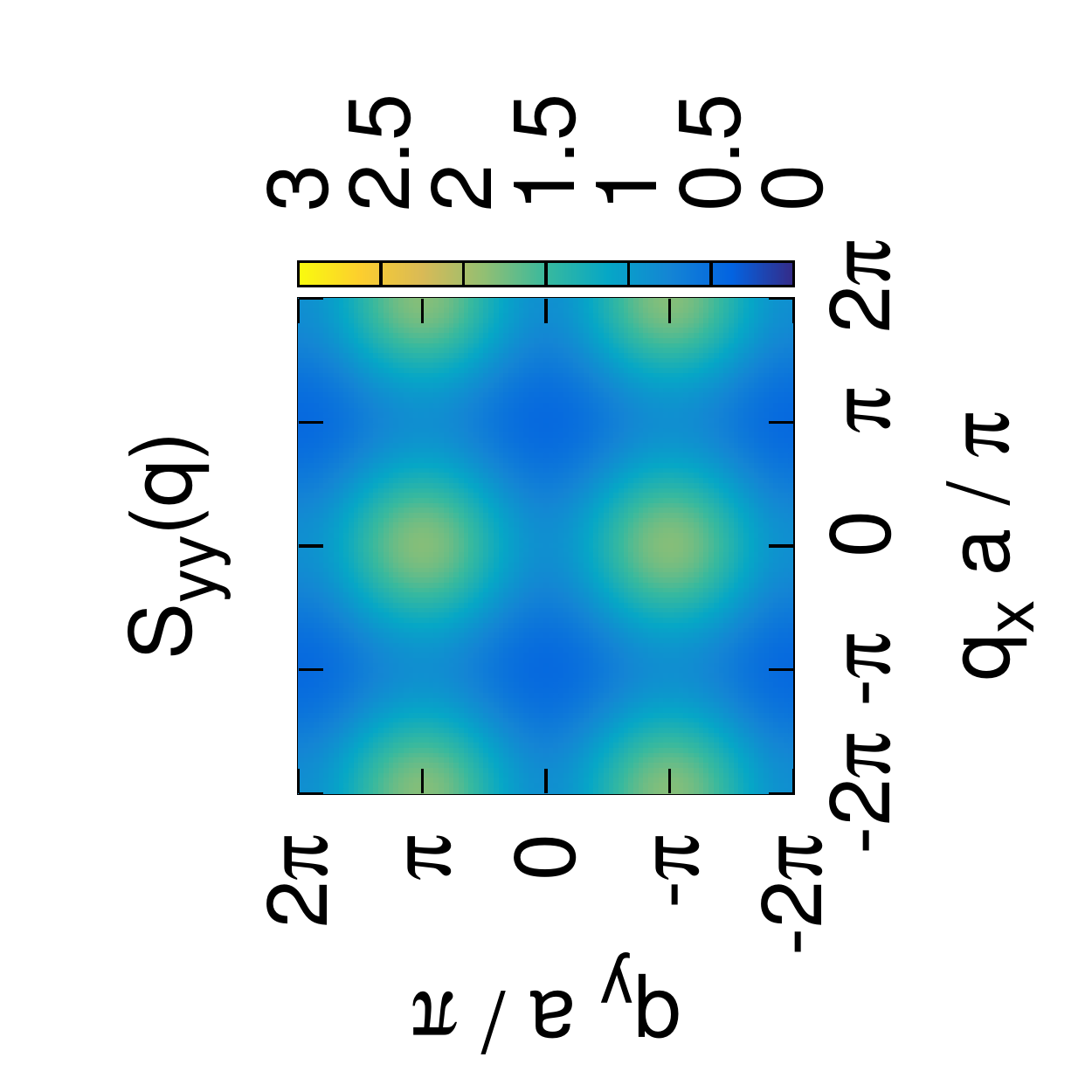}~
\includegraphics[width=0.42\columnwidth,angle=-90]{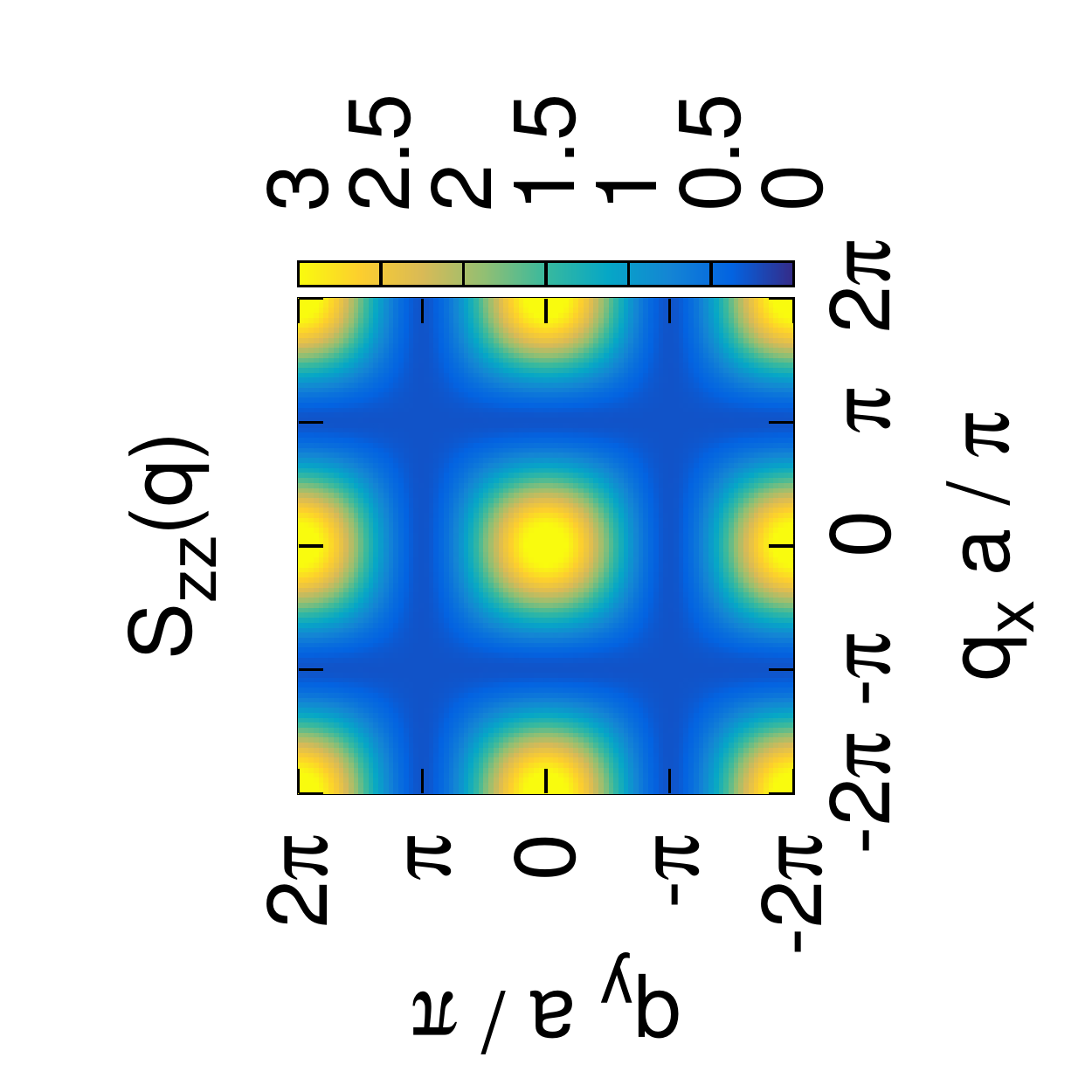}~
\includegraphics[width=0.42\columnwidth,angle=-90]{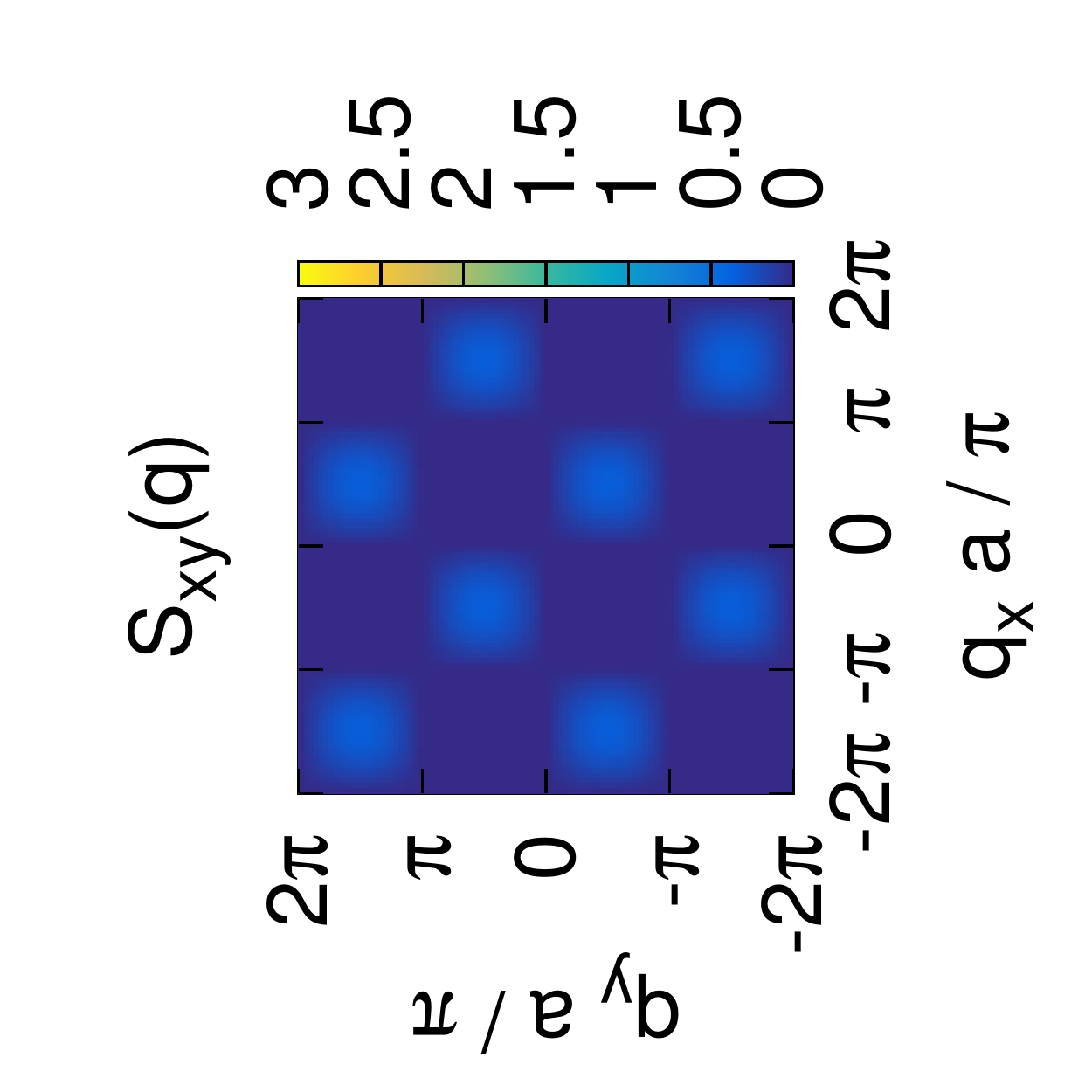}\\
(b) \\ 
\includegraphics[width=0.42\columnwidth,angle=-90]{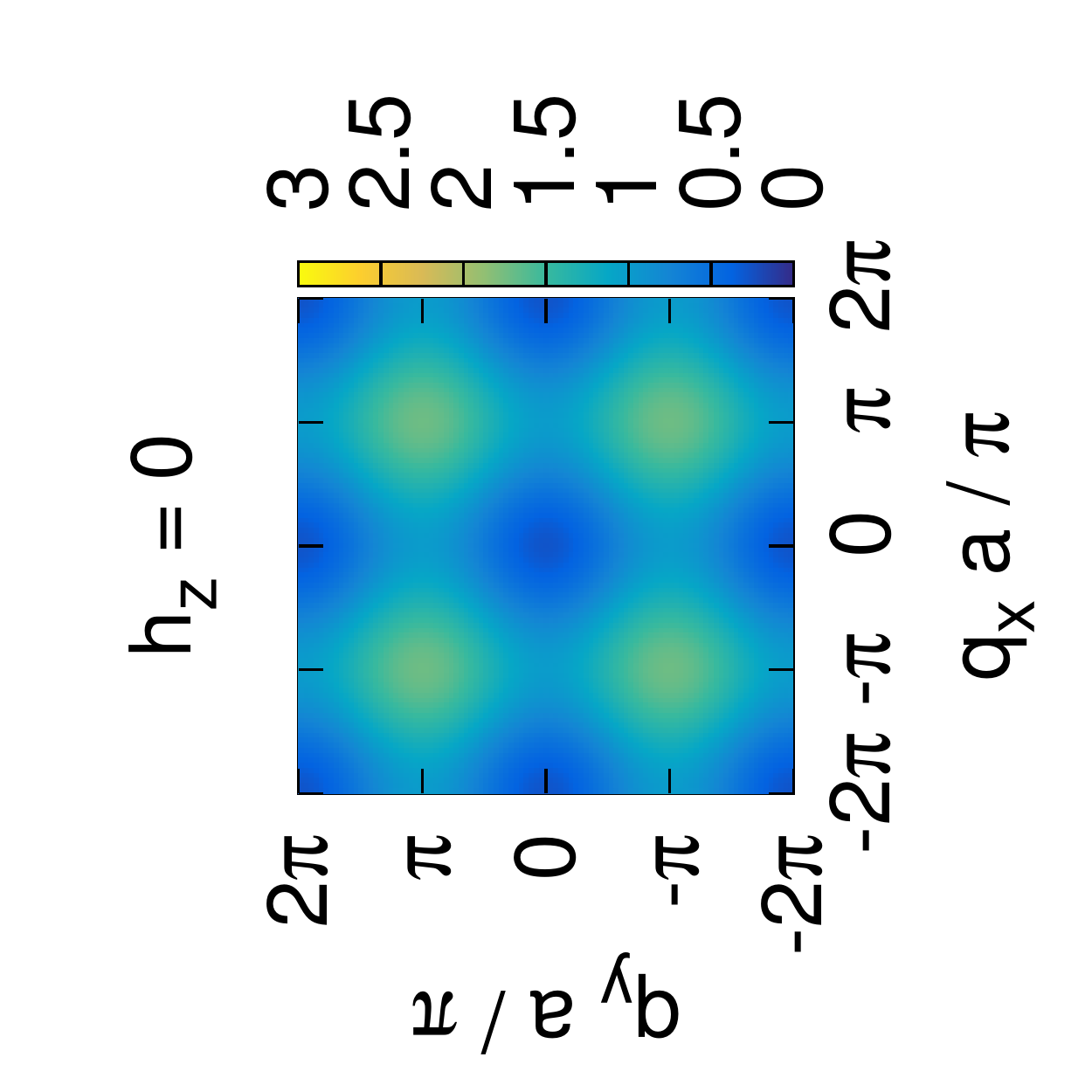}~
\includegraphics[width=0.42\columnwidth,angle=-90]{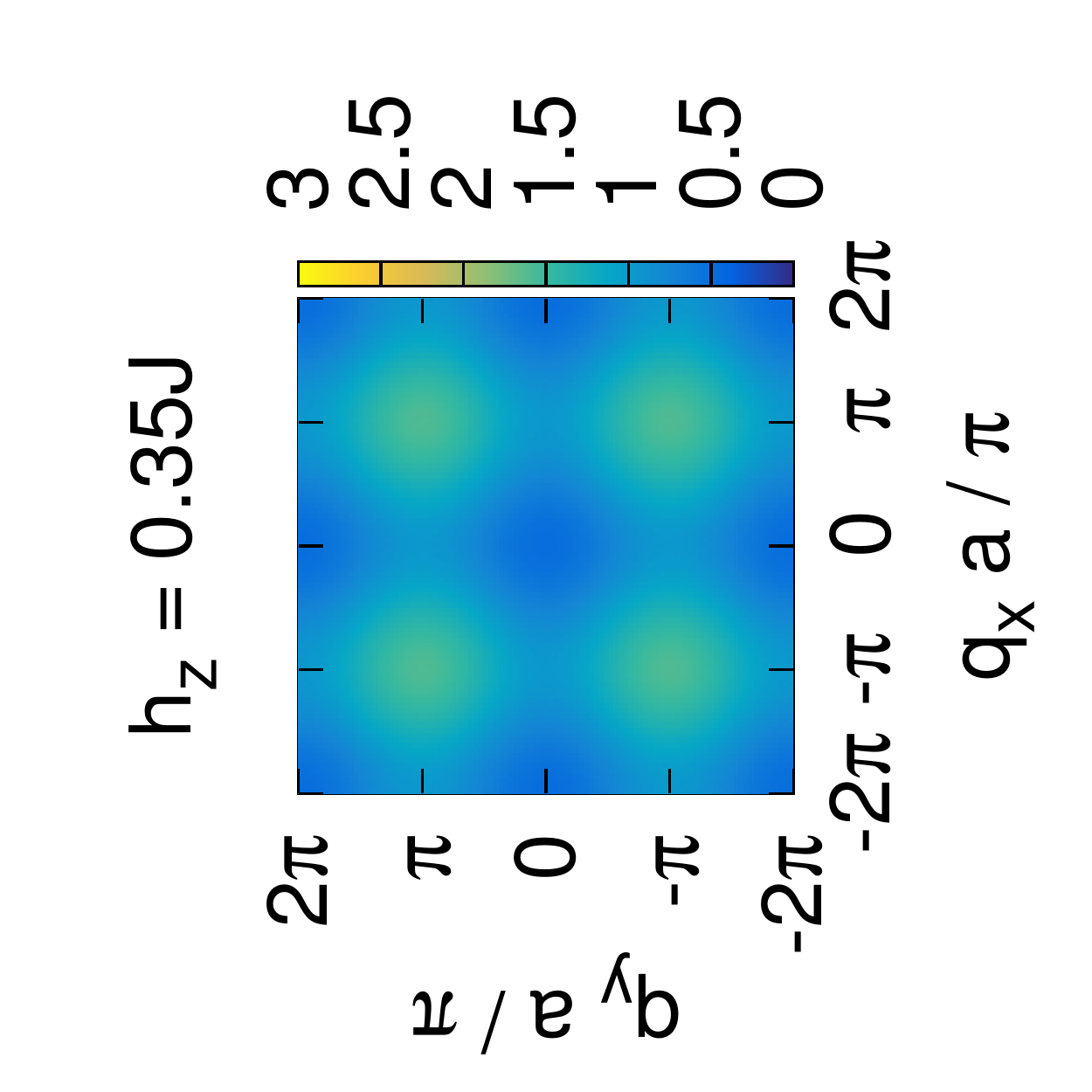}~
\includegraphics[width=0.42\columnwidth,angle=-90]{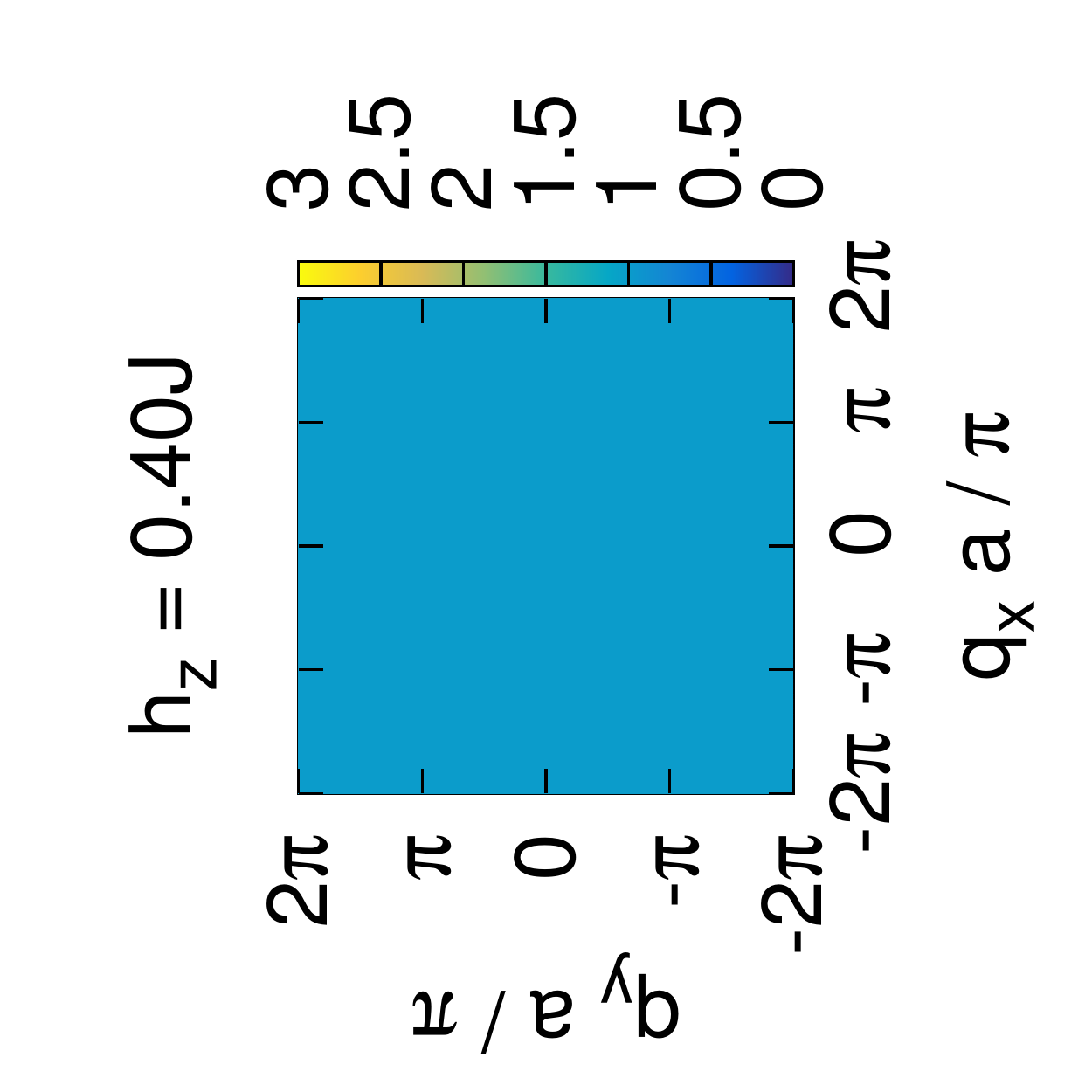}~
\includegraphics[width=0.42\columnwidth,angle=-90]{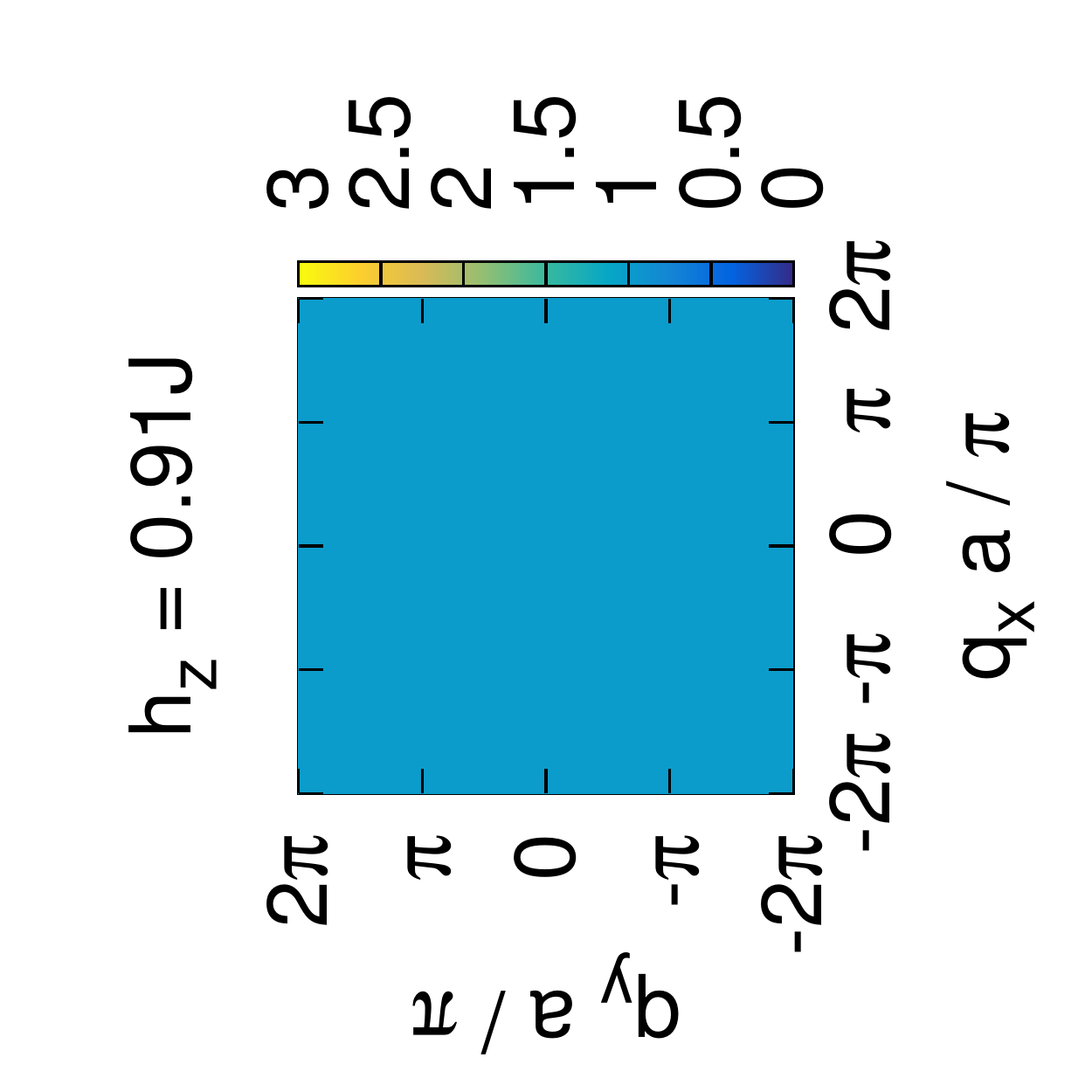}~
\includegraphics[width=0.42\columnwidth,angle=-90]{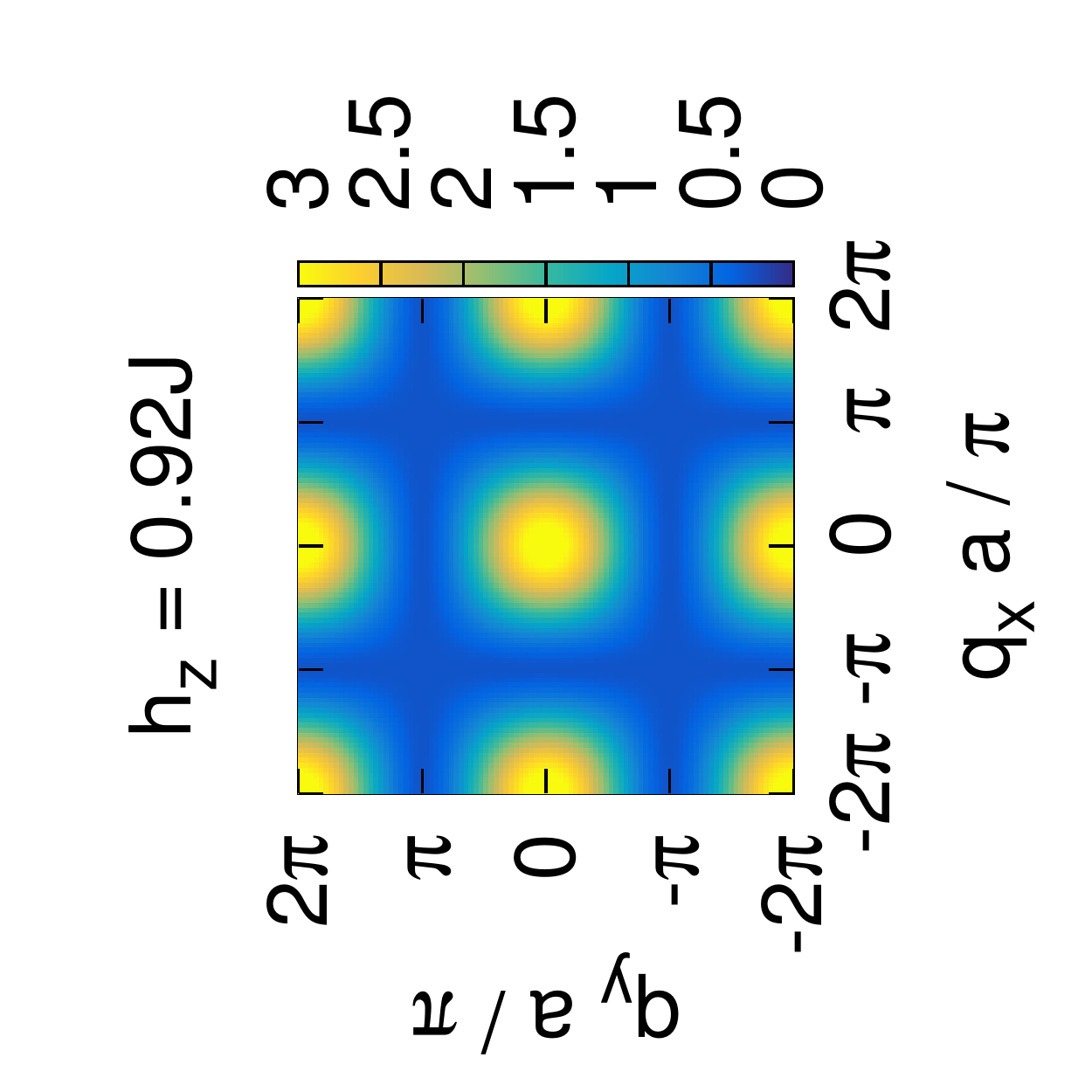}~
\end{centering}
\caption{
\label{fig:Neutron-scattering-function-breakdown}
(a) Field dependence of the spin-resolved correlators across the entanglement transition for a cluster with $N=4$ spins and anisotropy parameter $\gamma=0.4$. The top panels show the correlators $S_{xx}({\bf q})$, $S_{yy}({\bf q})$, $S_{zz}({\bf q})$, and $S_{xy}({\bf q})$, as indicated, for magnetic field $h=0.900J$, which is just below the factorisation field $h_f \approx 0.916J$. The bottom panels show the same correlators at a slightly higher field, $h=0.920J$, which is just above $h_f$. 
(b) The $S_{zz}({\bf q})$ correlator over a broader range of fields, as indicated. The two leftmost panels correspond to fields below the first gap closing, the third and fourth panels are between the firs gap closing and the factorisation field, and the last panel is above the factorisation field. The results are discussed in the main text. The definitions of the correlators are given in Appendix \ref{sec:neutron-formalism}.
}
\end{figure*}
It is illuminating to plot the individual correlation functions $S_{\alpha\beta}({\bf q})$, between different components of the spins which contribute to the scattering function $S({\bf q})$ (see Appendix \ref{sec:neutron-formalism}; note that in our geometry $S_{xz}({\bf q})$ and $S_{yz}({\bf q})$ do not contribute to the scattering function). Such spin-resolved correlators can be accessed experimentally {\it via} polarisation analysis. Alternatively, they can be obtained by observing, in a crystal, different regions of reciprocal space and exploiting the magnetic neutron scattering selection rules. Our predictions are shown in Fig.~\ref{fig:Neutron-scattering-function-breakdown} (a) for the ground state of the $N=4$ model with $\gamma =0.4$. The top panel shows how $S_{xx}({\bf q})$, $S_{yy}({\bf q})$, $S_{zz}({\bf q})$, and $S_{xy}({\bf q})$ change as we cross the factorisation field $h_f$. The latter is essentially unchanged by the entanglement transition. The $xx$ and $yy$ correlators have two sets of anti-ferromagnetic peaks: some are very intense and are unaffected by crossing the entanglement transition; others are much weaker and are suppressed as $h_z$ goes from just below to just above $h_f$. It is these latter peaks whose disappearance we noticed in our discussion of Fig.~\ref{fig:Neutron-scattering-function}. The stronger peaks are not accessible in the combined scattering function $S({\bf q})$ because they are suppressed by the selection rules. Their persistence indicates that anti-ferromagnetic correlations overall change very little at the entanglement transition. Clearly, the suppression of anti-ferromagnetic correlations is \emph{not} the dominant phenomenon at $h_f$. This sets a clear distinction between the entanglement transition and the quantum critical point known to exist in the bulk ($N\to\infty$) phase diagram of these models. In contrast, the $zz$ correlator changes dramatically at $h_f$: it goes from being featureless just below $h_f$ to showing very strong ferromagnetic peaks. This is consistent with the jump in magnetisation discussed above. Fig.~\ref{fig:Neutron-scattering-function-breakdown} (b) shows the $zz$ correlator over a broader range of fields. At low fields the $z$ components of the spins are anti-ferromagnetically correlated [$S_{zz}({\bf q})$ has peaks at ${\bf q}=(\pi,\pi)$ and equivalent reciprocal-space points]. At the first closing of the gap the system goes into the state where there are no correlations between the $z$ components of different spins [$S_{zz}({\bf q})$ is ${\bf q}$-independent], before emerging into the ferromagnetically-correlated state above $h_f$ [peaks at ${\bf q}=(0,0),~(2\pi,0),$ etc.] A detailed discussion of the structure of these ground states for $N=2$ and $N=4$ is offered in Appendix \ref{sec:gs_wf}.  Interestingly the first state is an adiabatic continuation of the third one, the only difference being the relative amplitudes of ferro- and antiferromagnetic configurations (see Fig.~\ref{fig:wf} in the Appendix). A similar pattern is found for other values of $N$. 

\begin{figure} %[h!]
   \centering
   \includegraphics[width=1\columnwidth,angle=-90]{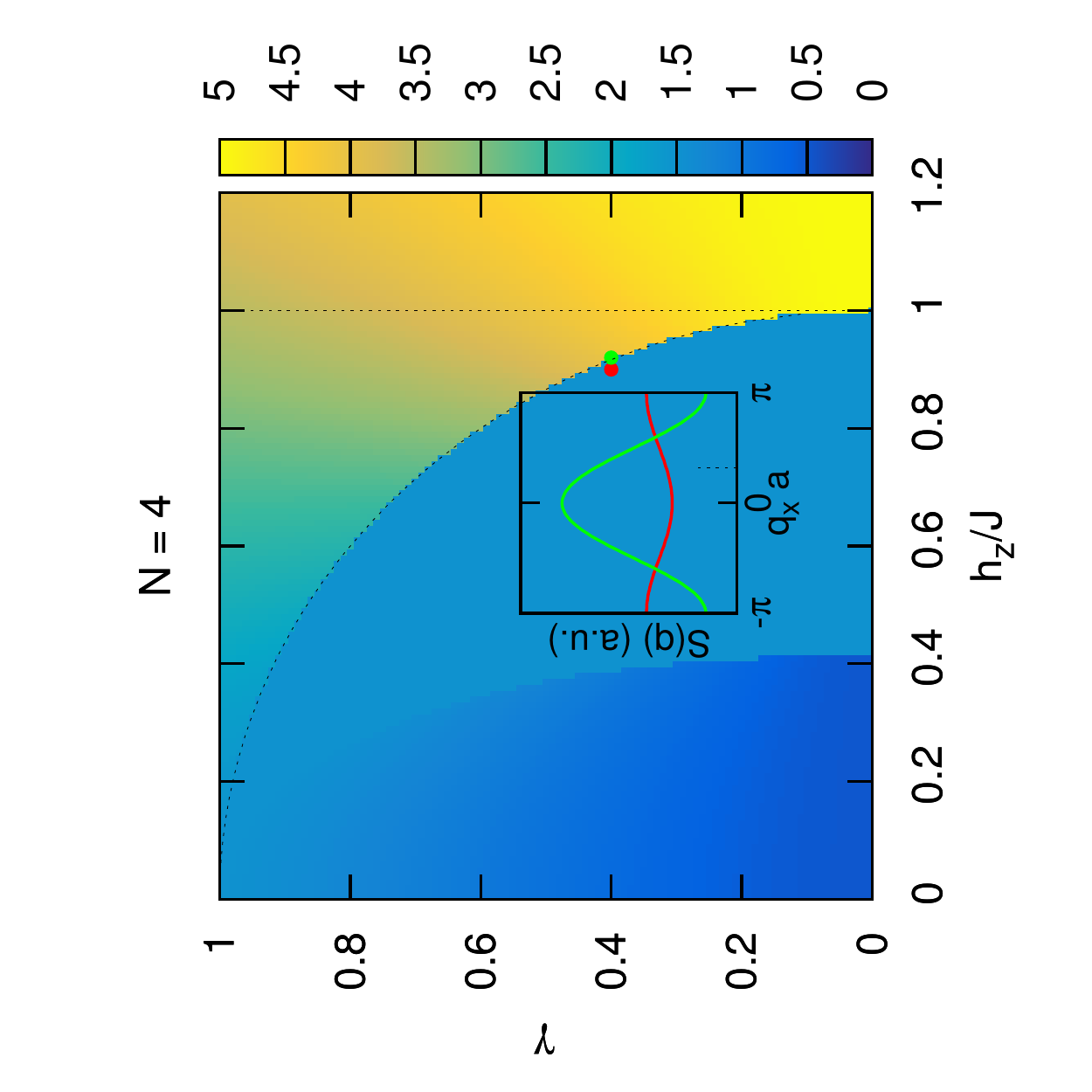}
\includegraphics[width=1\columnwidth,angle=-90]{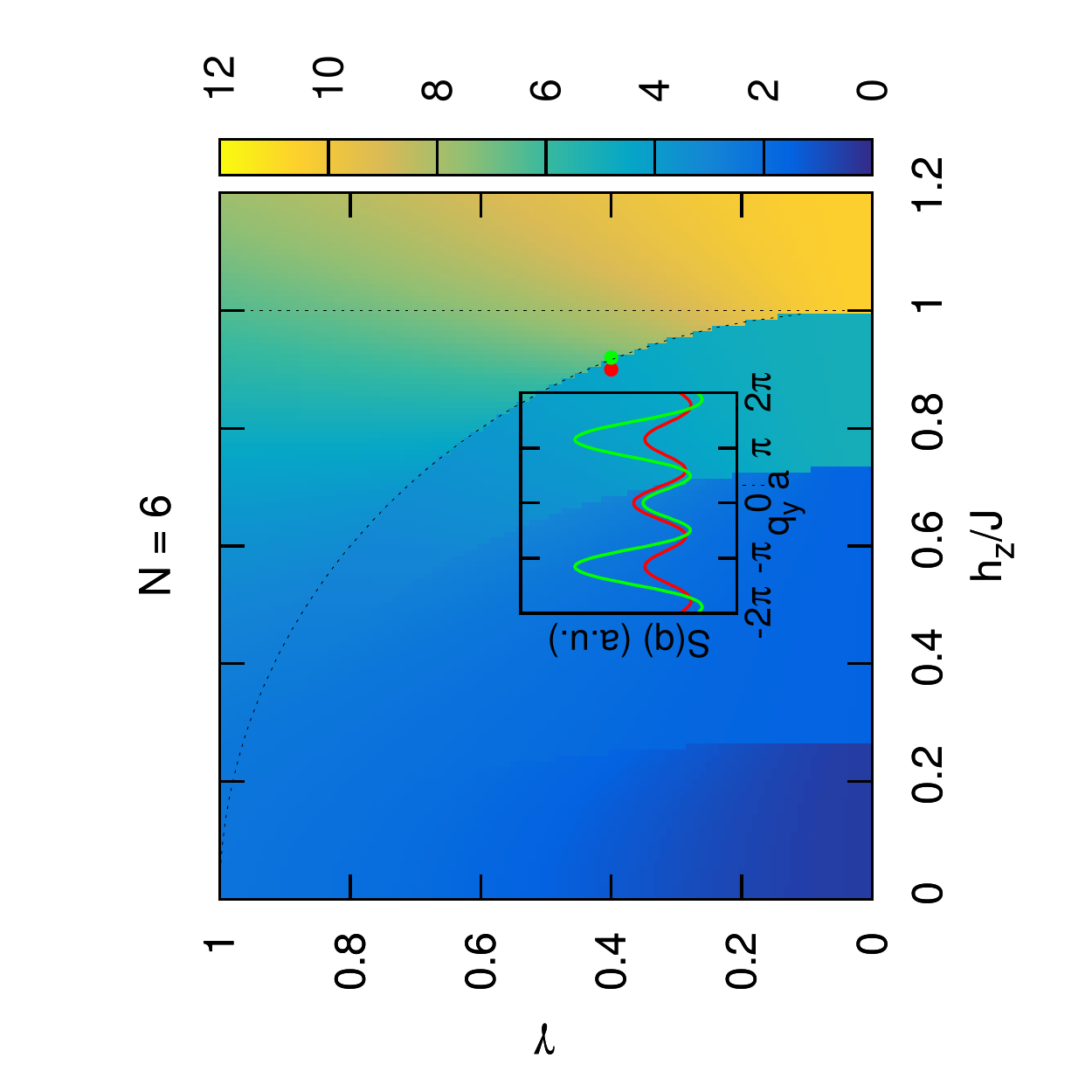}
\caption{\label{fig:Ground-state-value-at}
Ground-state value at $\mathbf{q}=0$ of the magnetic neutron scattering function, $S({\bf q})$, as a function of the anisotropy parameter $\gamma$ and the applied field $h_{z}$ for $N=4$ (top) and $N=6$ (bottom). The dotted lines indicate the factorisation field $h_{f}$
 and the quantum critical field in the limit $N\to \infty$, given by Eq.~(\ref{eq:hf}) and $h_{c}=1$ \cite{Barouch1970,Barouch1971}, respectively. Insets: dependence of $S({\bf q})$ on wave vector ${\bf q}$ for $h=0.90$ (red line) and $0.92$ (green line) with $\gamma = 0.4$.  These parameters lay on either side of the entanglement transition as indicated by the red and green dots on the main panels. The insets represent scans along the two directions in ${\bf q}$-space marked by the dashed lines in the corresponding panels of Fig.~\ref{fig:Neutron-scattering-function}.
}
\end{figure}

The re-organisation of correlations
occurs very suddenly at $h_{f}$. This is emphasized by Fig.~\ref{fig:Ground-state-value-at}, which shows the intensity of $S({\bf q})$ at ${\bf q}=0$ in the ground state as a function of the field $h_z$ and the anisotropy parameter $\gamma$. The sharp transition occurs at a value of the field that is $N-$independent and given by the Kurmann et al. formula (\ref{eq:hf}) (the cyan line in Fig.~\ref{fig:Ground-state-value-at}). The insets show a scan of the neutron scattering function through particular directions in reciprocal space, namely ${\bf q}=(q_x,0,0)$ for $N=4$ and ${\bf q}=(\pi,q_y,0)$ for $N=6$, on either side of the entanglement transition, emphasizing the sudden re-organisation of the magnetic scattering on crossing that boundary. Note in particular that the transition we have identified does \emph{not
} correspond with the quantum critical point (QCP) known to occur at
$h_{c}=1$ in the thermodynamic limit $N\to\infty$ (the black line in the same figure). 

It is clear from the above results that a diffuse neutron scattering
experiment on such finite-size magnets can be used to determine a
``phase diagram'' of the entanglement transition. Specifically, a sudden jump in $S({\bf q}=0)$ reflects the sudden change of correlations occurring at $h_z=h_f$. At finite temperatures, the neutron scattering functions look similar
to those in the ground state, as Fig.~\ref{fig:Neutron-scattering-function} also shows. The broadening of the entanglement transition with temperature is further discussed below.

%In a finite-size system, there cannot be a phase transition. Nevertheless, some broad crossovers may be interpreted as finite-size precursors of a critical point \cite{Fisher1972}. For instance, an order parameter, such as staggered magnetisation, may have a super-linear increase as a function of an applied field near the critical field value. Such smooth crossovers become steeper as the system size is increased. Indeed t
The region above the factorisation line in Fig.~\ref{fig:Ground-state-value-at} shows a smooth increase of $S(0)$ as a function of $h_z$. This increase is approximately independent of $\gamma$, consistent with the $\gamma$-independence of the critical field. Such finite-size precursors of criticality \cite{Fisher1972} are in sharp contrast to the behaviour of signatures of the entanglement transition and other gap closings described here, which are very sharp, in the low-temperature limit, even for the smallest system sizes. The latter are thus clearly not long-wavelength phenomena. We conjecture that unlike a QCP, an entanglement transition is not characterised by scale-invariance and
cannot, therefore, be understood within a picture based on universality classes
and the renormalisation group. Indeed as shown in Fig.~\ref{fig:Ground-state-value-at} the smoothed QCP is only apparent outside the dome defined by the factorisation field, indicating that factorisation, not criticality, dominates the phase diagram for clustered magnets. %The QCP also has a subtle signature in the ground state energy gap of our model at a distinct value of the field to that of the entanglement transition - see the Appendix,
%Sec.~\ref{sec:qcp_in_energies}. {\color{red}Do we need this last sentence?} 
A similar conclusion was reached by Campbell \emph{et al.} on the basis of their calculations of quantum discord, fidelity, entanglement of formation and the spectrum of the anisotropic XY model \cite{Campbell2013} (see also the related work \cite{Huang2014}). 

\begin{figure*} %[h!]
   \centering   
\includegraphics[width=0.65\columnwidth,angle=-90]{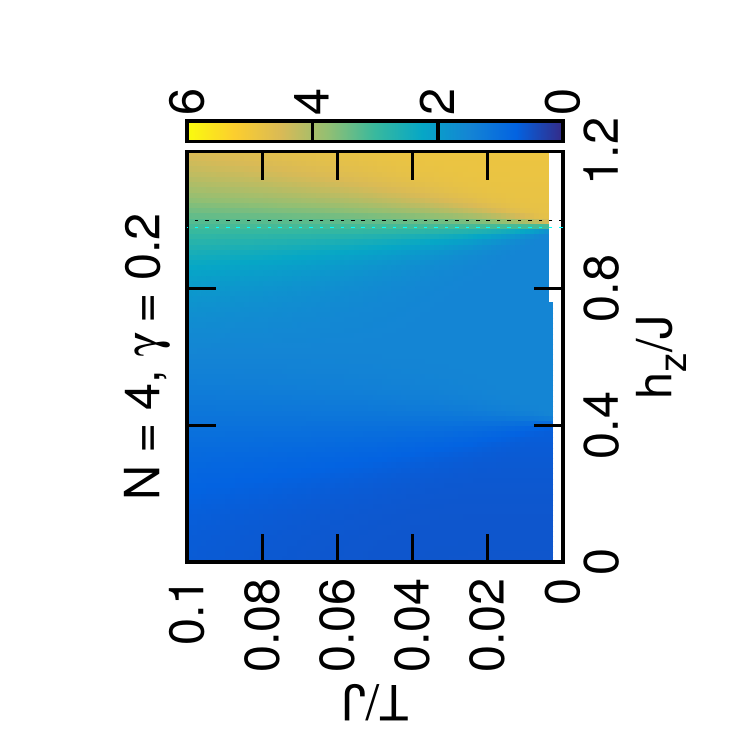}
\includegraphics[width=0.65\columnwidth,angle=-90]{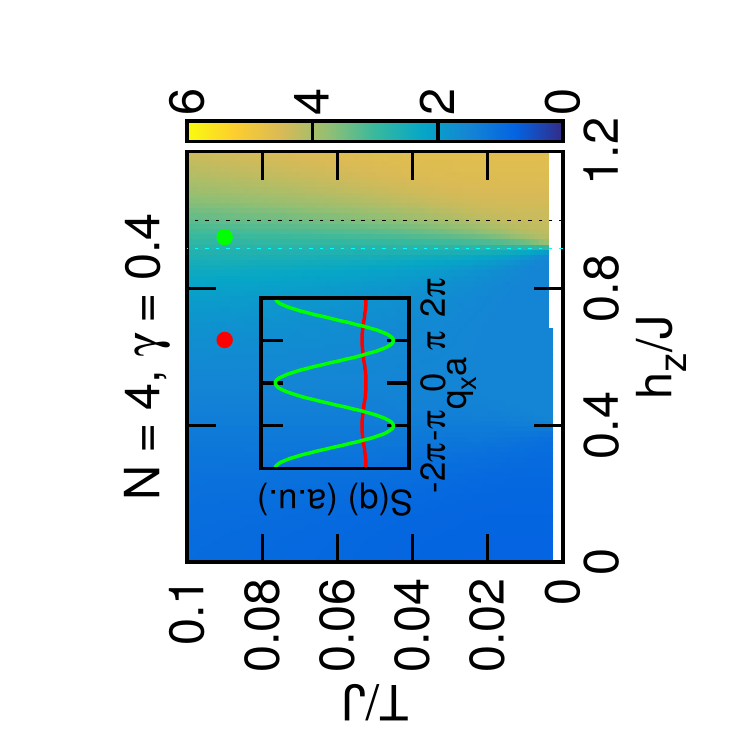}
\includegraphics[width=0.65\columnwidth,angle=-90]{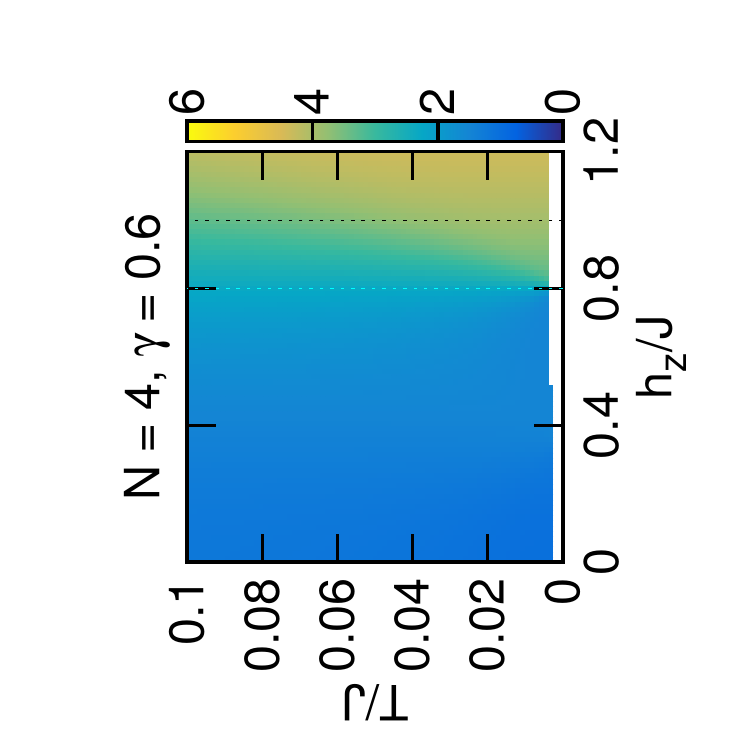}

\includegraphics[width=0.65\columnwidth,angle=-90]{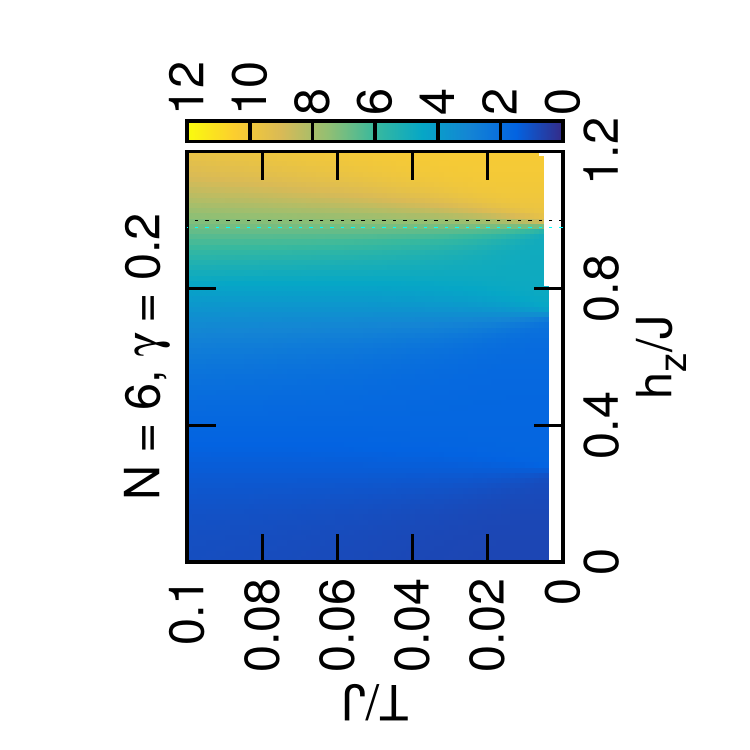}
\includegraphics[width=0.65\columnwidth,angle=-90]{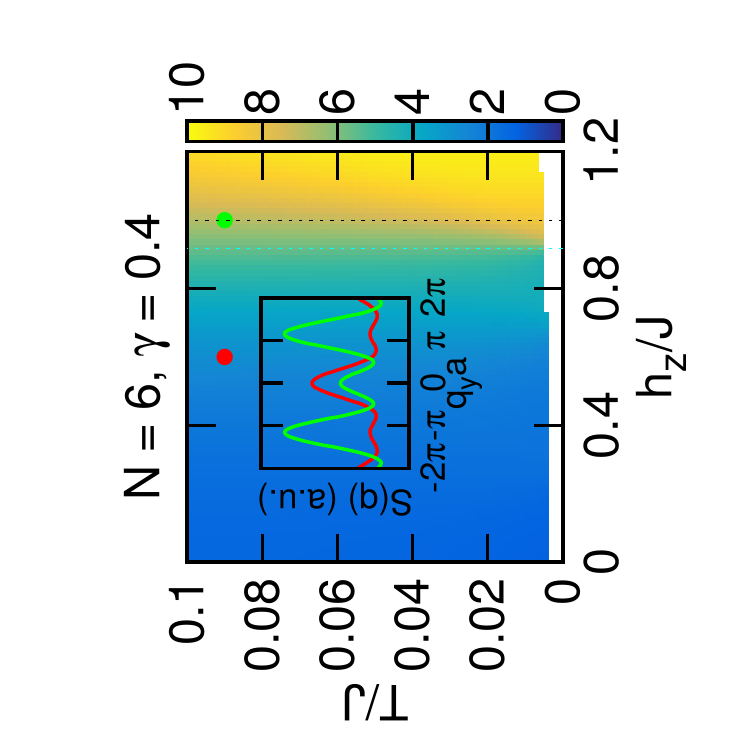}
\includegraphics[width=0.65\columnwidth,angle=-90]{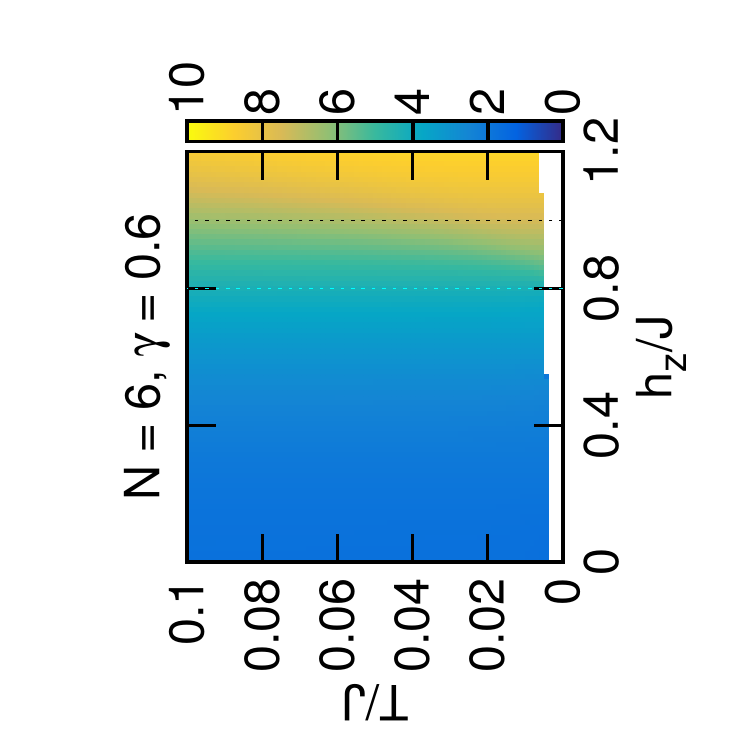}

   \caption{\label{fig:finite-T}Temperature-dependence of the quantity plotted
in Fig.~\ref{fig:Ground-state-value-at} for $N=4$ (top) and $N=6$ (bottom). The in-plane anisotropy is $\gamma=0.2,0.4,0.6$ (left to right). The factorisation field $h_{f}$ is indicated in each case by the cyan dotted line. The bulk value of the critical field is the black dotted line. Data for very low temperatures have been excluded as they suffer from unavoidable numerical round-off errors in evaluating the partition function ($T=0$ data shown in Figs.~\ref{fig:Neutron-scattering-function},\ref{fig:Neutron-scattering-function-breakdown},\ref{fig:Ground-state-value-at} are not affected). Insets: dependence of $S({\bf q})$ on wave vector ${\bf q}$ for the particular values of field and temperature indicated by the red and green filled circles on the main panels. The plots represent scans along the two directions in ${\bf q}$-space marked by the dashed lines in the corresponding panels of Fig.~\ref{fig:Neutron-scattering-function}. 
}
\end{figure*}
At finite temperatures, the signature of the entanglement transition
is less sharp, but still clearly visible for temperatures $\sim 5\%$
of the exchange constant $J$. This is clear from the finite-temperature panels in Fig.~\ref{fig:Neutron-scattering-function}. In addition, Fig.~\ref{fig:finite-T} shows the same quantity depicted in Fig.~\ref{fig:Ground-state-value-at} as a function of field and temperature for three particular values of the anisotropy parameter, $\gamma=0.2,0.4$ and $0.6$. Clearly, the rapid change of $S({\bf q}=0)$ with $h_z$ near $h_f$ persists. The insets to the $\gamma=0.4$ panels also show very similar re-arrangements of the ${\bf q}$-dependence of the scattering function to those shown in Fig.~\ref{fig:Ground-state-value-at}, albeit they occur over a wider field range.

\begin{figure}
   \centering
   \includegraphics[width=0.75\columnwidth]{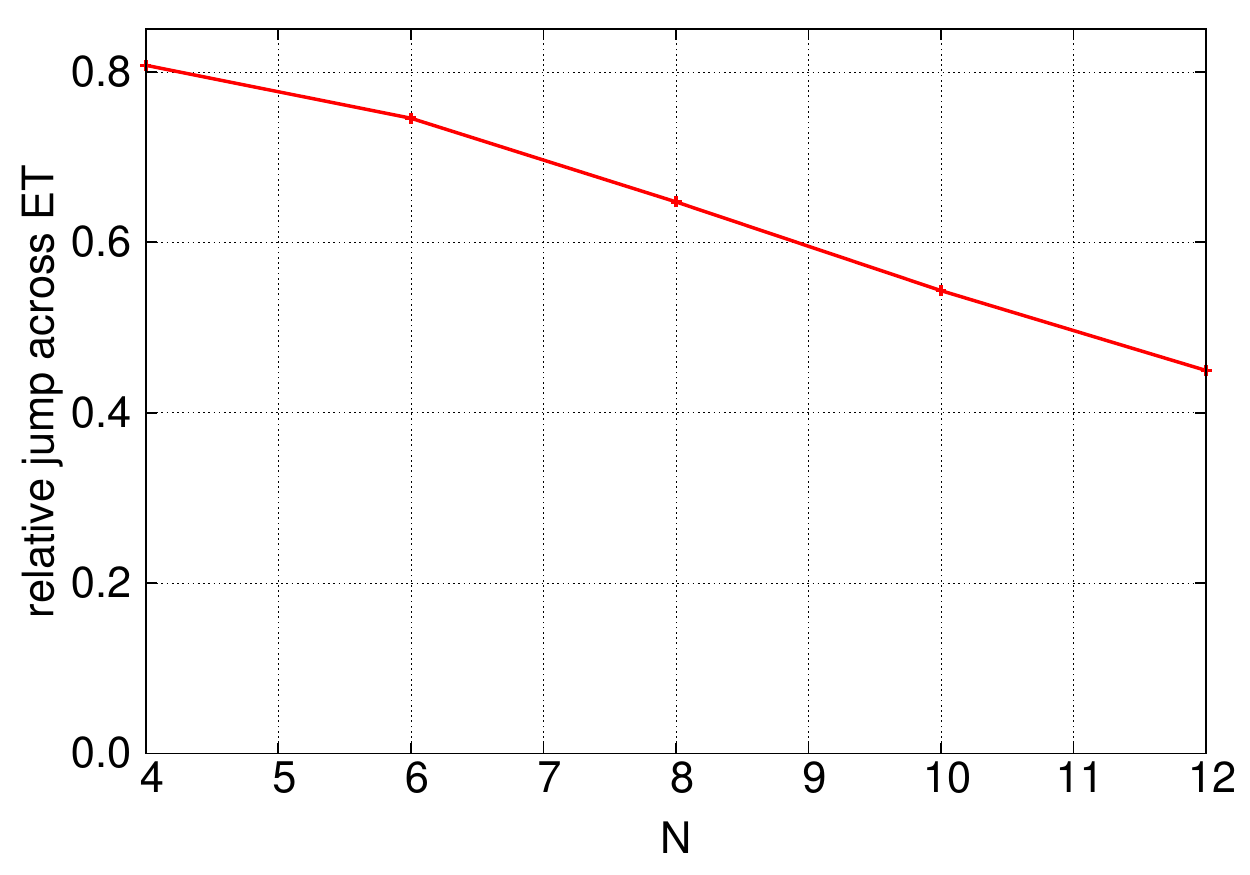}
\caption{\label{fig:size_effects}
Dependence of an entanglement transition signature
on cluster size, $N$. The plot shows the 
size of the jump in the quantity shown in Fig.~\ref{fig:Ground-state-value-at} as the entanglement transition
boundary is crossed, $\Delta S({\bf q}=0)$.
The in-plane anisotropy is fixed at $\gamma=0.4$.}
\end{figure}
\section{\label{sec:bulk}Approach to the bulk regime}

As the number of spins per cluster $N\to \infty$ our model approaches the limit of an infinite quantum spin chain. Interestingly, when the number of spins per cluster increases the
phenomena we have described become weaker, and it seems safe to predict that some of them cease to be useful to detect the entanglement transition
in the thermodynamic limit. Specifically, this is the case for the energy gap between the two lowest-lying states and the value of the magnetisation, as illustrated by the insets to Fig.~\ref{fig:gap}. The inset to the top panel shows the size of the energy gap between the cluster's ground state and first excited state, $\left|E_2-E_1\right|$, at $h_z=0$ as a function of $N$. Clearly, this energy gap vanishes rapidly as $N$ increases. The inset to the lower panel shows the rate of change of this gap with the applied field just above the factorisation value (red curve) and at a larger field (blue curve). Clearly, the field-dependence of this gap becomes flatter as the cluster size increases in the region near the factorisation field. In contrast, for larger fields the gradient is constant. This is consistent with the known fact that in the bulk limit ($N\to\infty$) the ground state is non-degenerate for $h_z>h_c=J$. For $h_z \leq h_c$, in contrast, the ground state has a two-fold degeneracy corresponding to $Z_2$ symmetry. Our results indicate that the way this degeneracy is achieved is quite different in the two sub-domains $0\leq h_z \leq h_f$ and $h_f \leq h_z < h_c$: in the former interval, the alternation between the two ground states, $\ket{1}$ and $\ket{2}$, becomes faster, and the energy gap separating them weaker (the number of closings of the gap in that interval is $N/2\to\infty$ as $N\to\infty$); in the latter interval, state $\ket{2}$ always has lower energy, and the gap increases monotonically with $h_z$, but the slope of that increase, $d\left|E_2-E_1\right|/dh_z,$ tends to zero as $N\to\infty$. In contrast for $h>h_c$ the slope remains finite as $N\to\infty$. Thus in the thermodynamic limit the quantity $d\left|E_2-E_1\right|/dh_z$ has a sudden jump from zero to a finite value at $h_c$, but is $h$-independent and equal to zero at $h_f$. As a direct consequence of this the closing of the gap is no longer a viable way of detecting the entanglement transition for infinite-chain compounds. The same conclusion applies to the magnetisation. We emphasise that the gap $\left|E_2-E_1\right|$ discussed here is quite distinct from the bulk gap separating the 2-fold degenerate ground state from the lowest-lying exctied states. The latter closes at the quantum critical point, not at the entanglement transition, whose signatures are quite different in the $N \to \infty$ limit.

The neutron-scattering signatures of the entanglement transition that we have discussed here are much clearer in smaller systems. This is already suggested by  Fig.~\ref{fig:Ground-state-value-at}, where the jump in $S({\bf q}=0)$ at $h_f$ for $N=6$ is somewhat less sharp than for $N=4$. Fig.~\ref{fig:size_effects} shows the dependence of the size of this jump $\Delta S({\bf q}=0)$ %difference in the values taken by $S({\bf q}=0)$ at fields just above and below $h_f$, respectively, 
on cluster size, $N$. Clearly, $\Delta S({\bf q}=0)$ decreases monotonically with $N$. This might suggest that it becomes negligible, making the entanglement transition undetectable by this method for very large clusters. However, we note that the $N$-dependence of this quantity is not nearly as fast as that of the gap $\left|E_2-E_1\right|$ (Fig.~\ref{fig:gap}, top panel inset). We cannot discard, from our finite-size calculations, the survival of a sharp feature in $\Delta S({\bf q})$ into the thermodynamic limit. In any case, in view of the discussion above it is clear that in infinite-chain compounds the situation is \textcolor{black}{overall quite different. 
%
%The above trends can be understood in terms of the correlation length $\xi$:
%at low enough temperatures, the system size may be smaller than $\xi$,
%placing the system effectively in the ground state. In contrast, for
%an infinite one-dimensional system any finite temperature leads to
%a system larger than $\xi$ and, therefore, to the proliferation of
%fluctuations (domains). 
%
%Although measures of entanglement evidence
%the entanglement transition at finite temperatures \cite{Amico2006}
%the spectrum, order parameter, and momentum-space correlations computed
%here seem not to reflect it in the large-$N$ limit. 
Our results suggest}
that whereas the quantum phase transition is the dominant phenomenon
in uniform systems, level crossings and the associated effects on entanglement dominate the phenomenology of clusters, where quantum critical effects are precluded by the finite system size. %That said, the entanglement transition does take place in infinite systems as well and we expect that experimentally-obtainable quantities may show signatures of it in the thermodynamic limit.
The neutron scattering signatures of the entanglement transition in infinite-chain compounds will be discussed elsewhere. 

\section{\label{sec:con}Conclusion}

We have predicted the experimental consequences of a field-tuned entanglement transition in clustered magnets, composed of independent units with a small number number $N$ of spins each, on the basis of a simple model. 
%As expected, we find that the quantum critical point existing in the $N\to \infty$ phase diagram is washed out by finite-size effects, even at zero temperature. In contrast, a
A number of ground-state crossings, culminating in the entanglement transition, lead to very sudden re-arrangements of the correlations, dramatically affecting the magnetisation and the neutron scattering cross-section. The latter effects survive at finite temperatures. %Our results suggest that the entanglement transition competes with the quantum critical point to determine the phenomenology of quantum magnets, becoming the dominant phenomenon in the case of clustered materials. This is in contrast with infinite-chain compounds, where the quantum critical point is dominant and the experimental consequences of the entanglement transition are much more subtle. 
The ability to observe and control the entanglement transition in clustered magnets opens the door to using the individual spins in such systems as qubits for quantum computation and the clusters themselves as multiqubit gates. The control of entanglement via a uniform (rather than local) magnetic field could be supplemented by uniform microwave irradiation to perform non-trivial multiqubit manipulations. %For example, tuning the field towards or away from $h_f$, where the range of entanglement diverges, could be used to propagate entanglement between different qubits. 

{\bf Acknowledgements:} The authors thank Miguel Angel Martin-Delgado for useful discussions and suggestions. HRI and JQ thank Sam T.~Carr, Greg Oliver, Paul Strange, Silvia Ramos and Chris Hooley for more useful discussions and Ewan Clark and Emma McCabe for bringing to their attention some relevant experimental literature. 

%%%%%%%%%%%%%%%%%%%%%%%%%%%%%%%%%%%%%%%%%%%%%%%%%%%%%%%%%%%%%
% APPENDIX
%

\appendix

\section{\label{sec:xyz}Anisotropic Heisenberg model}

\begin{figure} [h!]
    \centering
    \includegraphics[width=1.0\columnwidth]{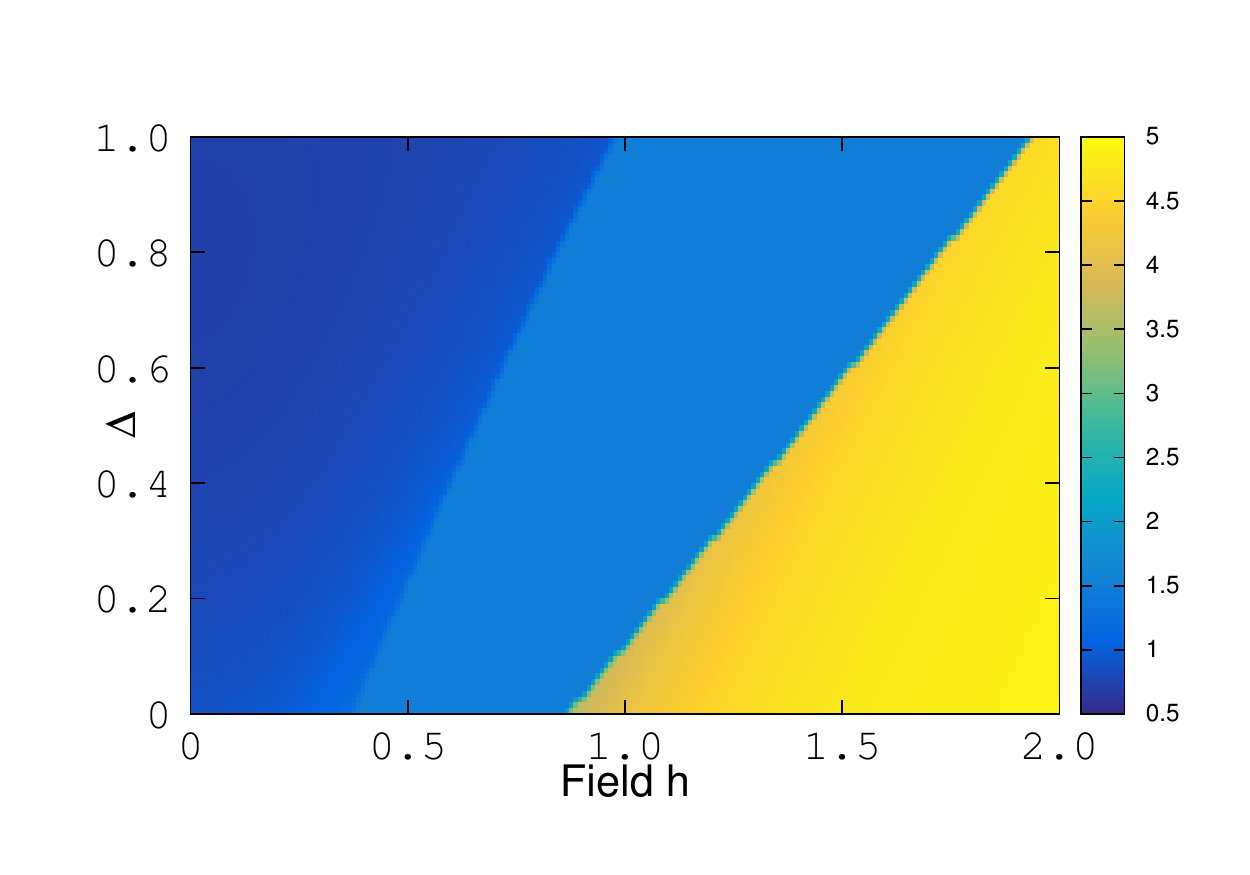}
     
    \caption{\label{fig:XYZphase}Ground-state value at $\mathbf{q}=0$
of the magnetic neutron scattering function, $S({\bf q})$, 
as a function
of $\Delta$ and $h_z$ for the Hamiltonian given by Eq.~(\ref{eq:H}) and Fig.~\ref{fig:model-1} with $\gamma=0.5$ and $N=4$.}
\end{figure}

The anisotropic Heisenberg model, or XYZ model, resulting when $\Delta > 0$ in Eq.~(\ref{eq:H}), behaves in much the same way as the anisotropic XY model discussed in the main text. The factorisation field depends on both ${\gamma}$ and ${\delta}$ and is given by \cite{Muller1982}
\begin{equation}
h_f=\sqrt{(1+{\Delta})^2-{\gamma}^2}
\label{eq:hf_xyz}
\end{equation}  
The same techniques employed for the anisotropic XY model can be employed here. As with the former model, the energy spectrum shows a level crossing between the two lowest-lying states at $h_z=h_f$ preceded by $N/2-1$ more crossings at lower fields. As in the XY model the last crossing indicates the entanglement transition where the ground state can be factorised. A phase diagram can be constructed in the same manner as Fig.~\ref{fig:Ground-state-value-at} and is given by Fig.~\ref{fig:XYZphase}. We find that the boundary between the yellow and purple regions is accurately given by (\ref{eq:hf_xyz}).

\section{\label{sec:gs_wf}Ground-state wave functions for $N=2$ and $N=4$}

It is straight-forward to obtain the wave functions of of our model analytically for $N=2$. For $h\leq h_f$, the ground state is (up to a normalisation factor) the anti-ferromagnetic singlet
\(
\left| 1 \right\rangle
=
%\frac{1}{\sqrt{2}}
%\left\lbrace
\left| \uparrow\downarrow \right\rangle
-
\left| \downarrow\uparrow \right\rangle
%\right\rbrace
%\equiv \left| \boxed{
%                                \begin{array}{cc}
%                                        \, & \,
%                                \end{array}
%                        } \right\rangle
\). For $h\geq h_f$ the ground state is ferromagnetic:
\(
\left| 2 \right\rangle
=
%\frac{1}{\sqrt{1-\delta^2}}
%\left\lbrace
\left| \uparrow\uparrow \right\rangle
+
\delta \left| \downarrow\downarrow \right\rangle
%\right\rbrace
.
\)
The parameter $\delta$ controls the amount of parallel entanglement in this state. It has the form 
\(
\delta
=
\sqrt{1+h_z^2/\gamma^2}-h_z/\gamma
\)
and evidently $\delta\to 0$ as $h_z\to \infty$. At $h=h_f$ any linear combination 
\begin{equation}
        \ket{\Psi}=A\left| 1\right\rangle+B\left| 2\right\rangle
        \label{eq:lc}
\end{equation} 
of these two states is a valid ground state. Remarkably, the coefficients $A$ and $B$ can be chosen so that the ground state factorises:
%\begin{equation}
\(
\ket{\Psi}=
\left(a_1\left|\uparrow\right\rangle+b_1\left|\downarrow\right\rangle\right)
\otimes
\left(a_2\left|\uparrow\right\rangle+b_2\left|\downarrow\right\rangle\right).
\)
%\end{equation}
Thus at exactly $h_f$ there is no entanglement. This is the factorisation field, given by the same formula (\ref{eq:hf}) that applies to an infinite chain. 

\begin{figure} [h!]
    \centering
    \includegraphics[width=1.0\columnwidth]{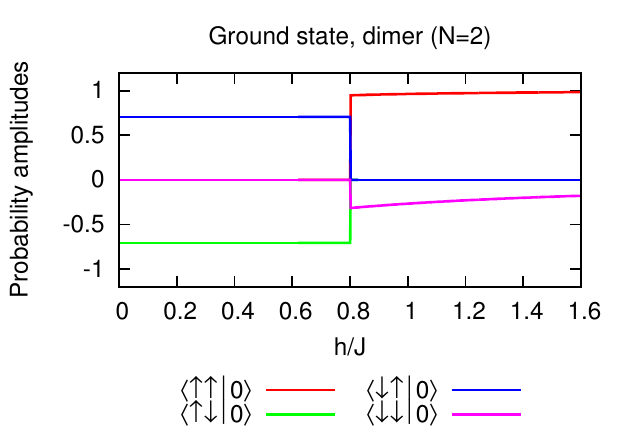} 
\includegraphics[width=1.0\columnwidth]{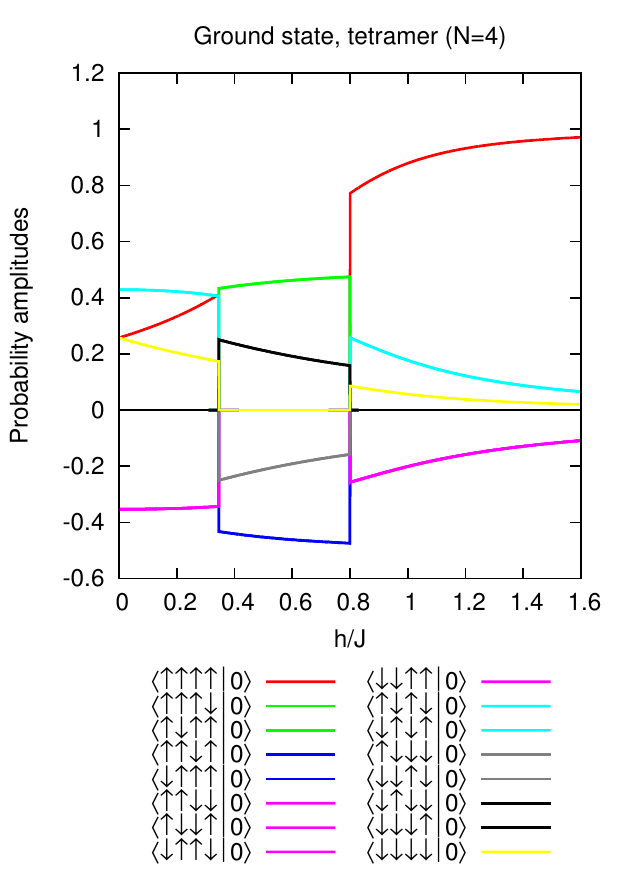} 

    \caption{\label{fig:wf}Field-dependence of the normalised ground state wave functions of our model for $\gamma = 0.6$, determined by exact diagonalisation. Top panel: $N=2$. Bottom panel: $N=4$. When more than one curve is assigned the same colour the curves coincide everywhere.}
\end{figure}

We have investigated higher values of $N=4,6,8,10,12$ by exact diagonalisation. We always find two lowest-lying states, $\left| 1 \right\rangle$ and $\left| 2 \right\rangle$, whose energies cross $N/2$ times as $h_z$ is increased. Unlike the $N=2$ case in general both states have finite magnetisation. However, $\ket{2}$ has non-zero amplitude of probability for the state in the basis corresponding to fully-saturated magnetisation, while for $\ket{1}$ the probability that all the spins are fully aligned is strictly zero. For instance, for $N=4$ the wave functions take the form (up to normalisation factors)
\begin{widetext}
\begin{eqnarray}
        \left| 1 \right\rangle
        &=&
        \frac{\alpha_1}{2}\kfst{\uw}{\uw}+\kfsr{\uw}{\uw}+\kfsb{\uw}{\uw}+\kfsl{\uw}{\uw}
        \nonumber
        \\
        &&
        +\frac{\alpha_2}{2}
        \left(
                \kfst{\dw}{\dw}+\kfsr{\dw}{\dw}+\kfsb{\dw}{\dw}+\kfsl{\dw}{\dw}        
        \right)
        \label{eq:state1N4}
        \\
        \left| 2 \right\rangle
        \nonumber
        &=&
        \beta_{1} \kfsn{\uw}{\uw}{\uw}{\uw}+\beta_{4} \kfsn{\dw}{\dw}{\dw}{\dw}
        \\
        &&
        +\beta_{2}  \left( \kfsn{\uw}{\dw}{\uw}{\dw} + \kfsn{\dw}{\uw}{\dw}{\uw} \right)
        \nonumber
        \\
        &&
        + \beta_{3} \left( \kfsn{\dw}{\uw}{\uw}{\dw}+\kfsn{\uw}{\dw}{\dw}{\uw}+\kfsn{\dw}{\dw}{\uw}{\uw}+\kfsn{\uw}{\uw}{\dw}{\dw}
        \right)
        \label{eq:state2N4}
\end{eqnarray}
\end{widetext}
where we have used the standard shorthand $\boxed{\begin{array}{cc} \, & \, \end{array}}$ for singlets. The parameters $\alpha_1,\alpha_2$, and $\beta_{1-4}$ are positive. $\alpha_2$ and $\beta_{2-4}$ are  monotonically-decreasing functions of $h_z$. The field-evolution of these wave functions is plotted in Fig.~\ref{fig:wf} alongside the $N=2$ case. Note that both ground states, $|1\rangle$ and $|2\rangle$, feature both parallel and anti-parallel entanglement. 

For all values of $N$ we investigated, the last crossing between the two ground states is at $h_f$. The ground state is $\ket{1}$ for $h_z \to h_f^-$ and $\ket{2}$ for any $h_z>h_f$. 

At $h_f$ the coefficients $A$ and $B$ in the linear combination (\ref{eq:lc}) can be chosen to produce an unentangled state, i.e. one of the from
\begin{eqnarray}
\hspace{-1cm}\ket{\Psi}&=&
\left(a_1\left|\uparrow\right\rangle+b_1\left|\downarrow\right\rangle\right)
\otimes
\left(a_2\left|\uparrow\right\rangle+b_2\left|\downarrow\right\rangle\right)
\otimes \nonumber
\\ && 
\hspace{-0.5cm}
\ldots
\otimes
\left(a_{N-1}\left|\uparrow\right\rangle+b_{N-1}\left|\downarrow\right\rangle\right)
\otimes
\left(a_N\left|\uparrow\right\rangle+b_N\left|\downarrow\right\rangle\right).
\label{eq:facst}
\end{eqnarray}
Indeed Kurmann, Thomas and Muller proved \cite{Muller1982} that the particular factorised state obtained by choosing $a_{2n+1}=a_1,b_{2n+1}=b_1,a_{2n}=a_2,b_{2n}=b_2$ for all $n=1,2,\ldots$ is realised at $h_f$ but not at any other value of the field (we note that the proof in \cite{Muller1982} is $N$-independent). 
In particular, the Kurmann-Thomas-Muller state is not realised at the other crossings occurring at lower values of $h_z$. 
One could ask, however, whether the more general factorised state in Eq.~(\ref{eq:facst}) could be achieved by an appropriate choice of the coefficients $A$ and $B$ at the other values of the field where there is a ground-state degeneracy. We have checked this explicitly in the $N=4$ case by examining the numerically-determined wave functions. 

Evidently, in view of structure of the ground state wave functions, given in Eqs.~(\ref{eq:state1N4},\ref{eq:state2N4}) and also shown in Fig.~\ref{fig:wf}, factorisation cannot be achieved unless there is degeneracy between $|1\rangle$ and $|2\rangle$. This still leaves open the possibility of factorisation at the field $h_1<h_f$ where the first gap closing occurs. To examine this possibility, we equate the linear superposition in (\ref{eq:lc}) to the factorsied state given in (\ref{eq:facst}). For $N$ spins, this leads to $2^N$ equations (one for each spin) in $2N+2$ unknowns ($A$, $B$ and the $a$ and $b$ coefficients). The variables are therefore over-determined for $N\geq 4$. Writing $\ket{1}$, $\ket{2}$ and $\ket{\Psi}$ in the basis $\left\lbrace\ket{\uparrow\uparrow\uparrow\uparrow},\ket{\uparrow\uparrow\uparrow\downarrow},\ket{\uparrow\uparrow\downarrow\uparrow},\ldots\right\rbrace$ used in Fig.~\ref{fig:wf} 
%we obtain
%\begin{equation}
%\left|1\right\rangle =\left(\begin{array}{c}
%0\\
%\alpha_{1}\\
%-\alpha_{1}\\
%0\\
%\alpha_{1}\\
%0\\
%0\\
%-\alpha_{2}\\
%-\alpha_{1}\\
%0\\
%0\\
%\alpha_{2}\\
%0\\
%-\alpha_{2}\\
%\alpha_{2}\\
%0
%\end{array}\right)~;~
%\left|2\right\rangle =
%\left(\begin{array}{c}
%-\beta_{1}\\
%0\\
%0\\
%\beta_{2}\\
%0\\
%-\beta_{3}\\
%\beta_{2}\\
%0\\
%0\\
%\beta_{2}\\
%-\beta_{3}\\
%0\\
%\beta_{2}\\
%0\\
%0\\
%-\beta_{4}
%\end{array}\right);
%\end{equation}
and equating amplitudes we arrive at the following set of equations: 
%while the factorised state $\ket{\Psi}$ takes the form 
%\begin{equation}
%        \left|\Psi\right\rangle =\left(\begin{array}{c}
%        a_1a_2a_3a_4\\
%        ...\\
%        ...\\
%        ...\\
%        ...\\
%        ...
%        \end{array}
%        \right)
%\end{equation}
%
%
%
%\textcolor{black}{Any linear combination of the two lowest states when they are degenerate are in turn valid %eigenstates. Using $\alpha$ and $\beta$ we show the vector configuration of the states for $N=4$ that are plotted %in Fig. {9}. It is easily seen that when compared to the factorised state given by equation {ref A4} that the 16 %simultaneous equations that are formed are decoupled into two subsets each dependent on one of the states. }
%
\[
\begin{array}{ccccccc}
A\alpha_{1} & = & a_{1}a_{2}a_{3}b_{4} & , & B\beta_{1} & = & a_{1}a_{2}a_{3}a_{4}\\
-A\alpha_{1} & = & a_{1}a_{2}b_{3}a_{4} & , & -B\beta_{2} & = & a_{1}a_{2}b_{3}b_{4}\\
A\alpha_{1} & = & a_{1}b_{2}a_{3}a_{4} & , & B\beta_{3} & = & a_{1}b_{2}a_{3}b_{4}\\
-A\alpha_{2} & = & a_{1}b_{2}b_{3}b_{4} & , & -B\beta_{2} & = & a_{1}b_{2}b_{3}a_{4}\\
-A\alpha_{1} & = & a_{1}b_{2}b_{3}b_{4} & , & -B\beta_{2} & = & b_{1}a_{2}a_{3}b_{4}\\
A\alpha_{2} & = & b_{1}a_{2}b_{3}b_{4} & , & B\beta_{3} & = & b_{1}a_{2}b_{3}a_{4}\\
-A\alpha_{2} & = & b_{1}b_{2}a_{3}b_{4} & , & -B\beta_{2} & = & b_{1}b_{2}a_{3}a_{4}\\
A\alpha_{2} & = & b_{1}b_{2}b_{3}a_{4} & , & B\beta_{4} & = & b_{1}b_{2}b_{3}b_{4}
\end{array}
\]
The coefficients $\alpha_1,\alpha_2,\beta_1,\beta_2,\beta_3$ and $\beta_4$ are determined by our exact diagonalisation calculation. It is easy to show that, given the values of these coefficients, the above system of equations in $A,B,a_1,a_2,a_3,a_4,b_1,b_2,b_3,b_4$ can have a solution only if 
\begin{equation}
\frac{\alpha_{1}^{2}\beta_{4}}{\alpha_{2}^{2}\beta_{1}}\frac{}{} = 1.
\end{equation}
For all values of the parameters we tested, this relation is obeyed to very high accuracy at $h=h_f$, but not at $h=h_1$. For example, for $\gamma=0.6$ we find that the ratio on the LHS of this equation equals 1 with a precision of 16 significant digits at $h_f = 0.8$, while at the other degeneracy field $h_1 \approx 0.345$ the same ratio is found to be 1.26.

\begin{figure} [h!]
    \centering
    \includegraphics[width=1.0\columnwidth]{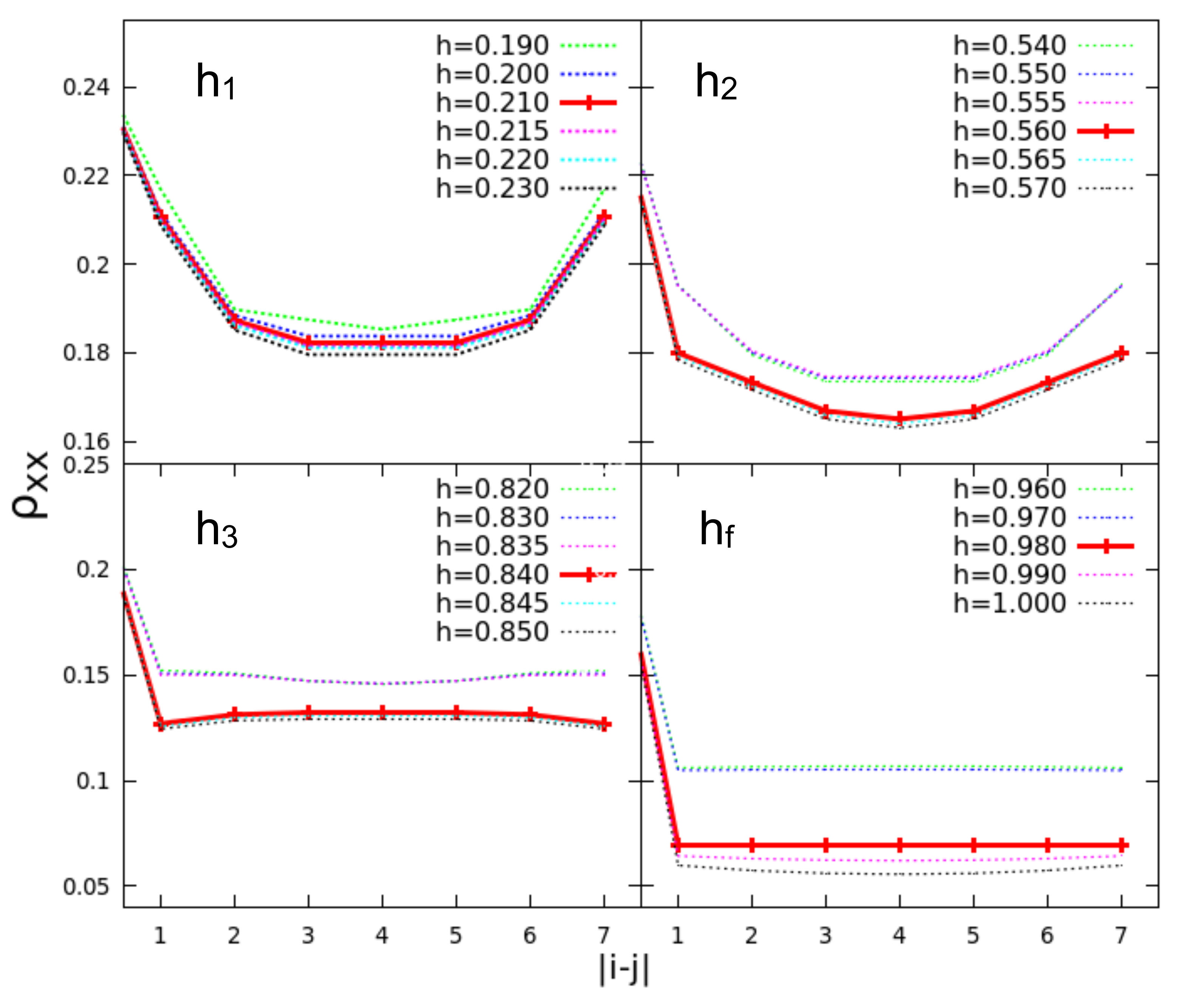} 

    \caption{\label{fig:flatcorr}Correlator between the $x$ components of the spins at sites $i$ and $j$ as a function of ``chain distance'' $|i-j|$ for the model in Eq.~(\ref{eq:H}) with $\gamma=0.2,\Delta=0$ and $N=8$. The results have been obtained by exact diagonalisation. Each panel shows the correlator for a number of values of the transverse field $h_z$ around each of the four special values where we find a ground state degeneracy, namely $h_1 \approx 0.210$, $h_2 \approx 0.560$, $h_3 \approx 0.840$ and $h_f = \sqrt{1-\gamma^2} \approx 0.980$, as indicated. The correlator is flat for $h_z = h_f$ only.  %\textcolor{black}{[Remove unnecessary labels "LC1, LC2, LC3, hf". %remove purple line in bottom-right panel and replace "0.9798" with "0.980". %Change "displacement" to "chain distance".]}
    }
\end{figure}

One of the most remarkable features of the factorised ground state is that the correlator $\rho_{xx}(i,j)$ between the $x$ components of the spins at two sites $i$ and $j$ becomes independent of the ``chain distance'' $|i-j|$ between the two sites \cite{Baroni2007} (as long as $i \neq j$). This is due to the special nature of the factorised ground state, which has long-range order and no quantum fluctuations. Additional evidence for the absence of exact foactorisation at other ground state degeneracies than the one at $h_f$ can be obtained by examining these correlators. Interestingly, for all instances of the model we have investigated the completely flat correlator is obtained for these finite systems too, but only at $h_f$. At the other ground state degeneracy fields the correlators are never flat. This is illustrated by Fig.~\ref{fig:flatcorr} which shows $\rho_{xx}(i,j)$ for $N=8$ and $\gamma=0.2$ (each panel shows the correlator for a range of fields near each of the four ground state degeneracy fields). Note also that the higher the field, the flatter the correlator is at degeneracy.

\section{\label{sec:neutron-formalism}Calculation of the neutron-scattering cross-section}

In the model defined by Fig.~\ref{fig:model-1} and Eq.~(\ref{eq:H}), $\hat{S}_{i}^{x}$ and $\hat{S}_{i}^{y}$ are the first two components of the spin
at site $i$, measured along axes contained in the $xy$ plane but forming an angle $\phi_{i}$
with the $x$ and $y$ axes, respectively. 
Let $\textcolor{black}{\hat{\tilde{S}}}_{\mathbf{R}_{i}}^{\alpha}$ be the $\alpha^{\underline{\mbox{th}}}$
component of the spin at site $i$ with respect to the global axes
$x,y$ depicted in Fig.~\ref{fig:model-1}, which are site-independent. These are global axes fixed to the orientation of the crystal. In the case of a neutron
scattering experiment, they could equivalently be taken to be the axes of the instrument. The neutron scattering cross-section is \citep{Lovesey1987b}
\begin{equation}
\frac{\partial^{2}\sigma}{\partial\Omega\partial E'}=\frac{k'}{k}\left(\gamma r_{e}\right)^{2}\left|\frac{g}{2}F\left(\mathbf{q}\right)\right|^{2}e^{-2W\left(\mathbf{q}\right)}S\left(\mathbf{q},\omega\right),\label{eq:cross-section}
\end{equation}
Here $\sigma$ is just a standard notation for cross-section. The
total scattering function is 
\begin{equation}
S\left(\mathbf{q},\omega\right)=\sum_{\alpha,\beta}\left(\delta_{\alpha,\beta}-\hat{\mathbf{q}}_{\alpha}\hat{\mathbf{q}}_{\beta}\right)S_{\alpha\beta}\left(\mathbf{q},\omega\right)
\label{S_breakdown}
\end{equation}
where the spin-resolved scattering function is given by 
\begin{equation}
S_{\alpha\beta}\left(\mathbf{q},\omega\right)=\frac{1}{2\pi\hbar}\int dte^{-i\omega t}\left\langle \textcolor{black}{\hat{\tilde{S}}}_{\mathbf{q}}^{\alpha}\left(0\right)\textcolor{black}{\hat{\tilde{S}}}_{-\mathbf{q}}^{\beta}\left(t\right)\right\rangle .\label{eq:S}
\end{equation}
Here, 
\begin{equation}
\textcolor{black}{\hat{\tilde{\mathbf{S}}}}_{\mathbf{q}}=\frac{1}{\sqrt{N}}\sum_{\mathbf{R}_{j}}e^{i\mathbf{q}.\mathbf{R}_{j}}\textcolor{black}{\hat{\tilde{\mathbf{S}}}}_{j}\label{eq:FTs}
\end{equation}
is the Fourier transform of the spin operator expressed in terms
of the global axes. Assuming we know the magnetic form factor, Debye-Waller
factor, etc. and that we detect all neutrons regardless of the energy
exchanged with the sample, $\hbar\omega$, our experiment gives the integral 
\(
S\left(\mathbf{q}\right) \equiv \int d\omega S\left(\mathbf{q},\omega\right),
\)
which can be straight-forwardly related {\it via} (\ref{S_breakdown}) to the energy-integrated scattering function,
\begin{equation}
S_{\alpha\beta}\left(\mathbf{q}\right)=\int d\omega S_{\alpha\beta}\left(\mathbf{q},\omega\right).
\end{equation}
Inserting (\ref{eq:FTs})
into (\ref{eq:S}) and integrating w.r.t. $\omega$ we obtain 
\begin{equation}
S_{\alpha\beta}\left(\mathbf{q}\right)=\frac{1}{N\hbar}\sum_{i,j}e^{i\mathbf{q}.\left(\mathbf{R}_{i}-\mathbf{R}_{j}\right)}\left\langle \textcolor{black}{\hat{\tilde{S}}}_{i}^{\alpha}\textcolor{black}{\hat{\tilde{S}}}_{j}^{\beta}\right\rangle.
\label{eq:Salphabeta}
\end{equation}
Here $\mathbf{R}_{i}$ denotes the position vector of the $i$\uth magnetic site in the cluster. 
The problem of predicting the neutron scattering experiment therefore
reduces to expressing the correlators $\left\langle \textcolor{black}{\hat{\tilde{S}}}_{i}^{\alpha}\textcolor{black}{\hat{\tilde{S}}}_{j}^{\beta}\right\rangle $
in terms of those in terms of the local axes, $\left\langle \hat{S}_{i}^{\alpha}\hat{S}_{j}^{\beta}\right\rangle $.
We do this using the rotations
\begin{equation}
\begin{array}{rcll}
\textcolor{black}{\hat{\tilde{S}}}_{i}^{x} & = & \cos\phi_{i}\hat{S}_{i}^{x}  -\sin\phi_{i}\hat{S}_{i}^{y} & ;\\
\textcolor{black}{\hat{\tilde{S}}}_{i}^{y} & = & \sin\phi_{i}\hat{S}_{i}^{x}  +\cos\phi_{i}\hat{S}_{i}^{y} & ;\\
\textcolor{black}{\hat{\tilde{S}}}_{i}^{z} & = &  \hat{S}_{i}^{z}  &. 
\end{array}
\end{equation}
Thus 
\begin{equation}
\textcolor{black}{\hat{\tilde{S}}}_{i}^{\alpha}=\sum_{\mu}\Lambda_{\alpha,\mu}\left(\phi_{i}\right)\hat{S}_{i}^{\mu}
\end{equation}
where the matrix 
\begin{equation}
\left[\Lambda_{\alpha,\mu}\left(\phi\right)\right]_{\begin{array}{rcl}
\alpha & = & x,y,z\\
\mu & = & x,y,z
\end{array}}=\left(\begin{array}{ccc}
\cos\phi & -\sin\phi & 0\\
\sin\phi & \cos\phi & 0\\
0 & 0 & 1
\end{array}\right)
\end{equation}
and the correlators are 
\begin{equation}
\left\langle \textcolor{black}{\hat{\tilde{S}}}_{i}^{\alpha}\textcolor{black}{\hat{\tilde{S}}}_{j}^{\beta}\right\rangle =\sum_{\mu}\sum_{\gamma}\Lambda_{\alpha,\mu}\left(\phi_{i}\right)\Lambda_{\beta,\gamma}\left(\phi_{j}\right)\left\langle \hat{S}_{i}^{\mu}\hat{S}_{j}^{\gamma}\right\rangle ,\label{eq:sig_sig}
\end{equation}
which reduces the problem of calculating the correlators between components
of the spins defined with respect to the instrument's axes $\left\langle \textcolor{black}{\hat{\tilde{S}}}_{i}^{\alpha}\textcolor{black}{\hat{\tilde{S}}}_{j}^{\beta}\right\rangle $
to the correlators with respect to the local crystal axes $\left\langle \hat{S}_{i}^{\mu}\hat{S}_{j}^{\gamma}\right\rangle $. 

We can insert this into (\ref{eq:Salphabeta}) to calculate $S_{\alpha\beta}\left(\mathbf{q}\right)$.
Once we have $S_{\alpha\beta}\left(\mathbf{q}\right)$ it is easy
to get $S\left(\mathbf{q}\right)$.

%%%%%%%%%%%%%%%%%%%%%%%%%%%%%%%%%%%%%%%%%%%%%%%%%%%%%%%%%%%%%%%%%%%%%%%%
%
% CALL TO .bib FILE. COMMENT OUT TO USE .bbl OUTPUT INSTEAD.
%
%\bibliography{paper_bibliography_v36}

\begin{thebibliography}{55}%
\makeatletter
\providecommand \@ifxundefined [1]{%
 \@ifx{#1\undefined}
}%
\providecommand \@ifnum [1]{%
 \ifnum #1\expandafter \@firstoftwo
 \else \expandafter \@secondoftwo
 \fi
}%
\providecommand \@ifx [1]{%
 \ifx #1\expandafter \@firstoftwo
 \else \expandafter \@secondoftwo
 \fi
}%
\providecommand \natexlab [1]{#1}%
\providecommand \enquote  [1]{``#1''}%
\providecommand \bibnamefont  [1]{#1}%
\providecommand \bibfnamefont [1]{#1}%
\providecommand \citenamefont [1]{#1}%
\providecommand \href@noop [0]{\@secondoftwo}%
\providecommand \href [0]{\begingroup \@sanitize@url \@href}%
\providecommand \@href[1]{\@@startlink{#1}\@@href}%
\providecommand \@@href[1]{\endgroup#1\@@endlink}%
\providecommand \@sanitize@url [0]{\catcode `\\12\catcode `\$12\catcode
  `\&12\catcode `\#12\catcode `\^12\catcode `\_12\catcode `\%12\relax}%
\providecommand \@@startlink[1]{}%
\providecommand \@@endlink[0]{}%
\providecommand \url  [0]{\begingroup\@sanitize@url \@url }%
\providecommand \@url [1]{\endgroup\@href {#1}{\urlprefix }}%
\providecommand \urlprefix  [0]{URL }%
\providecommand \Eprint [0]{\href }%
\providecommand \doibase [0]{http://dx.doi.org/}%
\providecommand \selectlanguage [0]{\@gobble}%
\providecommand \bibinfo  [0]{\@secondoftwo}%
\providecommand \bibfield  [0]{\@secondoftwo}%
\providecommand \translation [1]{[#1]}%
\providecommand \BibitemOpen [0]{}%
\providecommand \bibitemStop [0]{}%
\providecommand \bibitemNoStop [0]{.\EOS\space}%
\providecommand \EOS [0]{\spacefactor3000\relax}%
\providecommand \BibitemShut  [1]{\csname bibitem#1\endcsname}%
\let\auto@bib@innerbib\@empty
%</preamble>
\bibitem [{\citenamefont {Mathur}\ \emph {et~al.}(1998)\citenamefont {Mathur},
  \citenamefont {Grosche}, \citenamefont {Julian}, \citenamefont {Walker},
  \citenamefont {Freye}, \citenamefont {Haselwimmer},\ and\ \citenamefont
  {Lonzarich}}]{Mathur1998}%
  \BibitemOpen
  \bibfield  {author} {\bibinfo {author} {\bibfnamefont {N.~D.}\ \bibnamefont
  {Mathur}}, \bibinfo {author} {\bibfnamefont {F.~M.}\ \bibnamefont {Grosche}},
  \bibinfo {author} {\bibfnamefont {S.~R.}\ \bibnamefont {Julian}}, \bibinfo
  {author} {\bibfnamefont {I.~R.}\ \bibnamefont {Walker}}, \bibinfo {author}
  {\bibfnamefont {D.~M.}\ \bibnamefont {Freye}}, \bibinfo {author}
  {\bibfnamefont {R.~K.~W.}\ \bibnamefont {Haselwimmer}}, \ and\ \bibinfo
  {author} {\bibfnamefont {G.~G.}\ \bibnamefont {Lonzarich}},\ }\href
  {http://dx.doi.org/10.1038/27838} {\bibfield  {journal} {\bibinfo  {journal}
  {Nature}\ }\textbf {\bibinfo {volume} {394}},\ \bibinfo {pages} {39}
  (\bibinfo {year} {1998})}\BibitemShut {NoStop}%
\bibitem [{\citenamefont {Saxena}\ \emph {et~al.}(2000)\citenamefont {Saxena},
  \citenamefont {Agarwal}, \citenamefont {Ahilan}, \citenamefont {Grosche},
  \citenamefont {Haselwimmer}, \citenamefont {Steiner}, \citenamefont {Pugh},
  \citenamefont {Walker}, \citenamefont {Julian}, \citenamefont {Monthoux},
  \citenamefont {Lonzarich}, \citenamefont {Huxley}, \citenamefont {Sheikin},
  \citenamefont {Braithwaite},\ and\ \citenamefont {Flouquet}}]{Saxena2000}%
  \BibitemOpen
  \bibfield  {author} {\bibinfo {author} {\bibfnamefont {S.~S.}\ \bibnamefont
  {Saxena}}, \bibinfo {author} {\bibfnamefont {P.}~\bibnamefont {Agarwal}},
  \bibinfo {author} {\bibfnamefont {K.}~\bibnamefont {Ahilan}}, \bibinfo
  {author} {\bibfnamefont {F.~M.}\ \bibnamefont {Grosche}}, \bibinfo {author}
  {\bibfnamefont {R.~K.~W.}\ \bibnamefont {Haselwimmer}}, \bibinfo {author}
  {\bibfnamefont {M.~J.}\ \bibnamefont {Steiner}}, \bibinfo {author}
  {\bibfnamefont {E.}~\bibnamefont {Pugh}}, \bibinfo {author} {\bibfnamefont
  {I.~R.}\ \bibnamefont {Walker}}, \bibinfo {author} {\bibfnamefont {S.~R.}\
  \bibnamefont {Julian}}, \bibinfo {author} {\bibfnamefont {P.}~\bibnamefont
  {Monthoux}}, \bibinfo {author} {\bibfnamefont {G.~G.}\ \bibnamefont
  {Lonzarich}}, \bibinfo {author} {\bibfnamefont {A.}~\bibnamefont {Huxley}},
  \bibinfo {author} {\bibfnamefont {I.}~\bibnamefont {Sheikin}}, \bibinfo
  {author} {\bibfnamefont {D.}~\bibnamefont {Braithwaite}}, \ and\ \bibinfo
  {author} {\bibfnamefont {J.}~\bibnamefont {Flouquet}},\ }\href
  {http://dx.doi.org/10.1038/35020500} {\bibfield  {journal} {\bibinfo
  {journal} {Nature}\ }\textbf {\bibinfo {volume} {406}},\ \bibinfo {pages}
  {587} (\bibinfo {year} {2000})}\BibitemShut {NoStop}%
\bibitem [{\citenamefont {Coldea}\ \emph {et~al.}(2010)\citenamefont {Coldea},
  \citenamefont {Tennant}, \citenamefont {Wheeler}, \citenamefont {Wawrzynska},
  \citenamefont {Prabhakaran}, \citenamefont {Telling}, \citenamefont
  {Habicht}, \citenamefont {Smeibidl},\ and\ \citenamefont
  {Kiefer}}]{ColdeaE8}%
  \BibitemOpen
  \bibfield  {author} {\bibinfo {author} {\bibfnamefont {R.}~\bibnamefont
  {Coldea}}, \bibinfo {author} {\bibfnamefont {D.~A.}\ \bibnamefont {Tennant}},
  \bibinfo {author} {\bibfnamefont {E.~M.}\ \bibnamefont {Wheeler}}, \bibinfo
  {author} {\bibfnamefont {E.}~\bibnamefont {Wawrzynska}}, \bibinfo {author}
  {\bibfnamefont {D.}~\bibnamefont {Prabhakaran}}, \bibinfo {author}
  {\bibfnamefont {M.}~\bibnamefont {Telling}}, \bibinfo {author} {\bibfnamefont
  {K.}~\bibnamefont {Habicht}}, \bibinfo {author} {\bibfnamefont
  {P.}~\bibnamefont {Smeibidl}}, \ and\ \bibinfo {author} {\bibfnamefont
  {K.}~\bibnamefont {Kiefer}},\ }\href@noop {} {\bibfield  {journal} {\bibinfo
  {journal} {Science}\ }\textbf {\bibinfo {volume} {327}},\ \bibinfo {pages}
  {177} (\bibinfo {year} {2010})}\BibitemShut {NoStop}%
\bibitem [{\citenamefont {Anderson}(1987)}]{Anderson1987}%
  \BibitemOpen
  \bibfield  {author} {\bibinfo {author} {\bibfnamefont {P.~W.}\ \bibnamefont
  {Anderson}},\ }\href@noop {} {\bibfield  {journal} {\bibinfo  {journal}
  {Science}\ }\textbf {\bibinfo {volume} {235}},\ \bibinfo {pages} {1196}
  (\bibinfo {year} {1987})}\BibitemShut {NoStop}%
\bibitem [{\citenamefont {R\"{u}egg}\ \emph {et~al.}(2003)\citenamefont
  {R\"{u}egg}, \citenamefont {Cavadini}, \citenamefont {Petrology},
  \citenamefont {Furrer}, \citenamefont {Ablation}, \citenamefont {G\"{u}del},
  \citenamefont {Mineralogy}, \citenamefont {Kr\"{a}mer}, \citenamefont
  {Mutka}, \citenamefont {Wildes}, \citenamefont {Habicht},\ and\ \citenamefont
  {Vorderwisch}}]{Ruegg2003}%
  \BibitemOpen
  \bibfield  {author} {\bibinfo {author} {\bibfnamefont {C.}~\bibnamefont
  {R\"{u}egg}}, \bibinfo {author} {\bibfnamefont {N.}~\bibnamefont {Cavadini}},
  \bibinfo {author} {\bibfnamefont {M.}~\bibnamefont {Petrology}}, \bibinfo
  {author} {\bibfnamefont {a.}~\bibnamefont {Furrer}}, \bibinfo {author}
  {\bibfnamefont {Y.}~\bibnamefont {Ablation}}, \bibinfo {author}
  {\bibfnamefont {H.-U.}\ \bibnamefont {G\"{u}del}}, \bibinfo {author}
  {\bibfnamefont {K.~J.}\ \bibnamefont {Mineralogy}}, \bibinfo {author}
  {\bibfnamefont {K.}~\bibnamefont {Kr\"{a}mer}}, \bibinfo {author}
  {\bibfnamefont {H.}~\bibnamefont {Mutka}}, \bibinfo {author} {\bibfnamefont
  {a.}~\bibnamefont {Wildes}}, \bibinfo {author} {\bibfnamefont
  {K.}~\bibnamefont {Habicht}}, \ and\ \bibinfo {author} {\bibfnamefont
  {P.}~\bibnamefont {Vorderwisch}},\ }\href {\doibase 10.1038/nature01567.1.}
  {\bibfield  {journal} {\bibinfo  {journal} {Nature}\ }\textbf {\bibinfo
  {volume} {423}},\ \bibinfo {pages} {62} (\bibinfo {year} {2003})}\BibitemShut
  {NoStop}%
\bibitem [{\citenamefont {Merchant}\ \emph {et~al.}(2014)\citenamefont
  {Merchant}, \citenamefont {Normand}, \citenamefont {Kramer}, \citenamefont
  {Boehm}, \citenamefont {McMorrow},\ and\ \citenamefont {Ruegg}}]{Ruegg2014}%
  \BibitemOpen
  \bibfield  {author} {\bibinfo {author} {\bibfnamefont {P.}~\bibnamefont
  {Merchant}}, \bibinfo {author} {\bibfnamefont {B.}~\bibnamefont {Normand}},
  \bibinfo {author} {\bibfnamefont {K.~W.}\ \bibnamefont {Kramer}}, \bibinfo
  {author} {\bibfnamefont {M.}~\bibnamefont {Boehm}}, \bibinfo {author}
  {\bibfnamefont {D.~F.}\ \bibnamefont {McMorrow}}, \ and\ \bibinfo {author}
  {\bibfnamefont {C.}~\bibnamefont {Ruegg}},\ }\href {\doibase
  10.1038/nphys2902} {\bibfield  {journal} {\bibinfo  {journal} {Nature
  Physics}\ }\textbf {\bibinfo {volume} {10}},\ \bibinfo {pages} {373}
  (\bibinfo {year} {2014})}\BibitemShut {NoStop}%
\bibitem [{\citenamefont {Sachdev}(2011)}]{Sachdev2011}%
  \BibitemOpen
  \bibfield  {author} {\bibinfo {author} {\bibfnamefont {S.}~\bibnamefont
  {Sachdev}},\ }\href@noop {} {\emph {\bibinfo {title} {{Quantum Phase
  Transitions}}}}\ (\bibinfo  {publisher} {Cambridge University Press},\
  \bibinfo {year} {2011})\BibitemShut {NoStop}%
\bibitem [{\citenamefont {Amico}\ \emph {et~al.}(2008)\citenamefont {Amico},
  \citenamefont {Fazio}, \citenamefont {Osterloh},\ and\ \citenamefont
  {Vedral}}]{Amico2008}%
  \BibitemOpen
  \bibfield  {author} {\bibinfo {author} {\bibfnamefont {L.}~\bibnamefont
  {Amico}}, \bibinfo {author} {\bibfnamefont {R.}~\bibnamefont {Fazio}},
  \bibinfo {author} {\bibfnamefont {A.}~\bibnamefont {Osterloh}}, \ and\
  \bibinfo {author} {\bibfnamefont {V.}~\bibnamefont {Vedral}},\ }\href@noop {}
  {\bibfield  {journal} {\bibinfo  {journal} {Reviews of Modern Physics}\
  }\textbf {\bibinfo {volume} {80}},\ \bibinfo {pages} {517} (\bibinfo {year}
  {2008})}\BibitemShut {NoStop}%
\bibitem [{\citenamefont {Kurmann}\ \emph {et~al.}(1982)\citenamefont
  {Kurmann}, \citenamefont {Thomas},\ and\ \citenamefont
  {Muller}}]{Muller1982}%
  \BibitemOpen
  \bibfield  {author} {\bibinfo {author} {\bibfnamefont {J.}~\bibnamefont
  {Kurmann}}, \bibinfo {author} {\bibfnamefont {H.}~\bibnamefont {Thomas}}, \
  and\ \bibinfo {author} {\bibfnamefont {G.}~\bibnamefont {Muller}},\
  }\href@noop {} {\bibfield  {journal} {\bibinfo  {journal} {Physica}\ }\textbf
  {\bibinfo {volume} {112A}},\ \bibinfo {pages} {235} (\bibinfo {year}
  {1982})}\BibitemShut {NoStop}%
\bibitem [{\citenamefont {Roscilde}\ \emph {et~al.}(2004)\citenamefont
  {Roscilde}, \citenamefont {Verrucchi}, \citenamefont {Fubini}, \citenamefont
  {Haas},\ and\ \citenamefont {Tognetti}}]{Roscilde2004}%
  \BibitemOpen
  \bibfield  {author} {\bibinfo {author} {\bibfnamefont {T.}~\bibnamefont
  {Roscilde}}, \bibinfo {author} {\bibfnamefont {P.}~\bibnamefont {Verrucchi}},
  \bibinfo {author} {\bibfnamefont {A.}~\bibnamefont {Fubini}}, \bibinfo
  {author} {\bibfnamefont {S.}~\bibnamefont {Haas}}, \ and\ \bibinfo {author}
  {\bibfnamefont {V.}~\bibnamefont {Tognetti}},\ }\href@noop {} {\bibfield
  {journal} {\bibinfo  {journal} {Phys.~Rev.~Lett.}\ }\textbf {\bibinfo
  {volume} {93}},\ \bibinfo {pages} {167203} (\bibinfo {year}
  {2004})}\BibitemShut {NoStop}%
\bibitem [{\citenamefont {Roscilde}\ \emph {et~al.}(2005)\citenamefont
  {Roscilde}, \citenamefont {Verrucchi}, \citenamefont {Fubini}, \citenamefont
  {Haas},\ and\ \citenamefont {Tognetti}}]{Roscilde2005}%
  \BibitemOpen
  \bibfield  {author} {\bibinfo {author} {\bibfnamefont {T.}~\bibnamefont
  {Roscilde}}, \bibinfo {author} {\bibfnamefont {P.}~\bibnamefont {Verrucchi}},
  \bibinfo {author} {\bibfnamefont {A.}~\bibnamefont {Fubini}}, \bibinfo
  {author} {\bibfnamefont {S.}~\bibnamefont {Haas}}, \ and\ \bibinfo {author}
  {\bibfnamefont {V.}~\bibnamefont {Tognetti}},\ }\href@noop {} {\bibfield
  {journal} {\bibinfo  {journal} {Phys.~Rev.~Lett.}\ }\textbf {\bibinfo
  {volume} {94}},\ \bibinfo {pages} {147208} (\bibinfo {year}
  {2005})}\BibitemShut {NoStop}%
\bibitem [{\citenamefont {Amico}\ \emph {et~al.}(2006)\citenamefont {Amico},
  \citenamefont {Baroni}, \citenamefont {Fubini}, \citenamefont {Patan\`{e}},
  \citenamefont {Tognetti},\ and\ \citenamefont {Verrucchi}}]{Amico2006}%
  \BibitemOpen
  \bibfield  {author} {\bibinfo {author} {\bibfnamefont {L.}~\bibnamefont
  {Amico}}, \bibinfo {author} {\bibfnamefont {F.}~\bibnamefont {Baroni}},
  \bibinfo {author} {\bibfnamefont {A.}~\bibnamefont {Fubini}}, \bibinfo
  {author} {\bibfnamefont {D.}~\bibnamefont {Patan\`{e}}}, \bibinfo {author}
  {\bibfnamefont {V.}~\bibnamefont {Tognetti}}, \ and\ \bibinfo {author}
  {\bibfnamefont {P.}~\bibnamefont {Verrucchi}},\ }\href@noop {} {\bibfield
  {journal} {\bibinfo  {journal} {Phys.~Rev.~A}\ }\textbf {\bibinfo {volume}
  {74}},\ \bibinfo {pages} {022322} (\bibinfo {year} {2006})}\BibitemShut
  {NoStop}%
\bibitem [{\citenamefont {Fubini}\ \emph {et~al.}(2006)\citenamefont {Fubini}
  \emph {et~al.}}]{Fubini2006}%
  \BibitemOpen
  \bibfield  {author} {\bibinfo {author} {\bibfnamefont {A.}~\bibnamefont
  {Fubini}} \emph {et~al.},\ }\href@noop {} {\bibfield  {journal} {\bibinfo
  {journal} {Eurs.~Phys.~J.~D}\ }\textbf {\bibinfo {volume} {38}},\ \bibinfo
  {pages} {563} (\bibinfo {year} {2006})}\BibitemShut {NoStop}%
\bibitem [{\citenamefont {Giampaolo}\ \emph {et~al.}(2009)\citenamefont
  {Giampaolo}, \citenamefont {Adesso},\ and\ \citenamefont
  {Illuminati}}]{Giampaolo2009}%
  \BibitemOpen
  \bibfield  {author} {\bibinfo {author} {\bibfnamefont {S.~M.}\ \bibnamefont
  {Giampaolo}}, \bibinfo {author} {\bibfnamefont {G.}~\bibnamefont {Adesso}}, \
  and\ \bibinfo {author} {\bibfnamefont {F.}~\bibnamefont {Illuminati}},\
  }\href {\doibase 10.1103/PhysRevB.79.224434} {\bibfield  {journal} {\bibinfo
  {journal} {Phys. Rev. B}\ }\textbf {\bibinfo {volume} {79}},\ \bibinfo
  {pages} {224434} (\bibinfo {year} {2009})}\BibitemShut {NoStop}%
\bibitem [{\citenamefont {Giampaolo}\ \emph {et~al.}(2010)\citenamefont
  {Giampaolo}, \citenamefont {Adesso},\ and\ \citenamefont
  {Illuminati}}]{Giampaolo2010}%
  \BibitemOpen
  \bibfield  {author} {\bibinfo {author} {\bibfnamefont {S.~M.}\ \bibnamefont
  {Giampaolo}}, \bibinfo {author} {\bibfnamefont {G.}~\bibnamefont {Adesso}}, \
  and\ \bibinfo {author} {\bibfnamefont {F.}~\bibnamefont {Illuminati}},\
  }\href@noop {} {\bibfield  {journal} {\bibinfo  {journal} {Phys.~Rev.~Lett.}\
  }\textbf {\bibinfo {volume} {104}},\ \bibinfo {pages} {207202} (\bibinfo
  {year} {2010})}\BibitemShut {NoStop}%
\bibitem [{\citenamefont {Ghosh}\ \emph {et~al.}(2003)\citenamefont {Ghosh},
  \citenamefont {Rosenbaum}, \citenamefont {Aeppli},\ and\ \citenamefont
  {Coppersmith}}]{Ghosh2003}%
  \BibitemOpen
  \bibfield  {author} {\bibinfo {author} {\bibfnamefont {S.}~\bibnamefont
  {Ghosh}}, \bibinfo {author} {\bibfnamefont {T.~F.}\ \bibnamefont
  {Rosenbaum}}, \bibinfo {author} {\bibfnamefont {G.}~\bibnamefont {Aeppli}}, \
  and\ \bibinfo {author} {\bibfnamefont {S.~N.}\ \bibnamefont {Coppersmith}},\
  }\href@noop {} {\bibfield  {journal} {\bibinfo  {journal} {Nature}\ }\textbf
  {\bibinfo {volume} {425}},\ \bibinfo {pages} {48} (\bibinfo {year}
  {2003})}\BibitemShut {NoStop}%
\bibitem [{\citenamefont {Brukner}\ \emph {et~al.}(2006)\citenamefont
  {Brukner}, \citenamefont {Vedral},\ and\ \citenamefont
  {Zeilinger}}]{Brukner2004}%
  \BibitemOpen
  \bibfield  {author} {\bibinfo {author} {\bibfnamefont {C.}~\bibnamefont
  {Brukner}}, \bibinfo {author} {\bibfnamefont {V.}~\bibnamefont {Vedral}}, \
  and\ \bibinfo {author} {\bibfnamefont {A.}~\bibnamefont {Zeilinger}},\ }\href
  {\doibase 10.1103/PhysRevA.73.012110} {\bibfield  {journal} {\bibinfo
  {journal} {Phys. Rev. A}\ }\textbf {\bibinfo {volume} {73}},\ \bibinfo
  {pages} {012110} (\bibinfo {year} {2006})}\BibitemShut {NoStop}%
\bibitem [{\citenamefont {Bose}\ and\ \citenamefont
  {Tribedi}(2005)}]{Bose2005}%
  \BibitemOpen
  \bibfield  {author} {\bibinfo {author} {\bibfnamefont {I.}~\bibnamefont
  {Bose}}\ and\ \bibinfo {author} {\bibfnamefont {A.}~\bibnamefont {Tribedi}},\
  }\href {\doibase 10.1103/PhysRevA.72.022314} {\bibfield  {journal} {\bibinfo
  {journal} {Phys. Rev. A}\ }\textbf {\bibinfo {volume} {72}},\ \bibinfo
  {pages} {022314} (\bibinfo {year} {2005})}\BibitemShut {NoStop}%
\bibitem [{\citenamefont {Christensen}\ \emph {et~al.}(2007)\citenamefont
  {Christensen}, \citenamefont {Ronnow}, \citenamefont {McMorrow},
  \citenamefont {Harrison}, \citenamefont {Perring}, \citenamefont {Enderle},
  \citenamefont {Coldea}, \citenamefont {Regnault},\ and\ \citenamefont
  {Aeppli}}]{Christensen2007}%
  \BibitemOpen
  \bibfield  {author} {\bibinfo {author} {\bibfnamefont {N.}~\bibnamefont
  {Christensen}}, \bibinfo {author} {\bibfnamefont {H.}~\bibnamefont {Ronnow}},
  \bibinfo {author} {\bibfnamefont {D.}~\bibnamefont {McMorrow}}, \bibinfo
  {author} {\bibfnamefont {A.}~\bibnamefont {Harrison}}, \bibinfo {author}
  {\bibfnamefont {T.}~\bibnamefont {Perring}}, \bibinfo {author} {\bibfnamefont
  {M.}~\bibnamefont {Enderle}}, \bibinfo {author} {\bibfnamefont
  {R.}~\bibnamefont {Coldea}}, \bibinfo {author} {\bibfnamefont
  {L.}~\bibnamefont {Regnault}}, \ and\ \bibinfo {author} {\bibfnamefont
  {G.}~\bibnamefont {Aeppli}},\ }\href@noop {} {\bibfield  {journal} {\bibinfo
  {journal} {Proc.~Nat.~Acad.~Sci.}\ }\textbf {\bibinfo {volume} {104}},\
  \bibinfo {pages} {15264} (\bibinfo {year} {2007})}\BibitemShut {NoStop}%
\bibitem [{\citenamefont {Sahling}\ \emph {et~al.}(2015)\citenamefont
  {Sahling}, \citenamefont {Remenyi}, \citenamefont {Paulsen}, \citenamefont
  {Monceau}, \citenamefont {Saligrama}, \citenamefont {Marin}, \citenamefont
  {Revcolevschi}, \citenamefont {Regnault}, \citenamefont {Raymond},\ and\
  \citenamefont {Lorenzo}}]{Sahling2015}%
  \BibitemOpen
  \bibfield  {author} {\bibinfo {author} {\bibfnamefont {S.}~\bibnamefont
  {Sahling}}, \bibinfo {author} {\bibfnamefont {G.}~\bibnamefont {Remenyi}},
  \bibinfo {author} {\bibfnamefont {C.}~\bibnamefont {Paulsen}}, \bibinfo
  {author} {\bibfnamefont {P.}~\bibnamefont {Monceau}}, \bibinfo {author}
  {\bibfnamefont {V.}~\bibnamefont {Saligrama}}, \bibinfo {author}
  {\bibfnamefont {C.}~\bibnamefont {Marin}}, \bibinfo {author} {\bibfnamefont
  {A.}~\bibnamefont {Revcolevschi}}, \bibinfo {author} {\bibfnamefont {L.~P.}\
  \bibnamefont {Regnault}}, \bibinfo {author} {\bibfnamefont {S.}~\bibnamefont
  {Raymond}}, \ and\ \bibinfo {author} {\bibfnamefont {J.~E.}\ \bibnamefont
  {Lorenzo}},\ }\href {\doibase 10.1038/nphys3186} {\bibfield  {journal}
  {\bibinfo  {journal} {Nature Physics}\ }\textbf {\bibinfo {volume} {11}},\
  \bibinfo {pages} {255} (\bibinfo {year} {2015})}\BibitemShut {NoStop}%
\bibitem [{\citenamefont {Giorgi}(2009)}]{Giorgi2009}%
  \BibitemOpen
  \bibfield  {author} {\bibinfo {author} {\bibfnamefont {G.~L.}\ \bibnamefont
  {Giorgi}},\ }\href {\doibase 10.1103/PhysRevB.79.060405} {\bibfield
  {journal} {\bibinfo  {journal} {Phys. Rev. B}\ }\textbf {\bibinfo {volume}
  {79}},\ \bibinfo {pages} {060405(R)} (\bibinfo {year} {2009})},\ \bibinfo
  {note} {erratum: Ibid. {\bf 80}, 019901 (2009)}\BibitemShut {NoStop}%
\bibitem [{\citenamefont {Marty}\ \emph {et~al.}(2014)\citenamefont {Marty},
  \citenamefont {Epping}, \citenamefont {Kampermann}, \citenamefont {Bru\ss},
  \citenamefont {Plenio},\ and\ \citenamefont {Cramer}}]{Marty2014}%
  \BibitemOpen
  \bibfield  {author} {\bibinfo {author} {\bibfnamefont {O.}~\bibnamefont
  {Marty}}, \bibinfo {author} {\bibfnamefont {M.}~\bibnamefont {Epping}},
  \bibinfo {author} {\bibfnamefont {H.}~\bibnamefont {Kampermann}}, \bibinfo
  {author} {\bibfnamefont {D.}~\bibnamefont {Bru\ss}}, \bibinfo {author}
  {\bibfnamefont {M.~B.}\ \bibnamefont {Plenio}}, \ and\ \bibinfo {author}
  {\bibfnamefont {M.}~\bibnamefont {Cramer}},\ }\href@noop {} {\bibfield
  {journal} {\bibinfo  {journal} {Phys.~Rev.~B}\ }\textbf {\bibinfo {volume}
  {89}},\ \bibinfo {pages} {125117} (\bibinfo {year} {2014})}\BibitemShut
  {NoStop}%
\bibitem [{\citenamefont {Belik}\ \emph {et~al.}(2007)\citenamefont {Belik},
  \citenamefont {Koo}, \citenamefont {Whangbo}, \citenamefont {Tsujii},
  \citenamefont {Naumov},\ and\ \citenamefont
  {Takayama-Muromachi}}]{Belik2007}%
  \BibitemOpen
  \bibfield  {author} {\bibinfo {author} {\bibfnamefont {A.~A.}\ \bibnamefont
  {Belik}}, \bibinfo {author} {\bibfnamefont {H.-J.}\ \bibnamefont {Koo}},
  \bibinfo {author} {\bibfnamefont {M.-H.}\ \bibnamefont {Whangbo}}, \bibinfo
  {author} {\bibfnamefont {N.}~\bibnamefont {Tsujii}}, \bibinfo {author}
  {\bibfnamefont {P.}~\bibnamefont {Naumov}}, \ and\ \bibinfo {author}
  {\bibfnamefont {E.}~\bibnamefont {Takayama-Muromachi}},\ }\href {\doibase
  10.1021/ic7008418} {\bibfield  {journal} {\bibinfo  {journal} {Inorganic
  Chemistry}\ }\textbf {\bibinfo {volume} {46}},\ \bibinfo {pages} {8684}
  (\bibinfo {year} {2007})}\BibitemShut {NoStop}%
\bibitem [{\citenamefont {Engelhardt}\ \emph {et~al.}(2009)\citenamefont
  {Engelhardt}, \citenamefont {Martin}, \citenamefont {Prozorov}, \citenamefont
  {Luban}, \citenamefont {Timco},\ and\ \citenamefont
  {Winpenny}}]{Engelhardt2009}%
  \BibitemOpen
  \bibfield  {author} {\bibinfo {author} {\bibfnamefont {L.}~\bibnamefont
  {Engelhardt}}, \bibinfo {author} {\bibfnamefont {C.}~\bibnamefont {Martin}},
  \bibinfo {author} {\bibfnamefont {R.}~\bibnamefont {Prozorov}}, \bibinfo
  {author} {\bibfnamefont {M.}~\bibnamefont {Luban}}, \bibinfo {author}
  {\bibfnamefont {G.}~\bibnamefont {Timco}}, \ and\ \bibinfo {author}
  {\bibfnamefont {R.}~\bibnamefont {Winpenny}},\ }\href {\doibase
  10.1103/PhysRevB.79.014404} {\bibfield  {journal} {\bibinfo  {journal} {Phys.
  Rev. B}\ }\textbf {\bibinfo {volume} {79}},\ \bibinfo {pages} {014404}
  (\bibinfo {year} {2009})}\BibitemShut {NoStop}%
\bibitem [{\citenamefont {Baker}\ \emph
  {et~al.}(2012{\natexlab{a}})\citenamefont {Baker}, \citenamefont {Waldmann},
  \citenamefont {Piligkos}, \citenamefont {Bircher}, \citenamefont {Cador},
  \citenamefont {Carretta}, \citenamefont {Collison}, \citenamefont
  {Fernandez-Alonso}, \citenamefont {McInnes}, \citenamefont {Mutka},
  \citenamefont {Podlesnyak}, \citenamefont {Tuna}, \citenamefont {Ochsenbein},
  \citenamefont {Sessoli}, \citenamefont {Sieber}, \citenamefont {Timco},
  \citenamefont {Weihe}, \citenamefont {G\"{u}del},\ and\ \citenamefont
  {Winpenny}}]{Baker2012}%
  \BibitemOpen
  \bibfield  {author} {\bibinfo {author} {\bibfnamefont {M.~L.}\ \bibnamefont
  {Baker}}, \bibinfo {author} {\bibfnamefont {O.}~\bibnamefont {Waldmann}},
  \bibinfo {author} {\bibfnamefont {S.}~\bibnamefont {Piligkos}}, \bibinfo
  {author} {\bibfnamefont {R.}~\bibnamefont {Bircher}}, \bibinfo {author}
  {\bibfnamefont {O.}~\bibnamefont {Cador}}, \bibinfo {author} {\bibfnamefont
  {S.}~\bibnamefont {Carretta}}, \bibinfo {author} {\bibfnamefont
  {D.}~\bibnamefont {Collison}}, \bibinfo {author} {\bibfnamefont
  {F.}~\bibnamefont {Fernandez-Alonso}}, \bibinfo {author} {\bibfnamefont
  {E.~J.~L.}\ \bibnamefont {McInnes}}, \bibinfo {author} {\bibfnamefont
  {H.}~\bibnamefont {Mutka}}, \bibinfo {author} {\bibfnamefont
  {A.}~\bibnamefont {Podlesnyak}}, \bibinfo {author} {\bibfnamefont
  {F.}~\bibnamefont {Tuna}}, \bibinfo {author} {\bibfnamefont {S.}~\bibnamefont
  {Ochsenbein}}, \bibinfo {author} {\bibfnamefont {R.}~\bibnamefont {Sessoli}},
  \bibinfo {author} {\bibfnamefont {A.}~\bibnamefont {Sieber}}, \bibinfo
  {author} {\bibfnamefont {G.~A.}\ \bibnamefont {Timco}}, \bibinfo {author}
  {\bibfnamefont {H.~g.}\ \bibnamefont {Weihe}}, \bibinfo {author}
  {\bibfnamefont {H.~U.}\ \bibnamefont {G\"{u}del}}, \ and\ \bibinfo {author}
  {\bibfnamefont {R.~E.~P.}\ \bibnamefont {Winpenny}},\ }\href {\doibase
  10.1103/PhysRevB.86.064405} {\bibfield  {journal} {\bibinfo  {journal}
  {Phys.~Rev.~B}\ }\textbf {\bibinfo {volume} {86}},\ \bibinfo {pages} {064405}
  (\bibinfo {year} {2012}{\natexlab{a}})}\BibitemShut {NoStop}%
\bibitem [{\citenamefont {Baker}\ \emph
  {et~al.}(2012{\natexlab{b}})\citenamefont {Baker}, \citenamefont {Guidi},
  \citenamefont {Carretta}, \citenamefont {Ollivier}, \citenamefont {Mutka},
  \citenamefont {G\"udel}, \citenamefont {Timco}, \citenamefont {McInnes},
  \citenamefont {Amoretti}, \citenamefont {Winpenny},\ and\ \citenamefont
  {Santini}}]{Baker2012b}%
  \BibitemOpen
  \bibfield  {author} {\bibinfo {author} {\bibfnamefont {M.~L.}\ \bibnamefont
  {Baker}}, \bibinfo {author} {\bibfnamefont {T.}~\bibnamefont {Guidi}},
  \bibinfo {author} {\bibfnamefont {S.}~\bibnamefont {Carretta}}, \bibinfo
  {author} {\bibfnamefont {J.}~\bibnamefont {Ollivier}}, \bibinfo {author}
  {\bibfnamefont {H.}~\bibnamefont {Mutka}}, \bibinfo {author} {\bibfnamefont
  {H.~U.}\ \bibnamefont {G\"udel}}, \bibinfo {author} {\bibfnamefont {G.~A.}\
  \bibnamefont {Timco}}, \bibinfo {author} {\bibfnamefont {E.~J.~L.}\
  \bibnamefont {McInnes}}, \bibinfo {author} {\bibfnamefont {G.}~\bibnamefont
  {Amoretti}}, \bibinfo {author} {\bibfnamefont {R.~E.~P.}\ \bibnamefont
  {Winpenny}}, \ and\ \bibinfo {author} {\bibfnamefont {P.}~\bibnamefont
  {Santini}},\ }\href {\doibase 10.1038/nphys2431} {\bibfield  {journal}
  {\bibinfo  {journal} {Nature Physics}\ }\textbf {\bibinfo {volume} {8}},\
  \bibinfo {pages} {906} (\bibinfo {year} {2012}{\natexlab{b}})}\BibitemShut
  {NoStop}%
\bibitem [{\citenamefont {Furrer}\ and\ \citenamefont
  {Waldmann}(2013)}]{Furrer2013}%
  \BibitemOpen
  \bibfield  {author} {\bibinfo {author} {\bibfnamefont {A.}~\bibnamefont
  {Furrer}}\ and\ \bibinfo {author} {\bibfnamefont {O.}~\bibnamefont
  {Waldmann}},\ }\href {\doibase 10.1103/RevModPhys.85.367} {\bibfield
  {journal} {\bibinfo  {journal} {Rev. Mod. Phys.}\ }\textbf {\bibinfo {volume}
  {85}},\ \bibinfo {pages} {367} (\bibinfo {year} {2013})}\BibitemShut
  {NoStop}%
\bibitem [{\citenamefont {Timco}\ \emph {et~al.}(2013)\citenamefont {Timco},
  \citenamefont {McInnes},\ and\ \citenamefont {Winpenny}}]{Timco2013}%
  \BibitemOpen
  \bibfield  {author} {\bibinfo {author} {\bibfnamefont {G.~A.}\ \bibnamefont
  {Timco}}, \bibinfo {author} {\bibfnamefont {E.~J.~L.}\ \bibnamefont
  {McInnes}}, \ and\ \bibinfo {author} {\bibfnamefont {R.~E.~P.}\ \bibnamefont
  {Winpenny}},\ }\href {\doibase 10.1039/C2CS35232J} {\bibfield  {journal}
  {\bibinfo  {journal} {Chem. Soc. Rev.}\ }\textbf {\bibinfo {volume} {42}},\
  \bibinfo {pages} {1796} (\bibinfo {year} {2013})}\BibitemShut {NoStop}%
\bibitem [{\citenamefont {Khajetoorians}\ \emph {et~al.}(2012)\citenamefont
  {Khajetoorians}, \citenamefont {Wiebe}, \citenamefont {Chilian},
  \citenamefont {Lounis}, \citenamefont {Blugel},\ and\ \citenamefont
  {Wiesendanger}}]{Wiesendanger2012}%
  \BibitemOpen
  \bibfield  {author} {\bibinfo {author} {\bibfnamefont {A.~A.}\ \bibnamefont
  {Khajetoorians}}, \bibinfo {author} {\bibfnamefont {J.}~\bibnamefont
  {Wiebe}}, \bibinfo {author} {\bibfnamefont {B.}~\bibnamefont {Chilian}},
  \bibinfo {author} {\bibfnamefont {S.}~\bibnamefont {Lounis}}, \bibinfo
  {author} {\bibfnamefont {S.}~\bibnamefont {Blugel}}, \ and\ \bibinfo {author}
  {\bibfnamefont {R.}~\bibnamefont {Wiesendanger}},\ }\href {\doibase
  10.1038/nphys2299} {\bibfield  {journal} {\bibinfo  {journal} {Nat Phys}\
  }\textbf {\bibinfo {volume} {8}},\ \bibinfo {pages} {497} (\bibinfo {year}
  {2012})}\BibitemShut {NoStop}%
\bibitem [{\citenamefont {Heinrich}\ \emph {et~al.}(2013)\citenamefont
  {Heinrich}, \citenamefont {Braun}, \citenamefont {Pascual},\ and\
  \citenamefont {Franke}}]{Heinrich2013}%
  \BibitemOpen
  \bibfield  {author} {\bibinfo {author} {\bibfnamefont {B.~W.}\ \bibnamefont
  {Heinrich}}, \bibinfo {author} {\bibfnamefont {L.}~\bibnamefont {Braun}},
  \bibinfo {author} {\bibfnamefont {J.~I.}\ \bibnamefont {Pascual}}, \ and\
  \bibinfo {author} {\bibfnamefont {K.~J.}\ \bibnamefont {Franke}},\ }\href
  {http://dx.doi.org/10.1038/nphys2794} {\bibfield  {journal} {\bibinfo
  {journal} {Nat Phys}\ }\textbf {\bibinfo {volume} {9}},\ \bibinfo {pages}
  {765} (\bibinfo {year} {2013})}\BibitemShut {NoStop}%
\bibitem [{\citenamefont {Feldman}\ \emph {et~al.}(2017)\citenamefont
  {Feldman}, \citenamefont {Randeria}, \citenamefont {Li}, \citenamefont
  {Jeon}, \citenamefont {Xie}, \citenamefont {Wang}, \citenamefont {Drozdov},
  \citenamefont {Andrei~Bernevig},\ and\ \citenamefont
  {Yazdani}}]{Feldman2017}%
  \BibitemOpen
  \bibfield  {author} {\bibinfo {author} {\bibfnamefont {B.~E.}\ \bibnamefont
  {Feldman}}, \bibinfo {author} {\bibfnamefont {M.~T.}\ \bibnamefont
  {Randeria}}, \bibinfo {author} {\bibfnamefont {J.}~\bibnamefont {Li}},
  \bibinfo {author} {\bibfnamefont {S.}~\bibnamefont {Jeon}}, \bibinfo {author}
  {\bibfnamefont {Y.}~\bibnamefont {Xie}}, \bibinfo {author} {\bibfnamefont
  {Z.}~\bibnamefont {Wang}}, \bibinfo {author} {\bibfnamefont {I.~K.}\
  \bibnamefont {Drozdov}}, \bibinfo {author} {\bibfnamefont {B.}~\bibnamefont
  {Andrei~Bernevig}}, \ and\ \bibinfo {author} {\bibfnamefont {A.}~\bibnamefont
  {Yazdani}},\ }\href {http://dx.doi.org/10.1038/nphys3947} {\bibfield
  {journal} {\bibinfo  {journal} {Nat Phys}\ }\textbf {\bibinfo {volume}
  {13}},\ \bibinfo {pages} {286} (\bibinfo {year} {2017})}\BibitemShut
  {NoStop}%
\bibitem [{\citenamefont {Kim}\ \emph {et~al.}(2010)\citenamefont {Kim},
  \citenamefont {Chang}, \citenamefont {Korenblit}, \citenamefont {Islam},
  \citenamefont {Edwards}, \citenamefont {Freericks}, \citenamefont {Lin},
  \citenamefont {Duan},\ and\ \citenamefont {Monroe}}]{Kim2010}%
  \BibitemOpen
  \bibfield  {author} {\bibinfo {author} {\bibfnamefont {K.}~\bibnamefont
  {Kim}}, \bibinfo {author} {\bibfnamefont {M.-S.}\ \bibnamefont {Chang}},
  \bibinfo {author} {\bibfnamefont {S.}~\bibnamefont {Korenblit}}, \bibinfo
  {author} {\bibfnamefont {R.}~\bibnamefont {Islam}}, \bibinfo {author}
  {\bibfnamefont {E.~E.}\ \bibnamefont {Edwards}}, \bibinfo {author}
  {\bibfnamefont {J.~K.}\ \bibnamefont {Freericks}}, \bibinfo {author}
  {\bibfnamefont {G.-D.}\ \bibnamefont {Lin}}, \bibinfo {author} {\bibfnamefont
  {L.-M.}\ \bibnamefont {Duan}}, \ and\ \bibinfo {author} {\bibfnamefont
  {C.}~\bibnamefont {Monroe}},\ }\href {\doibase 10.1038/nature09071}
  {\bibfield  {journal} {\bibinfo  {journal} {Nature}\ }\textbf {\bibinfo
  {volume} {465}},\ \bibinfo {pages} {590} (\bibinfo {year}
  {2010})}\BibitemShut {NoStop}%
\bibitem [{\citenamefont {Simon}\ \emph {et~al.}(2011)\citenamefont {Simon},
  \citenamefont {Bakr}, \citenamefont {Ma}, \citenamefont {Tai}, \citenamefont
  {Preiss},\ and\ \citenamefont {Greiner}}]{Simon2011}%
  \BibitemOpen
  \bibfield  {author} {\bibinfo {author} {\bibfnamefont {J.}~\bibnamefont
  {Simon}}, \bibinfo {author} {\bibfnamefont {W.~S.}\ \bibnamefont {Bakr}},
  \bibinfo {author} {\bibfnamefont {R.}~\bibnamefont {Ma}}, \bibinfo {author}
  {\bibfnamefont {M.~E.}\ \bibnamefont {Tai}}, \bibinfo {author} {\bibfnamefont
  {P.~M.}\ \bibnamefont {Preiss}}, \ and\ \bibinfo {author} {\bibfnamefont
  {M.}~\bibnamefont {Greiner}},\ }\href {\doibase 10.1038/nature09994}
  {\bibfield  {journal} {\bibinfo  {journal} {Nature}\ }\textbf {\bibinfo
  {volume} {472}},\ \bibinfo {pages} {307} (\bibinfo {year}
  {2011})}\BibitemShut {NoStop}%
\bibitem [{\citenamefont {Campbell}\ \emph {et~al.}(2013)\citenamefont
  {Campbell}, \citenamefont {Richens}, \citenamefont {Gullo},\ and\
  \citenamefont {Busch}}]{Campbell2013}%
  \BibitemOpen
  \bibfield  {author} {\bibinfo {author} {\bibfnamefont {S.}~\bibnamefont
  {Campbell}}, \bibinfo {author} {\bibfnamefont {J.}~\bibnamefont {Richens}},
  \bibinfo {author} {\bibfnamefont {N.~L.}\ \bibnamefont {Gullo}}, \ and\
  \bibinfo {author} {\bibfnamefont {T.}~\bibnamefont {Busch}},\ }\href@noop {}
  {\bibfield  {journal} {\bibinfo  {journal} {Phys. Rev. A}\ }\textbf {\bibinfo
  {volume} {88}},\ \bibinfo {pages} {062305} (\bibinfo {year}
  {2013})}\BibitemShut {NoStop}%
\bibitem [{\citenamefont {Barouch}\ and\ \citenamefont
  {McCoy}(1971)}]{Barouch1971}%
  \BibitemOpen
  \bibfield  {author} {\bibinfo {author} {\bibfnamefont {E.}~\bibnamefont
  {Barouch}}\ and\ \bibinfo {author} {\bibfnamefont {B.~M.}\ \bibnamefont
  {McCoy}},\ }\href {\doibase 10.1103/PhysRevA.3.786} {\bibfield  {journal}
  {\bibinfo  {journal} {Phys. Rev. A}\ }\textbf {\bibinfo {volume} {3}},\
  \bibinfo {pages} {786} (\bibinfo {year} {1971})}\BibitemShut {NoStop}%
\bibitem [{\citenamefont {Tomasello}\ \emph {et~al.}(2011)\citenamefont
  {Tomasello}, \citenamefont {Rossini}, \citenamefont {Hamma},\ and\
  \citenamefont {Amico}}]{Tomasello2011}%
  \BibitemOpen
  \bibfield  {author} {\bibinfo {author} {\bibfnamefont {B.}~\bibnamefont
  {Tomasello}}, \bibinfo {author} {\bibfnamefont {D.}~\bibnamefont {Rossini}},
  \bibinfo {author} {\bibfnamefont {A.}~\bibnamefont {Hamma}}, \ and\ \bibinfo
  {author} {\bibfnamefont {L.}~\bibnamefont {Amico}},\ }\href
  {http://stacks.iop.org/0295-5075/96/i=2/a=27002} {\bibfield  {journal}
  {\bibinfo  {journal} {EPL (Europhysics Letters)}\ }\textbf {\bibinfo {volume}
  {96}},\ \bibinfo {pages} {27002} (\bibinfo {year} {2011})}\BibitemShut
  {NoStop}%
\bibitem [{\citenamefont {Asoudeh}\ \emph {et~al.}(2007)\citenamefont
  {Asoudeh}, \citenamefont {Karimipour},\ and\ \citenamefont
  {Sadrolashrafi}}]{Asoudeh2007}%
  \BibitemOpen
  \bibfield  {author} {\bibinfo {author} {\bibfnamefont {M.}~\bibnamefont
  {Asoudeh}}, \bibinfo {author} {\bibfnamefont {V.}~\bibnamefont {Karimipour}},
  \ and\ \bibinfo {author} {\bibfnamefont {A.}~\bibnamefont {Sadrolashrafi}},\
  }\href@noop {} {\bibfield  {journal} {\bibinfo  {journal} {Phys.~Rev.~B}\
  }\textbf {\bibinfo {volume} {76}},\ \bibinfo {pages} {25} (\bibinfo {year}
  {2007})}\BibitemShut {NoStop}%
\bibitem [{\citenamefont {Paulinelli}\ \emph {et~al.}(2013)\citenamefont
  {Paulinelli}, \citenamefont {de~Souza},\ and\ \citenamefont
  {Rojas}}]{Paulinelli2013}%
  \BibitemOpen
  \bibfield  {author} {\bibinfo {author} {\bibfnamefont {H.~G.}\ \bibnamefont
  {Paulinelli}}, \bibinfo {author} {\bibfnamefont {S.~M.}\ \bibnamefont
  {de~Souza}}, \ and\ \bibinfo {author} {\bibfnamefont {O.}~\bibnamefont
  {Rojas}},\ }\href@noop {} {\bibfield  {journal} {\bibinfo  {journal} {J.
  Phys.: Condens. Matt.}\ }\textbf {\bibinfo {volume} {25}},\ \bibinfo {pages}
  {306003} (\bibinfo {year} {2013})}\BibitemShut {NoStop}%
\bibitem [{\citenamefont {Hou}\ \emph {et~al.}(2005)\citenamefont {Hou},
  \citenamefont {Chen},\ and\ \citenamefont {Hu}}]{Hou2005}%
  \BibitemOpen
  \bibfield  {author} {\bibinfo {author} {\bibfnamefont {X.~W.}\ \bibnamefont
  {Hou}}, \bibinfo {author} {\bibfnamefont {J.~H.}\ \bibnamefont {Chen}}, \
  and\ \bibinfo {author} {\bibfnamefont {B.}~\bibnamefont {Hu}},\ }\href
  {\doibase 10.1103/PhysRevA.71.034302} {\bibfield  {journal} {\bibinfo
  {journal} {Phys.~Rev.~A}\ }\textbf {\bibinfo {volume} {71}},\ \bibinfo
  {pages} {034302} (\bibinfo {year} {2005})}\BibitemShut {NoStop}%
\bibitem [{\citenamefont {Hong}\ \emph {et~al.}(2011)\citenamefont {Hong},
  \citenamefont {Gvasaliya}, \citenamefont {Herringer}, \citenamefont
  {Turnbull}, \citenamefont {Landee}, \citenamefont {Regnault}, \citenamefont
  {Boehm},\ and\ \citenamefont {Zheludev}}]{AndreZheludev}%
  \BibitemOpen
  \bibfield  {author} {\bibinfo {author} {\bibfnamefont {T.}~\bibnamefont
  {Hong}}, \bibinfo {author} {\bibfnamefont {S.~N.}\ \bibnamefont {Gvasaliya}},
  \bibinfo {author} {\bibfnamefont {S.}~\bibnamefont {Herringer}}, \bibinfo
  {author} {\bibfnamefont {M.~M.}\ \bibnamefont {Turnbull}}, \bibinfo {author}
  {\bibfnamefont {C.~P.}\ \bibnamefont {Landee}}, \bibinfo {author}
  {\bibfnamefont {L.-P.}\ \bibnamefont {Regnault}}, \bibinfo {author}
  {\bibfnamefont {M.}~\bibnamefont {Boehm}}, \ and\ \bibinfo {author}
  {\bibfnamefont {A.}~\bibnamefont {Zheludev}},\ }\href {\doibase
  10.1103/PhysRevB.83.052401} {\bibfield  {journal} {\bibinfo  {journal} {Phys.
  Rev. B}\ }\textbf {\bibinfo {volume} {83}},\ \bibinfo {pages} {052401}
  (\bibinfo {year} {2011})}\BibitemShut {NoStop}%
\bibitem [{\citenamefont {Giamarchi}(2008)}]{Giamarchi2008}%
  \BibitemOpen
  \bibfield  {author} {\bibinfo {author} {\bibfnamefont {T.}~\bibnamefont
  {Giamarchi}},\ }\href
  {http://www.nature.com/nphys/journal/vaop/ncurrent/full/nphys893.html}
  {\bibfield  {journal} {\bibinfo  {journal} {Nature Physics}\ }\textbf
  {\bibinfo {volume} {4}},\ \bibinfo {pages} {198} (\bibinfo {year}
  {2008})}\BibitemShut {NoStop}%
\bibitem [{\citenamefont {B\"arwinkel}\ \emph {et~al.}(2000)\citenamefont
  {B\"arwinkel}, \citenamefont {Schmidt},\ and\ \citenamefont
  {Schnack}}]{Barwinkel2000}%
  \BibitemOpen
  \bibfield  {author} {\bibinfo {author} {\bibfnamefont {K.}~\bibnamefont
  {B\"arwinkel}}, \bibinfo {author} {\bibfnamefont {H.-J.}\ \bibnamefont
  {Schmidt}}, \ and\ \bibinfo {author} {\bibfnamefont {J.}~\bibnamefont
  {Schnack}},\ }\href {\doibase
  http://dx.doi.org/10.1016/S0304-8853(00)00481-9} {\bibfield  {journal}
  {\bibinfo  {journal} {Journal of Magnetism and Magnetic Materials}\ }\textbf
  {\bibinfo {volume} {220}},\ \bibinfo {pages} {227 } (\bibinfo {year}
  {2000})}\BibitemShut {NoStop}%
\bibitem [{\citenamefont {B\"arwinkel}\ \emph {et~al.}(2003)\citenamefont
  {B\"arwinkel}, \citenamefont {Hage}, \citenamefont {Schmidt},\ and\
  \citenamefont {Schnack}}]{Barwinkel2003}%
  \BibitemOpen
  \bibfield  {author} {\bibinfo {author} {\bibfnamefont {K.}~\bibnamefont
  {B\"arwinkel}}, \bibinfo {author} {\bibfnamefont {P.}~\bibnamefont {Hage}},
  \bibinfo {author} {\bibfnamefont {H.-J.}\ \bibnamefont {Schmidt}}, \ and\
  \bibinfo {author} {\bibfnamefont {J.}~\bibnamefont {Schnack}},\ }\href
  {\doibase 10.1103/PhysRevB.68.054422} {\bibfield  {journal} {\bibinfo
  {journal} {Phys. Rev. B}\ }\textbf {\bibinfo {volume} {68}},\ \bibinfo
  {pages} {054422} (\bibinfo {year} {2003})}\BibitemShut {NoStop}%
\bibitem [{\citenamefont {De~Pasquale}\ and\ \citenamefont
  {Facchi}(2009)}]{DePasquale2009}%
  \BibitemOpen
  \bibfield  {author} {\bibinfo {author} {\bibfnamefont {A.}~\bibnamefont
  {De~Pasquale}}\ and\ \bibinfo {author} {\bibfnamefont {P.}~\bibnamefont
  {Facchi}},\ }\href {\doibase 10.1103/PhysRevA.80.032102} {\bibfield
  {journal} {\bibinfo  {journal} {Phys. Rev. A}\ }\textbf {\bibinfo {volume}
  {80}},\ \bibinfo {pages} {032102} (\bibinfo {year} {2009})}\BibitemShut
  {NoStop}%
\bibitem [{\citenamefont {Affelck}(1989)}]{Affleck1989}%
  \BibitemOpen
  \bibfield  {author} {\bibinfo {author} {\bibfnamefont {I.}~\bibnamefont
  {Affelck}},\ }\href@noop {} {\bibfield  {journal} {\bibinfo  {journal}
  {Journal of Physics: Condensed Matter}\ }\textbf {\bibinfo {volume} {1}},\
  \bibinfo {pages} {3047} (\bibinfo {year} {1989})}\BibitemShut {NoStop}%
\bibitem [{\citenamefont {Waldmann}(2001)}]{Waldmann2001}%
  \BibitemOpen
  \bibfield  {author} {\bibinfo {author} {\bibfnamefont {O.}~\bibnamefont
  {Waldmann}},\ }\href {\doibase 10.1103/PhysRevB.65.024424} {\bibfield
  {journal} {\bibinfo  {journal} {Phys. Rev. B}\ }\textbf {\bibinfo {volume}
  {65}},\ \bibinfo {pages} {024424} (\bibinfo {year} {2001})}\BibitemShut
  {NoStop}%
\bibitem [{\citenamefont {Cheng}\ \emph {et~al.}(2010)\citenamefont {Cheng},
  \citenamefont {Shan}, \citenamefont {Huang}, \citenamefont {Liu},\ and\
  \citenamefont {Li}}]{Cheng2010}%
  \BibitemOpen
  \bibfield  {author} {\bibinfo {author} {\bibfnamefont {W.~W.}\ \bibnamefont
  {Cheng}}, \bibinfo {author} {\bibfnamefont {C.~J.}\ \bibnamefont {Shan}},
  \bibinfo {author} {\bibfnamefont {Y.~X.}\ \bibnamefont {Huang}}, \bibinfo
  {author} {\bibfnamefont {T.~K.}\ \bibnamefont {Liu}}, \ and\ \bibinfo
  {author} {\bibfnamefont {H.}~\bibnamefont {Li}},\ }\href {\doibase
  10.1016/j.physe.2010.07.012} {\bibfield  {journal} {\bibinfo  {journal}
  {Physica E: Low-Dimensional Systems and Nanostructures}\ }\textbf {\bibinfo
  {volume} {43}},\ \bibinfo {pages} {235} (\bibinfo {year} {2010})}\BibitemShut
  {NoStop}%
\bibitem [{\citenamefont {Siloi}\ and\ \citenamefont
  {Toriani}(2012)}]{Siloi2012}%
  \BibitemOpen
  \bibfield  {author} {\bibinfo {author} {\bibfnamefont {I.}~\bibnamefont
  {Siloi}}\ and\ \bibinfo {author} {\bibfnamefont {F.}~\bibnamefont
  {Toriani}},\ }\href@noop {} {\bibfield  {journal} {\bibinfo  {journal} {Phys.
  Rev. B}\ }\textbf {\bibinfo {volume} {86}},\ \bibinfo {pages} {224404}
  (\bibinfo {year} {2012})}\BibitemShut {NoStop}%
\bibitem [{\citenamefont {Christou}\ \emph {et~al.}(2000)\citenamefont
  {Christou}, \citenamefont {Gatteschi}, \citenamefont {Hendrickson},\ and\
  \citenamefont {Sessoli}}]{Christou2000}%
  \BibitemOpen
  \bibfield  {author} {\bibinfo {author} {\bibfnamefont {G.}~\bibnamefont
  {Christou}}, \bibinfo {author} {\bibfnamefont {D.}~\bibnamefont {Gatteschi}},
  \bibinfo {author} {\bibfnamefont {D.~N.}\ \bibnamefont {Hendrickson}}, \ and\
  \bibinfo {author} {\bibfnamefont {R.}~\bibnamefont {Sessoli}},\ }\href
  {http://journals.cambridge.org/abstract_S0883769400027925} {\bibfield
  {journal} {\bibinfo  {journal} {MRS Bulletin}\ }\textbf {\bibinfo {volume}
  {25}},\ \bibinfo {pages} {66} (\bibinfo {year} {2000})}\BibitemShut {NoStop}%
\bibitem [{\citenamefont {Edwards}\ and\ \citenamefont
  {Thouless}(1972)}]{Edwards1972}%
  \BibitemOpen
  \bibfield  {author} {\bibinfo {author} {\bibfnamefont {J.~T.}\ \bibnamefont
  {Edwards}}\ and\ \bibinfo {author} {\bibfnamefont {D.~J.}\ \bibnamefont
  {Thouless}},\ }\href {http://iopscience.iop.org/0022-3719/5/8/007} {\bibfield
   {journal} {\bibinfo  {journal} {J. Phys. C: Solid State Physics}\ }\textbf
  {\bibinfo {volume} {5}},\ \bibinfo {pages} {807} (\bibinfo {year}
  {1972})}\BibitemShut {NoStop}%
\bibitem [{\citenamefont {Barouch}\ \emph {et~al.}(1970)\citenamefont
  {Barouch}, \citenamefont {McCoy},\ and\ \citenamefont
  {Dresden}}]{Barouch1970}%
  \BibitemOpen
  \bibfield  {author} {\bibinfo {author} {\bibfnamefont {E.}~\bibnamefont
  {Barouch}}, \bibinfo {author} {\bibfnamefont {B.~M.}\ \bibnamefont {McCoy}},
  \ and\ \bibinfo {author} {\bibfnamefont {M.}~\bibnamefont {Dresden}},\ }\href
  {\doibase 10.1103/PhysRevA.2.1075} {\bibfield  {journal} {\bibinfo  {journal}
  {Phys. Rev. A}\ }\textbf {\bibinfo {volume} {2}},\ \bibinfo {pages} {1075}
  (\bibinfo {year} {1970})}\BibitemShut {NoStop}%
\bibitem [{\citenamefont {Fisher}\ and\ \citenamefont
  {Barber}(1972)}]{Fisher1972}%
  \BibitemOpen
  \bibfield  {author} {\bibinfo {author} {\bibfnamefont {M.~E.}\ \bibnamefont
  {Fisher}}\ and\ \bibinfo {author} {\bibfnamefont {M.~N.}\ \bibnamefont
  {Barber}},\ }\href@noop {} {\bibfield  {journal} {\bibinfo  {journal} {Phys.
  Rev. Lett.}\ }\textbf {\bibinfo {volume} {28}},\ \bibinfo {pages} {1516}
  (\bibinfo {year} {1972})}\BibitemShut {NoStop}%
\bibitem [{\citenamefont {Huang}(2014)}]{Huang2014}%
  \BibitemOpen
  \bibfield  {author} {\bibinfo {author} {\bibfnamefont {Y.}~\bibnamefont
  {Huang}},\ }\href {\doibase 10.1103/PhysRevB.89.054410} {\bibfield  {journal}
  {\bibinfo  {journal} {Phys. Rev. B}\ }\textbf {\bibinfo {volume} {89}},\
  \bibinfo {pages} {054410} (\bibinfo {year} {2014})}\BibitemShut {NoStop}%
\bibitem [{\citenamefont {Baroni}\ \emph {et~al.}(2007)\citenamefont {Baroni},
  \citenamefont {Fubini}, \citenamefont {Tognetti},\ and\ \citenamefont
  {Verrucchi}}]{Baroni2007}%
  \BibitemOpen
  \bibfield  {author} {\bibinfo {author} {\bibfnamefont {F.}~\bibnamefont
  {Baroni}}, \bibinfo {author} {\bibfnamefont {A.}~\bibnamefont {Fubini}},
  \bibinfo {author} {\bibfnamefont {V.}~\bibnamefont {Tognetti}}, \ and\
  \bibinfo {author} {\bibfnamefont {P.}~\bibnamefont {Verrucchi}},\ }\href
  {\doibase 10.1088/1751-8113/40/32/010} {\bibfield  {journal} {\bibinfo
  {journal} {Journal of Physics A: Mathematical and Theoretical}\ }\textbf
  {\bibinfo {volume} {40}},\ \bibinfo {pages} {9845} (\bibinfo {year}
  {2007})}\BibitemShut {NoStop}%
\bibitem [{\citenamefont {Lovesey}(1987)}]{Lovesey1987b}%
  \BibitemOpen
  \bibfield  {author} {\bibinfo {author} {\bibfnamefont {S.~W.}\ \bibnamefont
  {Lovesey}},\ }\href@noop {} {\emph {\bibinfo {title} {{Theory of Neutron
  Scattering from Condensed Matter, Vol. 2: Polarization Effects and Magnetic
  Scattering}}}}\ (\bibinfo  {publisher} {Oxford University Press},\ \bibinfo
  {year} {1987})\BibitemShut {NoStop}%
\end{thebibliography}
%
%%%%%%%%%%%%%%%%%%%%%%%%%%%%%%%%%%%%%%%%%%%%%%%%%%%%%%%%%%%%%%%%%%%%%%%%

%%%%%%%%%%%%%%%%%%%%%%%%%%%%%%%%%%%%%%%%%%%%%%%%%%%%%%%%%%%%%%%%%%%%%%%%
%
% .bbl OUTPUT. COMMENT OUT TO USE .bib FILE INSTEAD.

%
%%%%%%%%%%%%%%%%%%%%%%%%%%%%%%%%%%%%%%%%%%%%%%%%%%%%%%%%%%%%%%%%%%%%%%%%

\clearpage{}

%\part*{Supplemental material}

\end{document}